\documentclass[12pt]{article}
\usepackage{jheppub}

\usepackage{graphicx}
\usepackage{tikz,varwidth}
\usetikzlibrary{shapes,snakes}
\usetikzlibrary{arrows, decorations.markings}
\usetikzlibrary{matrix,decorations.pathreplacing,calc}
\usetikzlibrary{decorations.text,calc,arrows.meta}

\def\be{\begin{equation}}
\def\ee{\end{equation}}
\def\ba{\begin{eqnarray}}
\def\ea{\end{eqnarray}}

\def\e{\epsilon}

\def\C{\Chi}

\def\IR{\relax{\rm I\kern-.18em R}}

\def\inv{^{\raise.0ex\hbox{${\scriptscriptstyle -}$}\kern-.05em 1}}

\title{Holographic Aspects of Four Dimensional ${\cal N }=2$   SCFTs and their Marginal Deformations}

\author{Carlos N\'u\~nez$^1$,}
\author{Dibakar Roychowdhury$^2$,}
\author{Stefano Speziali$^1$ and}
 \author{Salom\'on Zacar\'{\i}as$^3$}

\affiliation{$^1$ Department of Physics, Swansea University, Swansea SA2 8PP, United Kingdom.}
\affiliation{$^2$ Department of Physics, Indian Institute of Technology Roorkee, Roorkee 247667 Uttaranchal, INDIA.}
\affiliation{$^3$ Shanghai Center for Complex Physics, Department of Physics and Astronomy,
Shanghai JiaoTong University, Shanghai 200240, China.}
\emailAdd{c.nunez@swansea.ac.uk} 
\emailAdd{dibakarphys@gmail.com} 
\emailAdd{stefano.speziali6@gmail.com} 
\emailAdd{szacarias@sjtu.edu.cn} 

\abstract{we study the holographic description of ${\cal N}=2$ Super Conformal Field Theories in four dimensions first given by Gaiotto and Maldacena. We present new expressions that holographically calculate characteristic numbers of the CFT and associated Hanany-Witten set-ups, or more dynamical observables, like the central charge. A number of examples of varying complexity are studied and some proofs for these new expressions are presented. We repeat this treatment for the case of the marginally deformed Gaiotto-Maldacena theories, presenting an infinite family of new solutions and compute some of its observables.  These new backgrounds rely on the solution of a Laplace equation and a  boundary condition, encoding the kinematics of the original conformal field theory.}


\begin{document}
\def\Tr{{\textrm{Tr}}}
\def\be{\begin{equation}}
\def\e{\end{equation}}
\def\bea{\begin{equation*}}
\def\ea{\end{equation*}}
\def\la{\label}
\def\bu{\bullet}
\maketitle 
\newpage



\section{Introduction and general idea of this paper}
In this work, we study holographic aspects of ${\cal N}=2$ and ${\cal N}=1$ Super Conformal Field Theories (SCFTs) in four dimensions. This is a very well explored topic from the SCFTs perspective and there was major progress on it in the last twenty years. In recent years, the work of Gaiotto \cite{Gaiotto:2009we} increased considerably the number of ${\cal N}=2$ SCFTs and the study of these systems gained a dominant position among the community's interests.

Our goal in this paper is to use the very extensive body of knowledge obtained with field theoretical tools and translate it into the language of holography 
\cite{Maldacena:1997re}, first presented in the work of Gaiotto and Maldacena \cite{Gaiotto:2009gz}.
Having both languages at our disposal is important as the calculation of various observables (correlation functions) may be more feasible to be done using the holographic approach.
Hence, having this mapping between descriptions clearly lay-out is both important and necessary. The main objective of this work is to start to explore this mapping or correspondence.

We shall do so for the case of ${\cal N}=2$  SCFTs in four dimensions and some of their marginal deformations. A very interesting project would be to extend the developments in this work to conformal field theories in different dimensions.

One possible way the reader may become interested on these holographic elaborations is by the study of non-Abelian T-duality, see for example \cite{Lozano:2016kum}. In fact, non-Abelian T-duality and other integrable deformations of the sigma model for the string theory on a given background, change the sigma model on $AdS_5\times S^5$ into one on a ${\cal N}=2$ preserving space-time  \cite{Sfetsos:2010uq}, that must belong to the class of backgrounds presented in  \cite{Gaiotto:2009gz}. The study of these backgrounds from the viewpoint of holography contributes to the field theoretical understanding of non-Abelian T-duality and other integrable deformations.

This paper and its contents are organised as follows: in Part 1, consisting of Sections \ref{GMgenerics}-\ref{sectionexamples}, we discuss the holographic aspects of ${\cal N}=2$ SCFTs in four dimensions. The starting point is the work of Gaiotto and Maldacena on which we elaborate. We shall present new solutions of a Laplace-like equation and a careful study of such solutions. We present compact expressions that calculate the charges, number of branes composing the associated Hanany-Witten set-up, a new formula for the linking numbers of these branes and central charge of the SCFTs, all of these in terms of the function that specify the boundary conditions for the Laplace equation defining the dynamics of the system.
We exemplify our new expressions using different field theories. In the appendixes we provide proofs of our expressions or more elaborated examples for the reader wishing to work on the topic.
We then present a field theoretical picture of the action of non-Abelian T-duality on $AdS_5\times S^5$--the Sfetsos-Thompson background \cite{Sfetsos:2010uq}, and extend this analysis to another particular solution.
\\
The Part 2, consists on a very extended and  dense Section \ref{sectionN=1}, we study the effect of applying a marginal deformation to the ${\cal N}=2$ SCFTs discussed above.
The approach is again of holographic nature. We present a proposal for the dual CFTs, the deformation that is acting and an  infinite family of new supergravity backgrounds.  The existence of these backgrounds rely only on a solution to a Laplace equation with a given boundary condition. We finally conclude indicating future research lines in Section \ref{conclusection}.
\\
The paper is complemented  by many  very detailed appendixes that  work-out technically elaborated examples, show explicit steps in the construction of new backgrounds, present explicit new solutions and  discuss the proofs and  workings of our new expressions for the CFT observables mentioned above.

\section{Part 1: ${\cal N}=2$ SCFTs and their dual backgrounds}\label{GMgenerics}
Let us summarise some aspects of the ${\cal N}=2$ field theories that occupy our attention in the subsequent sections.

The study of the strong ${\cal N}=2$ dynamics received an important push forward
with the work of Seiberg and Witten \cite{Seiberg:1994rs}. The
'Seiberg-Witten curve' (defined by a relation between two complex variables) encodes important information about the field theory. Some field theoretical results can also be  obtained
using  Hanany-Witten set-ups  \cite{Hanany:1996ie}. In the case at hand (${\cal N}=2 $ four-dimensional field theories), the set-up consists of D4, NS5 and D6 branes.

\begin{figure}[h!]
    \centering
    {{\includegraphics[width=12.5cm]{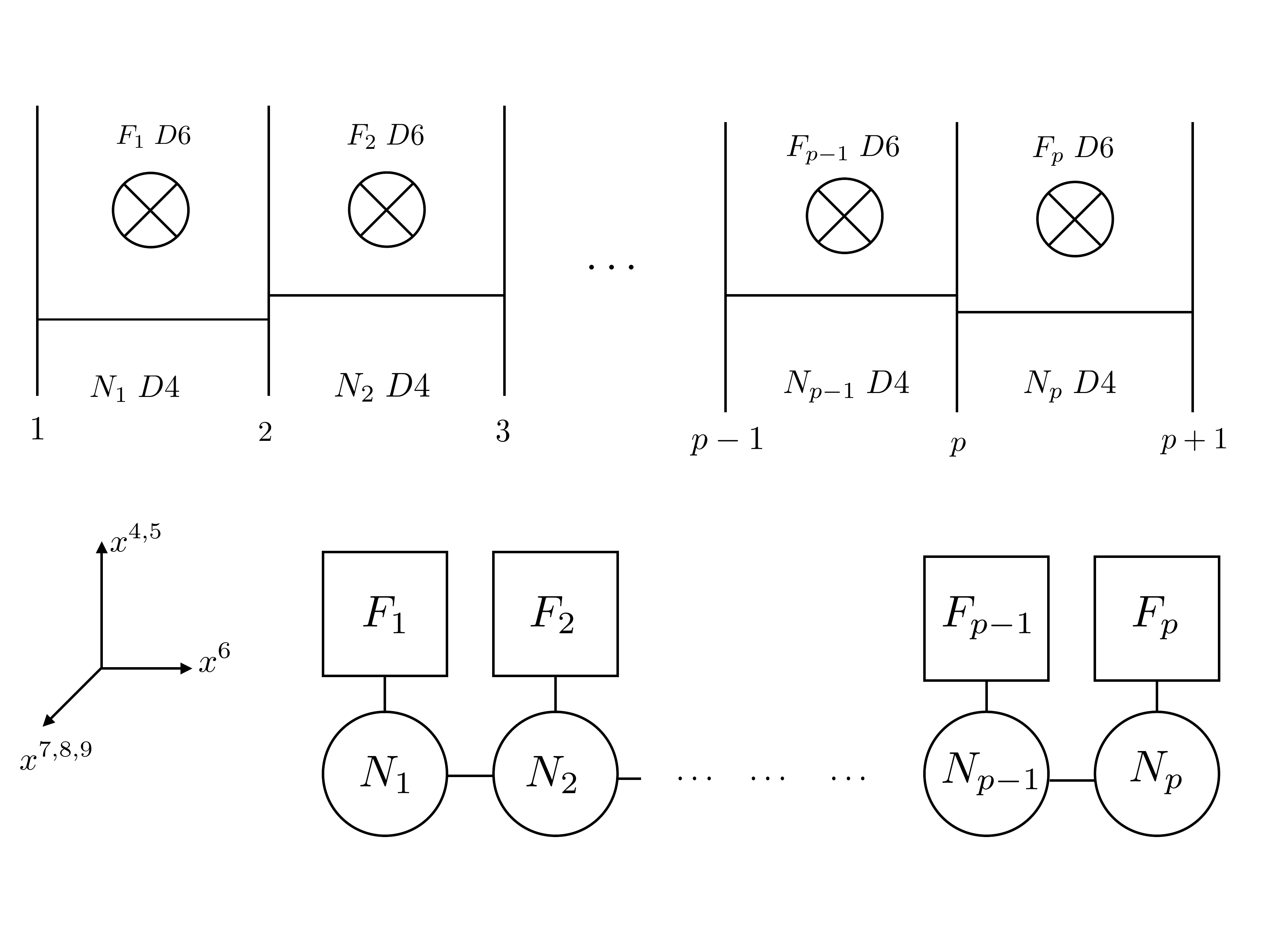} }}%
\caption{The quiver and Hanany-Witten set-up for a generic situation. The vertical lines denote individual Neveu-Schwarz branes extended on the $(x_4,x_5)$ space. The horizontal ones D4 branes, that extend on $x_6$, in between five branes and the crossed-circles D6  branes, that extend on the $(x_7,x_8,x_9)$ directions. All the branes share the Minkowski directions. This realises the isometries $SO(1,3) \times SO(3)\times SO(2)$.}
\label{hananywittensetup}
\end{figure}

These  branes all share four Minkowski directions. The NS five branes extend along the $(x_4,x_5)$ directions---realising $SO(2)\sim U(1)_r$ rotations. 
They are placed at fixed positions in the $x_6$-direction along which the D4 branes extend. This leads at low energies to an effective four dimensional field theory. The D6 branes extend along the $(x_7,x_8,x_9)$ directions---realising $SO(3)\sim SU(2)_R$ invariance. The $SU(2)_R\times U(1)_r$ is the R-symmetry of the CFT. If conformality is broken, the five branes bend in the ($x_4,x_5$) plane, breaking the $U(1)_r$. See the Figure \ref{hananywittensetup} for a generic quiver field theory and corresponding Hanany-Witten set-up.
The  associated eleven dimensional picture realises the field theories on different stacks of M5 branes wrapping a Riemann surface \cite{Witten:1997ep}, which encodes  the Seiberg-Witten curve.
This relates the problem to integrable systems in two dimensions \cite{Donagi:1995cf}.

In 2009, Gaiotto \cite{Gaiotto:2009we}  proposed a generalisation of these ideas for the conformal case. He used  that many ${\cal N}=2$ CFTs are realised  by compactification of  the ${\cal N}=(0,2)$ six dimensional theory on a  punctured Riemann surface.
In this way, the  usual description of ${\cal N}=2$ SCFTs in terms of the space of couplings $\tau_i\sim \frac{i}{g_i^2} + \theta_i$ turned into the study of the moduli space of Riemann surfaces with punctures.

Further investigations of these systems showed their richness. For example, one can obtain precise expressions for the central charges \cite{Shapere:2008zf}  or expressions for the Nekrasov partition function of these theories and correlators in a Liouville theory on the associated Riemann surface \cite{Alday:2009aq}. 

Another description of these CFTs  is obtained by constructing their holographic dual. The authors  of  \cite{Lin:2004nb} found the most generic eleven dimensional background preserving eight Poincare supercharges, with bosonic isometry group $SO(2,4)\times SU(2)_R\times U(1)_r$. In eleven dimensions the geometries have the form
\begin{eqnarray}
& & ds_{11}^2\sim  f_1 AdS_5 + f_2 d\Omega_2(\chi,\xi) + f_3 (d\beta-A_i dx^i)^2+ f_4 dy^2 + f_6 (dx_1^2+dx_2^2).\nonumber
\end{eqnarray}
The background is complemented with a four form field, respecting the isometries. All the functions $f_i(y, x_1,x_2)$ can be written in terms of a single function $D(y,x_1,x_2)$ that solves a Toda equation,
\begin{eqnarray}
\nabla^2_{(x_1,x_2)} D + \partial_y^2 e^{D}=0.\nonumber
\end{eqnarray}
The boundary conditions supplementing this non-linear partial differential equation are specified at $y=0$, where the two sphere $d\Omega_2(\chi,\xi)$ shrinks smoothly and at an arbitrary point $y=y_c$, where the circle $(d\beta-A_1)^2$ shrinks in a smooth fashion. The flux of $F_4$ on the two sub-manifolds $\Sigma_4=[y,x_1,x_2,\beta]$ and $\hat{\Sigma}_4=[S^2(\chi,\xi),x_1,x_2]$ define the number of 'colour' and 'flavour' M5 branes.

In Section \ref{holographicdescriptionx} and what follows, we shall consider the situation in which the flavour M5 branes (analogously the special punctures of the Riemann surface) are smeared in such a way that
we gain a $U(1)$ isometry in the $x_1$ direction.  This makes feasible a reduction to Type IIA. Below, we write the expression of the partial differential equation and boundary conditions in the Type IIA framework, that lead to a well defined geometry and dual field theory.
\\
We now move to the holographic description of the ${\cal N}=2$ SCFTs.

\subsection{The holographic description}\label{holographicdescriptionx}
Let us discuss briefly the holographic description that  emerged along various papers \cite{Lin:2004nb}, \cite{Gaiotto:2009gz}, 
\cite{ReidEdwards:2010qs}, \cite{Aharony:2012tz}.
The generic metric with the $SO(2,4)\times SU(2)\times U(1)$ isometries required to be a dual holographic description of ${\cal N}=2$ SCFTs reads,
\begin{equation} \label{10d}
\begin{split}
ds_{10}^2 = &\alpha' \mu^2 \left[ 4 f_1 ds_{AdS_5}^2 +   f_2 (d\sigma^2 + d\eta^2) +f_3 ds_{S^2}^2(\chi,\xi) + f_4 d\beta^2 \right].\\
\end{split}
\end{equation}
The quantity $\mu^2=\frac{L^2}{\alpha'}$ indicates the size of the space in units of $\alpha'$.  
The range of the ($\sigma,\eta$) coordinates is $0\leq \eta\leq N_5$ and $0\leq \sigma<\infty$. The coordinates $(\chi,\xi)$ parametrise the two sphere (as usual  we take $0\leq\chi\leq \pi$ and  $0\leq \xi\leq 2\pi$) and realise geometrically the $SU(2)_R$ isometry, while the coordinate $\beta $ in $ [0 , 2\pi]$, realises the $U(1)_r$ isometry. The $SO(2,4)$ isometries are realised by the $AdS_5$ spacetime, whose
coordinates we need not specify.
\\
The matter fields in the background are,
\begin{equation}\label{ns}
e^{2 \phi} = f_8, \qquad B_{2}= \mu^2\alpha'  f_5 d \Omega_2(\chi,\xi),\;\;\;\;
C_{1} = \mu^4\sqrt{\alpha'} f_6 d \beta, \qquad A_{3} = \mu^6\alpha'^{3/2} f_7 d \beta \wedge d\Omega_2.\end{equation}
 The functions $(f_1,....,f_8)$ depend only on the coordinates ($\sigma,\eta$). Imposing that eight Poincare supersymmetries are preserved, one finds after lengthy algebra  \cite{Lin:2004nb}, that
 these eight functions $f_i(\sigma,\eta)$ can be written in terms of a single function (we shall refer to it as 'potential') $V(\sigma,\eta)$.
 
In fact, defining the derivatives of the potential function and the function $\Delta(\sigma,\eta)$,
 \begin{eqnarray}
& &\dot{V}=\sigma \partial_\sigma V, \;\;\;\; V'(\sigma,\eta)=\partial_\eta V,\;\;\; \ddot{V}=\sigma \partial_\sigma \dot{V}, \;\;\;\; V''=\partial^2_\eta V,\;\;\;  \Delta = (2 \dot V - \ddot V)V'' + (\dot V')^2,\nonumber
\end{eqnarray}
 it was shown in \cite{Lin:2004nb}  that the functions $f_i(\sigma,\eta)$ are given by,
\begin{equation}
\begin{split}
 f_1=&\left( \frac{2 \dot V - \ddot V}{V''} \right)^{\frac{1}{2}}, \quad f_2=f_1\frac{2 V'' }{ \dot V},\quad f_3=f_1\frac{2 V'' \dot V}{ \Delta}, \quad f_4=f_1\frac{4 V''}{2\dot V - \ddot V} \sigma^2,\\
 f_5=&2 \left(\frac{\dot V \dot V'}{ \Delta} - \eta \right),\quad f_6=\frac{2\dot V \dot V'}{2 \dot V - \ddot V} ,\quad f_7= -\frac{4 \dot V^2 V''}{ \Delta},\quad f_8=\left(\frac{4(2 \dot V - \ddot V)^3}{\mu^{12} V'' \dot V^2  \Delta^2}\right)^{1/2}.\label{definitions1}
\end{split}
\end{equation}
We have checked that this background satisfies the Einstein, Maxwell, Bianchi
and dilaton equations  when the potential function $V(\sigma,\eta)$ solves the equation
 \begin{equation}\label{toda}
 \ddot{V}+\sigma^2 V''=0.
 \end{equation}
This differential equation should be supplemented by boundary conditions in the $(\sigma,\eta)$-space.
One such conditions is that $V(\sigma\to\infty,\eta)=0$. The other  boundary conditions at $\sigma=0$ are better expressed in terms of the function $\lambda(\eta)$, defined as
 \begin{equation}
 \lambda(\eta)=\sigma \partial_\sigma V|_{\sigma=0},\label{xxz}
 \end{equation}
for which we impose,
 \begin{equation}
 \lambda(\eta=0)=\lambda(\eta=N_5)=0.\label{boundaryconditions}
\end{equation}

The equation (\ref{toda}) is sometimes referred to as 'Laplace equation' and the function $\lambda(\eta)$ as 'charge density'. In Appendix \ref{physicalintlambda}  we  justify the terminology and clarify the physical interpretation of $\lambda(\eta)$.

 Any solution to eq.(\ref{toda}) satisfying the boundary conditions at $\sigma\to\infty$ and those in eq.(\ref{boundaryconditions}) can be used to calculate  the 
warp factors $f_i(\sigma,\eta)$  in eq.(\ref{definitions1}) and construct the matter fields and background in eqs.(\ref{10d})-(\ref{ns}). These solutions are conjectured to be dual to ${\cal N}=2$ SCFTs \cite{Lin:2004nb}, \cite{Gaiotto:2009gz}.
Below, we shall discuss  the correspondence between some observables of the conformal quiver field theory and the function $V(\sigma,\eta)$.

For future purposes, it is useful to lift the Type IIA solution to eleven dimensions. The solution in eqs.(\ref{10d})-(\ref{ns}) reads \cite{Gaiotto:2009gz}-\cite{Aharony:2012tz},
\begin{equation}
\begin{split} \label{11d}
ds_{11}^2 = & \kappa^{2/3}\left( 4 F_1 ds_{AdS_5}^2 + F_2 (d\sigma^2 + d\eta^2)+F_3d\Omega_2^2(\chi,\xi)+F_4 d\beta^2 +F_5 \left(dy + \tilde{A}d\beta \right)^2 \right),\\
&\qquad \qquad \qquad \qquad C_{3} =  \kappa \left( F_6 d\beta +F_7 dy \right)\wedge d\Omega_2.
\end{split}
\end{equation}
The functions $F_{i}=F_{i}(\sigma,\eta)$ and $\tilde{A}=\tilde{A}(\sigma,\eta)$ are given by       
\begin{equation}
\begin{split}
F_1=&\left(\frac{\dot V  \Delta}{2V''} \right)^{1/3},\quad F_2=F_1\frac{2V''}{\dot V},\quad F_3=F_1 \frac{2V'' \dot V}{ \Delta},\quad F_4=F_1 \frac{4V''}{2 \dot V - \ddot V} \sigma^2 \\
F_5=&F_1  \frac{2(2\dot V - \ddot V)}{\dot V \Delta},\quad F_6=-4 \frac{\dot V^2 V''}{ \Delta},\quad F_7=2 \left( \frac{\dot V \dot V'}{\ \Delta} - \eta \right),\quad \tilde{A}=\frac{2\dot V \dot V'}{2\dot V - \ddot V} .
\label{Fixx}\end{split}
\end{equation}
In Appendix \ref{appendixreduction} we comment on some subtleties of the IIA-M theory oxidation, like the precise correspondence between constants, dimension of the coordinates, etc. It obviously follows that the eleven dimensional supergravity equations are solved if the potential function $V(\sigma,\eta)$ solves eq.(\ref{toda}). 
\\
Let us now discuss generic solutions to eq.(\ref{toda}).
 \subsection{Generic solutions to the Laplace equation}
 We shall consider two different types of solutions to the Laplace equation (\ref{toda}). The first  type  of solutions, we call  $V_1(\sigma,\eta)$, is defined in the whole range of the $\sigma$-coordinate and was discussed in
 \cite{ReidEdwards:2010qs}, \cite{Aharony:2012tz}. These will be mostly used in the rest of this paper.
 The second type of solutions, labelled by $ V_2(\sigma,\eta)$ should be thought of as series expansion near $\sigma=0$ and is an extension of the expansions presented in
  \cite{Itsios:2017nou},  \cite{Nunez:2018qcj}. The potentials in each case read,
  \begin{eqnarray}
 & & V_1(\sigma, \eta)=-\sum_{n=1}^{\infty}\frac{c_n}{w_{n}}K_{0}(w_n \sigma)\sin (w_n \eta),\;\;\;\;\;\;w_n=\frac{n\pi}{N_5}.\label{Vk}\\
 & & V_2(\sigma,\eta)=F(\eta)+G(\eta)\log \sigma+\sum_{k=1}^{\infty}\sigma^{2 k}\left(h_{k}(\eta)+\hat{f}_{k}(\eta)\log\sigma\right).\label{vtay}
 \end{eqnarray}
 The numbers $c_n$ in eq.(\ref{Vk}) are related to the Fourier coefficients of the odd-extension of the function $\lambda(\eta)$---see eq.(\ref{xxz})-- in the interval $[-N_5,N_5]$. In more detail,
 \begin{equation}
 c_n=\frac{n \pi}{N_5^2}\int_{-N_5}^{N_5} \lambda(\eta)\sin(w_n\eta) d\eta,\;\;\;\;w_n=\frac{n\pi}{N_5}.\label{lalaxa}
 \end{equation}
On the other hand, the functions $h_k, \hat{f}_k$ in eq.(\ref{vtay}) can be written in terms of the input functions
$F(\eta), G(\eta)$ according to the recursive relations,
\begin{eqnarray}
& & h_1(\eta)=\frac{1}{4}(G''(\eta)- F''(\eta)),\;\;\; \hat{f}_1(\eta)=-\frac{1}{4}G''(\eta),\nonumber\\
& & \hat{f}_k(\eta)=-\frac{1}{4 k^2} \hat{f}''_{k-1}(\eta), \;\;\;\;\;k=2,3,4....\\
& &h_k(\eta)=-\frac{1}{4 k^2}\left(h_{k-1}''(\eta) -\frac{1}{k} \hat{f}_{k-1}''(\eta)\right)\;\;\;\;\;k=2,3,4....\label{relationsfh}
\end{eqnarray}
Using the  equation (\ref{Vk}), we obtain the  expression for $\lambda(\eta)$,
\begin{equation}
\lambda(\eta)=\sigma \partial_\sigma V_1(\sigma=0,\eta)=\sum_{n=1}^{\infty}\frac{c_n}{w_{n}} \sin (w_n \eta).
\label{lambdasum}\end{equation}
On the other hand, from  eq.(\ref{vtay}), we find
\begin{equation}
\lambda(\eta)=\sigma \partial_\sigma V_2(\sigma=0,\eta)= G(\eta).\label{lambdatay}
\end{equation}
Thanks to the asymptotic behaviour of the modified Bessel function $K_0(\sigma)$ the boundary condition at $\sigma\to\infty$ is satisfied by $V_1$.
The convergence properties of the expansion proposed for $V_2$ are less clear. For this reason, in the rest of this paper, we will discuss mostly solutions in the form of eq.(\ref{Vk}).
In the Appendix \ref{appendixc}, we quote the expansions for all the functions in the background, close to $\sigma=0$ and $\sigma\to\infty$, calculated with the solutions in eqs.(\ref{Vk})-(\ref{vtay}).
\\
We now comment on the  detailed correspondence between the backgrounds in eq.(\ref{10d})-(\ref{ns}) and the conformal field theories of interest.
\subsection{Correspondence with a conformal quiver field theory}
We consider ${\cal N}=2$ SCFTs with a product gauge group $SU(r_1)\times SU(r_2)\times....\times SU(r_n)$. The field theory has $n$ ${\cal N}=2$ vector multiplets,  $n-1$ hypermultiplets transforming in the bifundamental of each pair of consecutive gauge groups and a set of hypers transforming in the fundamental of each gauge group. The condition of zero beta-function, namely that for each gauge factor, the number of colours equals twice the number of flavours,  translates into,
\begin{equation}
2 r_i= f_i + r_{i+1}+ r_{i-1},\;\;\;\;\;\;\; i=1,....,n.\label{nf=2nc}
\end{equation}
We denoted by  $f_i$ the number of fundamental hypers in the i-th group and with $r_{i+1}$ and $r_{i-1}$ the ranks of the two adjacent gauge groups. Following
\cite{Cremonesi:2015bld}, we can define the forward and backwards 'lattice derivatives'
\begin{eqnarray}
\partial_+ r_i= r_{i+1}-r_i,\;\;\;\;\; \partial_- r_i= r_i- r_{i-1}.\nonumber
\end{eqnarray}
In terms of $\partial_\pm$, the vanishing beta function condition reads,
\begin{eqnarray}
f_i=2 r_i- r_{i+1}- r_{i-1}= - \partial_+ r_i + \partial_- r_i= - \partial_+\partial_- r_i.\label{llkk}
\end{eqnarray}
Since the number of fundamental fields $f_i$ is positive, we find that the function $r$ is convex. One can similarly define the slopes,
\begin{eqnarray}  
s_i= r_i-r_{i-1}=\partial_- r_i\;\;\;\to \;\;\; f_i=-\partial_+ s_i.\nonumber
\end{eqnarray}
This indicates that the slope is a decreasing function. We can  define a 'rank function' $R(\eta)$, where $\eta$ parametrises the 'theory space'.
 The derivatives of $R(\eta)$ will contain the slopes $R'=s$ and the second derivative the number of fundamentals $-R''= f$. Let us clarify this with a generic example.

Consider the quiver of Figure 2.
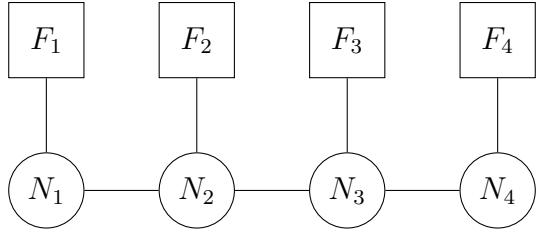
\begin{figure}
\centering
\label{vavacxx}
\begin{tikzpicture}
\node[circle,draw, minimum size=1cm] (A) at  (0,0) {$N_1$};
\node[rectangle,draw, minimum size=1cm] (B) at  (0,2)  {$F_1$};
\draw (A) -- (B);
\node[circle,draw, minimum size=1cm] (C) at  (2,0) {$N_2$};
\node[rectangle,draw, minimum size=1cm] (D) at  (2,2)  {$F_2$};
\draw (C) -- (D);
\draw (A) -- (C);
\node[circle,draw, minimum size=1cm] (C1) at  (4,0) {$N_3$};
\node[rectangle,draw, minimum size=1cm] (D1) at  (4,2)  {$F_3$};
\draw (C1) -- (D1);
\draw (C) -- (C1);
\node[circle,draw, minimum size=1cm] (C2) at  (6,0) {$N_4$};
\node[rectangle,draw, minimum size=1cm] (D2) at  (6,2)  {$F_4$};
\draw (C2) -- (D2);
\draw (C1) -- (C2);
\end{tikzpicture}

\caption{A generic quiver. The squares indicate flavour groups and the circles gauge groups.}
\end{figure}
 For this quiver to represent an ${\cal N}=2$ SCFT, the following conditions must be satisfied,
\begin{eqnarray}
& & 2N_1\!- \!N_2=\!F_1,\;\;\;\;\;\; 2 N_2-\! N_1-N_3 \!=\!F_2,\nonumber\\
& &  2 N_3- \! N_2-N_4=\!F_3,\;\;\;\;\;\; 2N_4-\!N_3\!=\! F_4.\label{rangosvv}
\end{eqnarray}  
We construct the rank-function
\[ R(\eta)= \begin{cases} 
      N_1\eta  &  0 \leq \eta\leq 1\\
      (N_2-N_1)(\eta-1)+N_1 & 1\leq \eta\leq 2 \\
      (N_3-N_2)(\eta-2)+N_2 &  2\leq \eta\leq 3\\
      (N_4-N_3)(\eta-3) + N_3 & 3\leq \eta \leq 4\\
     -N_4 (\eta-4) + N_4 & 4\leq\eta\leq 5. 
   \end{cases}
\]
Calculating $R'(\eta)$ we find a piecewise discontinuous function encoding the numbers to be accommodated as columns of  a Young diagram
\[ R'(\eta)= \begin{cases} 
     N_1  &  0  \leq \eta\leq 1\\
      (N_2-N_1) & 1  < \eta\leq 2 \\
    (N_3-N_2) &  2 < \eta\leq 3\\
     (N_4-N_3) & 3 <  \eta \leq 4\\
     -N_4 & 4 < \eta\leq 5. 
   \end{cases}
\]
The Young diagram contains all kinematic information of the CFT. On the other hand and in agreement with eq.(\ref{llkk}) calculating $-R''(\eta)$ we find the function that gives the number of fundamental hypermultiplets (localised in each gauge group ).  
In fact,
\begin{eqnarray}
F(\eta)= -R''(\eta)=& & 
      (2N_1 - N_2) \delta(\eta-1) +
      (2N_2-N_1-N_3)\delta(\eta-2)     \nonumber\\
      & & +  (2N_3-N_2-N_4)\delta(\eta-3)     +      (2N_4-N_3)\delta(\eta-4) . 
 \end{eqnarray}
This function agrees with the condition for the number of fundamentals $F_i$ in eq.(\ref{rangosvv}).

The connection between the gravitational picture of Section \ref{holographicdescriptionx}
with the field theory description in this section comes  from the identification of the functions
 \begin{equation}
\lambda(\eta)=R(\eta).\label{correspondencexx}
\end{equation}
This is a non-trivial step as it relates the 'field-theory space' with the space coordinate $\eta$ in IIA or M-theory background  \cite{Gaiotto:2009gz} . 

The logic to follow is then clear. First choose a conformal quiver field theory. Then write the rank function $R(\eta)$ and use this function as the boundary condition for the Laplace-like problem in eqs.(\ref{toda})-(\ref{boundaryconditions}) setting $\lambda(\eta)=R(\eta)$.  Then, we write solutions as in eq.(\ref{Vk}), calculating the Fourier coefficients as in eq.(\ref{lalaxa}). It is equivalent 
 to start from the Young diagram constructed using $R'(\eta)$, work out $R(\eta)$ imposing piecewise continuity and the conditions $R(0)= R(N_5)=0$, that is the same condition on the function $\lambda(\eta)$. Finally identify the function $R(\eta)=\lambda(\eta)$ and proceed as above. Let us discuss under which conditions the backgrounds capture the Physics of the dual CFT.

\subsubsection{Trustability of the holographic description}
The validity of the  supergravity solutions in eqs.(\ref{10d}), (\ref{ns}) and (\ref{11d}) was carefully analysed in  \cite{Aharony:2012tz}.
These backgrounds explicitly present D6 and NS-five branes, and close to those branes the curvatures in units of $\alpha'$ and the string coupling respectively become very large. We  can not trust holographic calculations in regions where  $g_s\sim e^{\phi} $ and/or $\frac{\alpha'}{R_{eff}} $ become large. In other words, our backgrounds are defined by a manifold $V(\sigma,\eta)$, the points at which the D6 of NS branes are placed are singular points of this manifold. The information obtained by holographic calculations close to these points is not trustable. 

The idea  is to 'localise' those regions to small patches of the manifold defined by $V(\sigma,\eta)$. To do this, it was suggested in  \cite{Aharony:2012tz} that one can take $N_5$ (the range of the $\eta$-coordinate)
 very large, hence dealing with a long-linear quiver. We can also
 scale the function $\lambda(\eta)\to N_c \lambda(\eta)$. In this way we change the number of D4 and D6 branes (but keep the number of NS five branes fixed) and we can have good control over string loop corrections (in a 't Hooft limit, with $g_s N_4$ fixed).  Similarly, scalings of the $\eta$-coordinate increase the number of five branes reducing curvatures.
 
 In summary, we shall consider in all of our comparisons between CFT results and holographic results that the range of the $\eta$-coordinate $N_5$ is large (this will turn out to be the number of five-branes) and that the function $\lambda(\eta)$ is scaled up by a (large) factor $N_c$, that will turn to be proportional to the number of D4 and D6 branes as we explain below.
 
Now, using the holographic description, we  calculate some  observables that characterise the CFT.
\subsection{Page Charges}\label{pagechargesII}
 In this section, we will calculate the Page charges for a generic Gaiotto-Maldacena background. These charges are identified with the number of branes in the associated Hanany-Witten set-up.
We define Page charges as,
  \begin{equation}
\hat{F}=Fe^{-B_2},\quad Q_{D_{p}}=\frac{1}{2\kappa_{10}^2 T_{Dp}}\int_{\Sigma}\hat{F}_{8-p},\quad 2\kappa_{10}^2 T_{D_{p}}=(2\pi)^{(7-p)} g_s (\alpha')^{\frac{7-p}{2}}.\label{pagedefinitions}
\end{equation}
Using the expressions for the fields in eq.(\ref{ns}), we derive
\begin{eqnarray}
& & H_3= d B_2=\mu^2\alpha' \left(\partial_\sigma f_5 d\sigma+\partial_\eta f_5 d\eta \right)\wedge d\Omega_2,\nonumber\\
& & F_2= dC_1= \mu^4 \sqrt{\alpha'} \left(  \partial_\sigma f_6 d\sigma+\partial_\eta f_6 d\eta \right)\wedge d\beta,\nonumber\\
& & \hat{F}_4= d A_3 - B_2\wedge F_2=\mu^6(\alpha')^{3/2}\left[ (\partial_\sigma f_7 - f_5\partial_\sigma f_6) d\sigma+   (\partial_\eta f_7 - f_5\partial_\eta f_6) d\eta   
  \right] d\Omega_2\wedge d\beta.\nonumber
\end{eqnarray}
We specify the cycles on which the integrals are to be performed. 
The two and four non-trivial cycles in the geometry  will be placed at $\sigma\to 0$ and  the three-cycle can be placed either at $\sigma\to\infty$ or at $\sigma=0$. The cycles are defined as,
\begin{equation}
\begin{split}
\Sigma_{2}=&(\beta,\eta)\vert_{\sigma=0},\quad \Sigma_{3}=(\eta,\chi,\xi)\vert_{\sigma=0}\;\;\;\tilde{\Sigma}_{3}=(\eta,\chi,\xi)\vert_{\sigma\to\infty},\quad \Sigma_{4}=(\beta,\eta,\chi,\xi)\vert_{\sigma=0}.\label{kalaca}
\end{split}
\end{equation}
We  then calculate, 
\begin{eqnarray}
& & Q_{NS}=\frac{1}{(2\pi)^2 g_s^2 \alpha'}\times \mu^2\alpha' \int d\Omega_2 \int_{0}^{\eta_f} \partial_\eta f_5(\sigma=0,\eta)d\eta=\frac{\mu^2}{g_s^2 \pi}[f_5(0,\eta_f)- f_5(0,0)],\nonumber\\
& & \tilde{Q}_{NS}=\frac{1}{(2\pi)^2 g_s^2 \alpha'}\times \mu^2\alpha' \int d\Omega_2 \int_{0}^{\eta_f} \partial_\eta f_5(\sigma=\infty,\eta)d\eta=\frac{\mu^2}{g_s^2 \pi}[f_5(\infty,\eta_f)- f_5(\infty,0)],\nonumber\\
& & Q_{D6}=\frac{1}{(2\pi) g_s \sqrt{\alpha'}}\times \mu^4\sqrt{\alpha'} \int d\beta \int_{0}^{\eta_f} \partial_\eta f_6(\sigma=0,\eta)d\eta=\frac{\mu^4}{g_s}[f_6(0,\eta_f)- f_6(0,0)].\nonumber\label{vbvbvb}
\end{eqnarray}
In what follows, we set $g_s=1$ and  use the expansion for the functions $f_5,f_6,f_7$ in Appendix \ref{appendixc} . We find,
\begin{equation}
{Q_{NS5}= \tilde{Q}_{NS5}=\frac{2\mu^2}{\pi}\eta_f ,\label{qns5good}}
\end{equation} 
\begin{equation}
  Q_{D6}=\mu^4 (\lambda'(0)-\lambda'(\eta_f)).\label{qd6good}    
\end{equation} 
Using that  $\eta_f=N_5$ (an integer), we impose $\mu^2=\frac{\pi}{2}$ to have a well-quantised  charge of NS-five branes. Defining $N_6=\frac{\pi^2}{4} N_c$--- the integer $N_c$ is  a global factor in the function $\lambda(\eta)$--
gives also a well quantised charge of D6 branes.

The calculation of the  D4 brane charge is  more subtle.
In fact,  the associated Page charge is,
\begin{eqnarray}
& & q_4=\frac{1}{(2\pi)^3 g_s (\alpha')^{3/2}}\int_{\Sigma_4} \hat{F}_4=\nonumber\\
& & \frac{1}{(2\pi)^3 g_s (\alpha')^{3/2}}\times \mu^6(\alpha')^{3/2} \int d\Omega_2 d\beta \Big[( f_7(0,\eta_f)- f_7(0,0) )- \int_{0}^{\eta_f} f_5 (0,\eta)\partial_\eta f_6(0,\eta)d\eta\Big]=\nonumber\\
& & \frac{2}{\pi}\mu^6 \eta_f \lambda'(\eta_f).\label{xxu}
\end{eqnarray}
The expression in eq.(\ref{xxu}), is not just the charge of D4 brane. In fact, on the D6 branes there is also charge of D4 induced by the $B_2$ field due to the Myers effect and those are counted by eq.(\ref{xxu}).

{To avoid this 'overcounting' we have found a nice new expression that calculates the total D4 brane charge.
The expression is  proven in Appendix \ref{appendixd4}. It reads, }
\begin{equation}
{ Q_{D4}=
\frac{2}{\pi}\mu^6 \int_{0}^{\eta_f}\lambda(\eta)d\eta.\label{qd4good}     }
\end{equation} 
This will be properly quantised when  $\mu^2=\frac{\pi}{2}$ and  $N_6=\frac{\pi^2}{4} N_c$.

In Section \ref{sectionexamples} we shall test equations (\ref{qns5good}), (\ref{qd6good}) and (\ref{qd4good}) in different examples.  Now, let us derive some general expressions for the linking numbers.
\subsection{Linking numbers}\label{sectionlinking}

The linking numbers in brane set-ups were defined by Hanany and Witten in \cite{Hanany:1996ie}.
In this paper, we are working with  set-ups of NS five branes and D6 branes in the presence of D4 branes. We define the linking numbers for the $i^{th}$ five brane ($K_{i}$) and for the $j^{th}$ D6 brane ($L_j$) by counting the number of the other branes to the left and to the right of a given one. The definitions of the linking numbers are,
\begin{eqnarray}
& &K_{i}=N_{D4}^{right}- N_{D4}^{left} - N_{D6}^{right},\nonumber\\
& & L_{j}=  N_{D4}^{right}- N_{D4}^{left} + N_{NS}^{left}.\label{linkingshw}
\end{eqnarray}
They must satisfy 
\begin{equation}
\sum_{i=1}^{N_5} K_i+ \sum_{j=1}^{N_6} L_j=0.\label{consistency}
\end{equation}
The linking numbers are topological invariants and they do not change under Hanany-Witten moves. They can easily be calculated with the brane set-up by simple counting of branes.

 With the dual supergravity background we can compute these invariants. In fact, for the case of the NS five branes, we find that in our generic backgrounds the linking number are all equal $K_1=K_2=....=K_{N5}$. We propose that they are calculated by,
\begin{equation}
K_i= \frac{2}{\pi}\mu^6 \lambda'(\eta_f).
\end{equation}
As a consequence of this the sum over NS five branes in eq.(\ref{consistency}) gives
\begin{equation}
{\sum_{i=1}^{N_5} K_i= \frac{2}{\pi}\mu^6 \lambda'(\eta_f)\eta_f=\frac{1}{2\kappa_{10}^2T_{D4}}\int_{\Sigma_4}F_4-B_2\wedge F_2.} \label{linkingNSgrav}
\end{equation}
Where we used that the manifold $\Sigma_4=[\eta,\Omega_2(\chi,\xi),\beta]_{\sigma=0}$ as specified in eq.(\ref{kalaca}).

Inspired by  \cite{Aharony:2012tz}, we  can obtain nice expressions for the linking number of the D6 branes using the supergravity background. In fact, for a general quiver, the Hanany-Witten set-up will have D6 branes placed at different points $\eta_1,\eta_2,....\eta_l$. The number of D6 branes in each stack will be given by the difference in slopes 'before and after' the $j^{th}$ stack. More explicitly the number of D6 branes in the j-stack is
$\lambda'(\eta_j-\epsilon)-\lambda'(\eta_j+\epsilon)$. Aside, all the branes in the j-stack have linking number $L_{j}=\eta_j$. The sum over D6 branes in eq.(\ref{consistency}) gives,
\begin{equation}
\sum_{j=1}^{N_6} L_j=-\frac{2\mu^6}{\pi}\sum_{j=1}^{N_6} \lambda'(\eta_j)\eta_j=-\frac{2\mu^6}{\pi}\lambda'(\eta_f)\eta_f.
\end{equation}
To calculate this explicitly in supergravity, we perform a large gauge transformation on the field $C_1$ at each point $\eta_i$ where the stacks of D6 branes are placed,
\begin{equation}
C_1\to C_1+ \mu^4\sqrt{\alpha'}\left(\lambda'(\eta_j-\epsilon)-\lambda'(\eta_j+\epsilon)\right) d\beta.
\end{equation}
We equate the D6 linking numbers with the flux that we calculate on the four manifold $\tilde{\Sigma}_4=[\eta,\Omega_2(\chi,\xi),\beta]_{\sigma\to\infty}$. We propose the formula,
\begin{equation}
{\sum_{i=1}^{N_6} L_i=
\frac{1}{2\kappa_{10}^2T_{D4}}\int_{\tilde{\Sigma}_4}F_4+C_1\wedge H_3=-\frac{2}{\pi}\mu^6 \lambda'(\eta_f)\eta_f. }\label{linkingD6grav}
\end{equation}

In Section \ref{sectionexamples} and in Appendix \ref{detailsCFT}, we evaluate the expressions of eqs.(\ref{linkingNSgrav}),(\ref{linkingD6grav})
in various examples and check them against the expressions derived from the Hanany-Witten set-up, finding a precise match.

Let us now discuss another observable characterising the CFT that has a nice holographic description, the central charge.

\subsection{Central charge for Gaiotto-Maldacena backgrounds}\label{centralcharge}
Our aim is to find an expression for the central charge of a generic CFT using the  solutions of eqs.(\ref{10d})-(\ref{ns}) or eqs.(\ref{11d}).
The calculation in this section uses the formalism of the papers in \cite{Klebanov:2007ws}. We consider the metric in eqs.(\ref{10d}),(\ref{11d}) and rewrite them in the form,
\be
ds^2=a(R,y^{i})(dx_{1,3}^2+b(R)d R^2)+g_{ij}(R,y^{i})dy^{i}dy^{j}.
\e
where $g_{ij}$ is the metric of the internal space.
Comparing  with eqs. (\ref{10d}) and (\ref{11d}) we identify 
\be
\begin{split}
a(R)=&4 \mu^2 \alpha' R^2 f_1,\qquad b(R)=1/R^4, \qquad \qquad \qquad \qquad \quad \textrm{for eq. (\ref{10d})}\\
a(R)=&4 \kappa^{2/3}R^2 F_1, \qquad b(R)=1/R^4 . \qquad \qquad \qquad \qquad \quad \textrm{for eq. (\ref{11d})}
\end{split}
\label{manaba}\e
Now, we compute the following auxiliary quantities, necessary for the holographic expression of the central charge in eq.(\ref{ppqq}) below. First we calculate,
\be
\begin{split}
\sqrt{e^{-4\phi}det\,g_{int}a^3}=&2^5\alpha'^4\mu^{14} R^3\sigma \sin\chi V''\dot V,\qquad \qquad \qquad \qquad ~\textrm{using eq. (\ref{10d})}\\
\sqrt{det\,g_{int}a^3}=&2^5\kappa^3 R^3\sigma \sin\chi V''\dot V.\qquad \qquad \qquad \qquad \qquad \textrm{using eq. (\ref{11d})}
\end{split} 
\e
 We continue this calculation only  in Type IIA (the case with the eleven-dimensional description is analogous). The internal volume is 
\begin{eqnarray}
& & {\cal V}_{int}=2^5 \alpha'^4 \mu^{14} R^3 \int^{\pi}_{0}\sin\chi d\chi\int_{0}^{2\pi}d\beta\int_{0}^{2\pi}d\xi  \int_{0}^{\infty}\int_{0}^{\eta_f}\sigma \dot V V'' d\sigma d\eta ={\cal N} R^3,\nonumber\\
& & {\cal N}=2^7\pi^2\alpha'^4 \mu^{14} \int_0^{\eta_f} \lambda^2(\eta)d\eta.\label{vintGM}
\end{eqnarray}
To arrive to the last expression we have used equation (\ref{toda}), the fact that\\ $\dot{V}(\sigma\to\infty,\eta)=0$ and the definition of $\lambda(\eta)$ in eq.(\ref{xxz}).
The above integral is explicitly evaluated for the generic solution in eq.(\ref{Vk}) as, 
\be
\int_{0}^{\eta_f}\dot V^2\vert_{\sigma=0}d\eta =\sum_{m=1}^{\infty}\sum_{l=1}^{\infty}\frac{c_m c_l}{m  l}\frac{N_5^2}{\pi^2}\int_{0}^{\eta_f}\sin\omega_m\eta \sin\omega_l\eta d\eta.
\e
We obtain 
\be
{\cal V}_{int}=2^7\pi^2 \alpha'^4 \mu^{14} R^3 \int_0^{\eta_f}\lambda^2(\eta)d\eta=2^6 N_5^3 R^3{\alpha'^4 \mu^{14}}\sum_{m=1}^{\infty}\frac{c_m^2}{m^2}\equiv \mathcal{N}R^3.
\label{vintxx}\e
Now, coming back to our original goal, we use the formula for the central charge derived  in \cite{Klebanov:2007ws},
\be
c=\frac{3^3}{G_{10}}\frac{b(R)^{3/2}\mathcal{H}^{7/2}}{\mathcal{H}'^3},\label{ppqq}
\e
where $\mathcal{H}={\cal V}_{int}^2$. Using that  $G_{10}=2^3\pi^6\alpha'^4 g_s^2$ (we chose $g_s=1$) and $\eta_f=N_5$. We arrive to our new expression,
\begin{equation}\label{eqcc}
{c= \frac{2\mu^{14} }{\pi^4}\int_0^{N_5} \lambda^2(\eta) d\eta=\frac{ N_5^3\mu^{14}}{\pi^6}\sum_{m=1}^{\infty}\frac{c_m^2}{m^2} .}
\end{equation} 
This  indicates that
the central charge is proportional  to the area under the function $\lambda^2(\eta)$. {These formulas are  similar to those derived in dual to six dimensional SCFTs with ${\cal N}=(1,0)$ SUSY, see eq.(2.14) of the paper \cite{Nunez:2018ags} }. 

On the CFT side, it was shown by the authors of \cite{Shapere:2008zf} that an expression for the  two central charges $a$ and $c$ can be written in terms of the number of  ${\cal N}=2$ vector multiplets ($n_v$) and hypermultiplets ($n_h$) in the quiver. The expressions read,
\begin{equation}
a=\frac{5n_v+n_h}{24\pi}, \;\;\;\; c=\frac{2 n_v+ n_h}{12 \pi}.\label{centralchargeCFT}
\end{equation}
The comparison with the holographic result in eq.(\ref{eqcc}) holds only when the IIA/M-theory background is trustable, that is when $N_5\to\infty$ and $N_c\to\infty$, in which case we also have $a=c$. 
In Section \ref{sectionexamples} and in Appendix \ref{detailsCFT}, we shall compare the result of eq.(\ref{eqcc}) with the explicit field theoretical counting of degrees of freedom  in eq.(\ref{centralchargeCFT}), for various examples. 

Along similar lines, we derive an expression for the Entanglement Entropy of a square region in a generic CFT, see Appendix \ref{appendixEE}.

To summarise, in this section we discussed  some observables of generic ${\cal N}=2$ SCFTs (brane charges, linking numbers, central charges) 
and presented new expressions to compute them using generic holographic dual backgrounds. In the next section we study  some particular CFTs and check the matching of these results for the observable when computed with the holographic and with the field theoretical description.

\section{Examples of $ {\cal N}=2$ CFTs}\label{sectionexamples}
In this section we work with two particularly simple and  interesting solutions for the potential function of the form given in eq.(\ref{Vk}). We will explicitly check that the field theoretical calculation and the
holographic calculation match precisely in the limit in which the supergravity description is trustable.  In Appendix \ref{detailsCFT} we will discuss more elaborated CFTs, again obtaining a precise  match.
\\
 Let us first present the two basic examples that occupy us in this section.
 \subsection{Two interesting solutions of the Laplace equation}
\begin{figure}[h!]
    \centering
    {{\includegraphics[width=12.5cm]{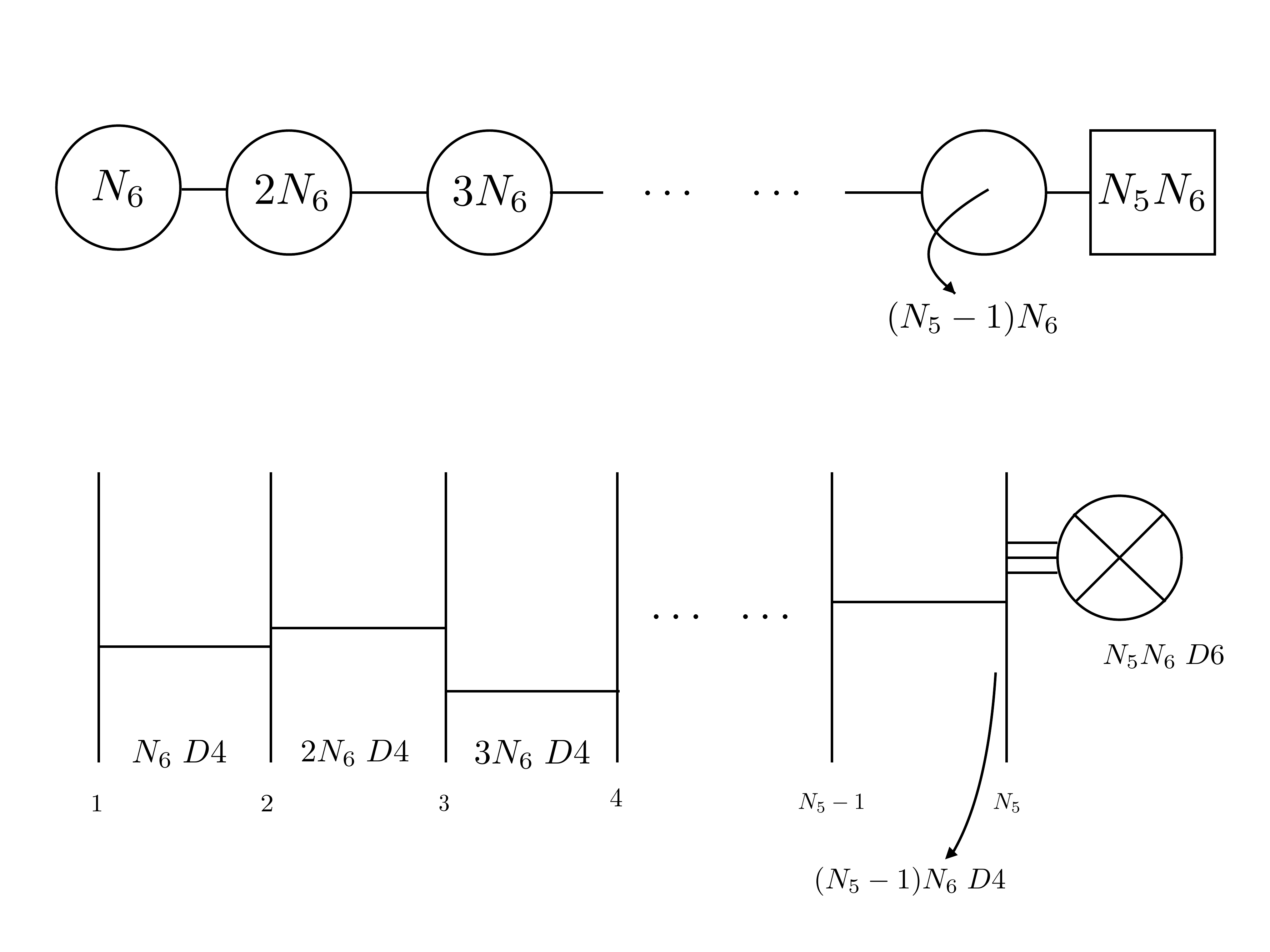} }}%
\caption{The quiver and Hanany-Witten set-up for the profile in eq.(\ref{profile1}). The vertical lines denote individual Neveu-Schwarz branes. The horizontal lines D4 branes and the crossed circles D6 branes.}
\label{Figure1}
\end{figure}

The first solution was used in \cite{Lozano:2016kum} in the study of the non-Abelian T-dual of  $AdS_{5}\times S^5$.
The  charge density or $\lambda$-profile is\footnote{Here and in the rest of the paper $\frac{\pi^2}{4}N_c=N_6$.}
\begin{equation} \label{profile1}
\lambda(\eta)
                    =N_c\left\{ \begin{array}{ccrcl}
                       \eta & 0\leq\eta\leq (N_5-1) \\
                       (N_5-1)(N_5-\eta) &(N_5-1) \leq\eta\leq N_5.
                                             \end{array}
\right.
\end{equation}
In this case the Fourier coefficients in eqs.(\ref{Vk}), (\ref{lalaxa}) are calculated to be,
\begin{equation}
c_m=\frac{2N_c N_5}{m \pi}\sin\left(\frac{m\pi(N_5-1)}{N_5}\right).\label{fourier1}
\end{equation}
The associated quiver and Hanany-Witten set-up are shown in Figure \ref{Figure1}.

\begin{figure}[h!]
    \centering
    {{\includegraphics[width=12.5cm]{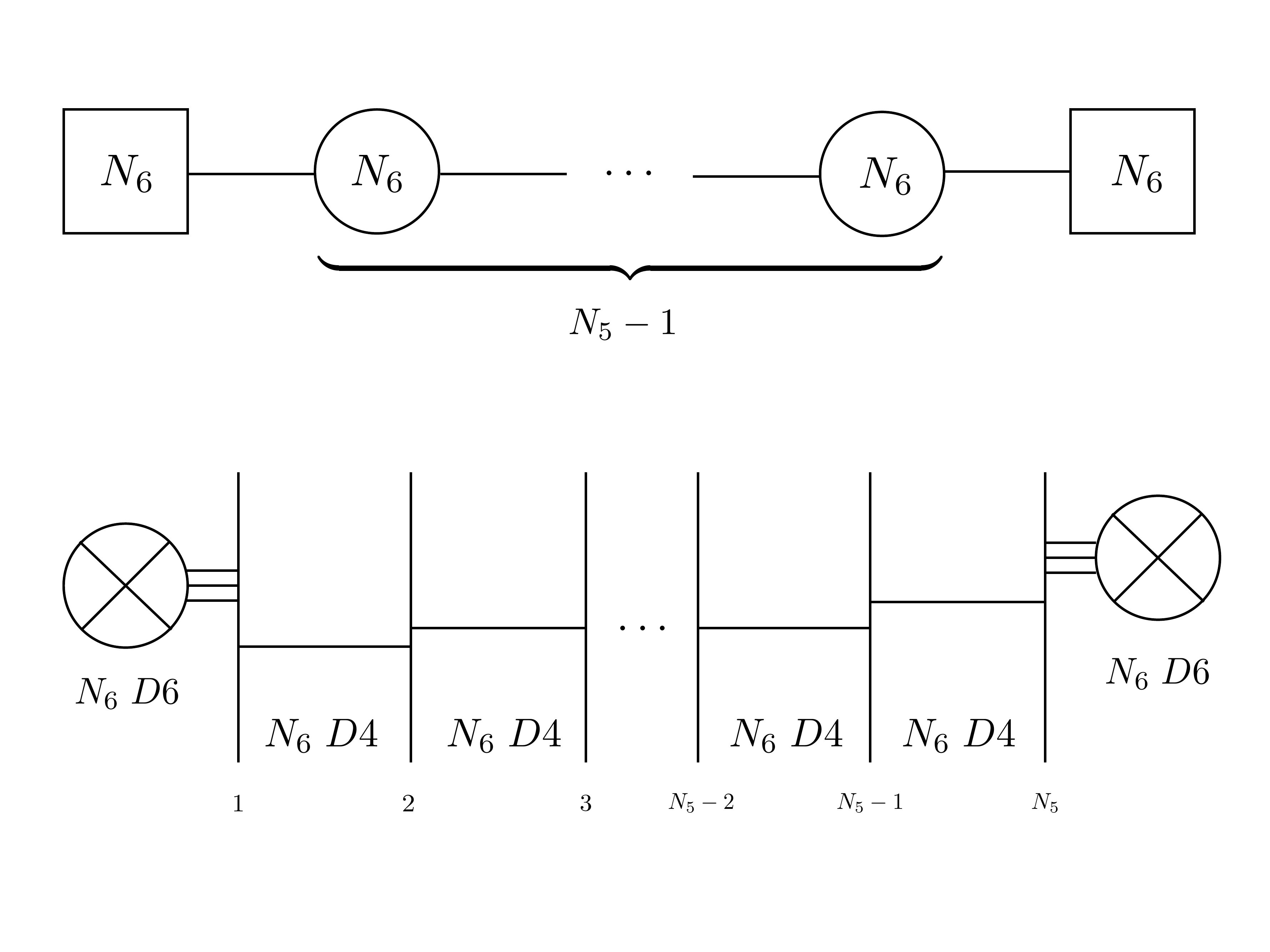} }}%
\caption{The quiver and Hanany-Witten set-up for the profile in eq.(\ref{profile2}).}
\label{Figure2}
\end{figure}

The second solution has a $\lambda$-profile given by,
  \begin{equation} \label{profile2}
\lambda(\eta)
                    =N_c\left\{ \begin{array}{ccrcl}
                       \eta & 0\leq\eta\leq 1 \\
                       1 &1 \leq\eta\leq (N_5-1) \\
                        N_5-\eta, \;\;& (N_5-1)\leq\eta\leq N_5
                                             \end{array}
\right.
\end{equation}
The Fourier coefficients are, 
\begin{equation}
c_n= \frac{2 N_c }{n \pi}\left[\sin\left(\frac{n\pi}{N_5}\right) +\sin\left(\frac{n\pi(N_5-1)}{N_5}\right)   \right].\label{fourier2}
\end{equation} 
 The quiver and Hanany-Witten set up are shown in Figure \ref{Figure2}.

In both examples, we proceed as described above: given the function $\lambda(\eta)$ and the Fourier expansion of its odd-extension, we construct the potential in eq.(\ref{Vk}).
With this we construct the full background in eqs.(\ref{10d})-(\ref{ns}). In the following we show details of the precise matching between field theoretical and holographic calculations of the observables in Section \ref{GMgenerics} for these cases.

 \subsection{Page charges and linking numbers}
 Let us evaluate the expressions for $Q_{NS5}, Q_{D6}, Q_{D4}$ in eqs.(\ref{qns5good}),(\ref{qd6good}),(\ref{qd4good}) for the two backgrounds obtained using  eqs.(\ref{profile1}),(\ref{profile2}).
 For the $\lambda(\eta)$ in eq.(\ref{profile1}) we have 
$\lambda'(\eta_f)=N_c(1-N_5),\;\lambda'(0)=N_c.$
This implies
\begin{equation}
Q_{NS5}=N_5,\;\;\;\;Q_{D6}=\mu^4 N_c N_5=\left(\frac{\pi^2 N_c}{4}\right) N_5= N_{6} N_5.
\label{cargasprofile11}\end{equation}
Finally, for the charge of D4 branes, we find using eq.(\ref{qd4good})
\begin{equation}
Q_{D4}=\frac{1}{2}N_{6}N_5(N_5-1).
\end{equation}
These precisely coincide with what we obtain by simply inspecting Figure \ref{Figure1},
\begin{equation}
N_{D4}= N_6\sum_{r=1}^{N_5-1} r= \frac{N_6}{2}  N_5 (N_5-1),\;\;\; N_{D6}= N_6 N_5,\;\; N_{NS5}=N_5.
\end{equation}
For the profile in eq.(\ref{profile2}) we find,
\begin{eqnarray}
& &N_{NS5}=N_5,\;\;\; Q_{D6}=2 \frac{\pi^2 N_c}{4} =2 N_6,\;\;Q_{D4}= N_6 (N_5-1).\nonumber
\end{eqnarray}
This coincides with the results obtained by simple inspection of the quiver and Hanany-Witten set-up displayed in Figure \ref{Figure2}.

 Analysing the linking numbers we use the expressions in eqs.(\ref{linkingNSgrav}),(\ref{linkingD6grav}). We find that the calculation on the gravity side for the profile in eq.(\ref{profile1}) gives,
 \begin{eqnarray}
 - \sum_i K_i=  \sum_j L_j= \frac{2}{\pi} \mu^6 N_5 (N_5-1) N_c= N_6 N_5(N_5-1).\label{linkingsprofile1}
 \end{eqnarray}
 This result is easily confirmed by studying the Hanany-Witten set-up in Figure \ref{Figure1}. In fact, we find
 \begin{eqnarray}
& &  K_i=N_6(1-N_5),\;\;\; \sum_i K_i= N_6 N_5(1-N_5).\nonumber\\
& &  L_i=N_5-1,\;\;\;\; \sum_j L_j= N_5 N_6 (N_5-1).\label{linkingsprofile1CFT}
 \end{eqnarray}
 The same match is found for the quiver associated with eqs.(\ref{profile2}) and Figure \ref{Figure2}. Counting with the Hanany-Witten set-up, we find
 \begin{eqnarray}
 & & L_1=L_2=...=L_{N6}=1,\;\;\; \tilde{L}_1=\tilde{L}_2=....=\tilde{L}_{N6}=N_5-1.\nonumber\\
 & & \sum_j L_j= N_6+ N_6(N_5-1)= N_6N_5,\nonumber\\
 & & K_1=K_2=...=K_{N5}=- N_6,\;\;\;\; \sum_i K_i= - N_5 N_6
 \end{eqnarray}
 We have denoted by $L_j$ ($\tilde{L}_j$) the D6 branes to the left (right) of the Hanany-Witten set-up of Figure {\ref{Figure2}}. These results are matched by the supergravity calculation with $\lambda(\eta)$ in eq.(\ref{profile2}). In fact, using eqs. (\ref{linkingNSgrav}),(\ref{linkingD6grav}) we find,
 \begin{equation}
 \sum_i K_i= - \sum_j L_j= -\frac{2}{\pi} \mu^6 N_c N_5 =-N_5 N_6.
 \end{equation}
 Let us now compare central charges calculated with eqs.(\ref{eqcc}) and (\ref{centralchargeCFT}).
\subsection{Central charge} 
We evaluate the holographic expressions of eq.(\ref{eqcc})
 and compare them (in the large $N_c, N_5$ limit) with the result of eq.(\ref{centralchargeCFT}). We start with the background obtained using the $\lambda$-profile in eq.(\ref{profile1}).
Using  eqs.(\ref{profile1})-(\ref{fourier1})  and eq.(\ref{eqcc}) we find,
\begin{eqnarray}
& & c=\frac{2\mu^{14}}{\pi^4}\int_{0}^{N_5}\lambda^2 d\eta=\frac{2\mu^{14}}{3\pi^4}N_c^2 N_5^3 (1-\frac{1}{N_5})^2\sim \frac{2\mu^{14}}{3\pi^4}N_c^2 N_5^3 .
\end{eqnarray}
We have used that  $N_5\to\infty$ and $N_c\to\infty$ to have a trustable holographic description. We can work with right hand side of eq.(\ref{eqcc}), which implies
\begin{eqnarray}
& & c= \frac{4 N_5^5 N_c^2\mu^{14}}{\pi^8}\sum_{m=1}^{\infty}\frac{1}{m^4}\left[\sin\left(\frac{m\pi(N_5-1)}{N_5}\right)\right]^2=\nonumber\\
& & =\frac{4 N_5^5 N_c^2 \mu^{14}}{\pi^8}\left[\frac{\pi^4}{180}-45(\textrm{Polylog}[4,e^{i2\pi/N_5}]+\textrm{Polylog}[4,e^{-i2\pi/N_5}])\right]\sim \frac{2\mu^{14}}{3\pi^4}N_c^2 N_5^3.\nonumber
\end{eqnarray}
Using that $\mu^2=\frac{\pi}{2}$ and $N_6=\frac{\pi^2}{4}N_c$, we find the holographic result,
\be
c= \frac{N_5^3N_6^2}{12\pi}+\mathcal{O}(1/N_5,1/N_6).\label{centralholoxx}
\e
This is precisely the central charge obtained by performing  a CFT calculation. Indeed, using the expression in  eq.(\ref{centralchargeCFT}) and the quiver in Figure \ref{Figure1}, we obtain
\begin{eqnarray}
& & n_v=\sum_{r=1}^{N_5-1} r^2 N_6^2-1= \frac{(N_5-1)}{6}(2 N_5^2 N_6- N_5 N_6^2-6),\label{nv1}\nonumber\\
& & n_h=\sum_{r=1}^{N_5-1} r(r+1)N_6^2=\frac{N_6^2}{3}N_5(N_5^2-1),\label{nh1}\nonumber\\
& &c=\frac{(N_5-1)(N_5^2 N_6^2 -2)}{12\pi}\sim\frac{N_6^2 N_5^3}{12\pi}.
\end{eqnarray}
Finding, in the large $N_5$ and large $N_6$ limit a precise matching with the holographic calculation of eq.(\ref{centralholoxx}).

The reader can check that eq.(\ref{eqcc}) applied to eqs.(\ref{profile2})-(\ref{fourier2})---for large $N_5$--leads to
\begin{equation}
c=\frac{2\mu^{14}}{\pi^4}N_c^2 N_5=\frac{N_6^2 N_5}{4\pi}.
\end{equation}
This expression is matched in the appropriate limit of the CFT calculation. In fact for the quiver associated with the profile in eq.(\ref{profile2}), we have 
\begin{eqnarray}
& & n_v=(N_6^2-1)(N_5-1),\;\;\;\;n_h=(N_5-1)N_6^2,\label{p2nv}\nonumber\\
& &c=\frac{N_6^2 N_5}{4\pi}(1-\frac{2}{3N_6^2} -\frac{1}{3 N_5} +\frac{2}{N_5 N_6^2})\sim\frac{N_6^2 N_5}{4\pi}.
\end{eqnarray}
The reader can verify that the same expressions are obtained for the $a$ central charge in the holographic limit (since $a=c$ in this case).

In the Appendix \ref{detailsCFT} we extend the precise matching  of Page charges, linking numbers and central charge to more general and elaborated CFTs. The interested reader is invited to study these nice agreements. Let us now study two solutions to the Laplace equation (\ref{toda}) that are qualitatively different from those discussed above.

\subsection{The Sfetsos-Thompson solution}
Let us discuss a particular solution  obtained by Sfetsos and Thompson in \cite{Sfetsos:2010uq},  that received attention in the last few years. 
The solution to eq.(\ref{toda}) with charge density as in eq.(\ref{xxz})
are given by,
\begin{equation}
V_{ST}= N_c\left[ \eta\log\sigma -\frac{\eta \sigma^2}{2} +\frac{\eta^3}{3}    \right],\;\;\;\; \lambda(\eta)= N_c\eta.\label{sfetsosthompson}
\end{equation}
In the language of eqs.(\ref{vtay}), (\ref{relationsfh})
the defining functions are, \begin{equation}
F(\eta)= N_c \frac{\eta^3}{3},\;\;\; G(\eta)=N_c\eta,\;\;\; h_1= -N_c\frac{\eta}{2},\;\;\;\; f_k=h_{k+1}=0, \;\; k>1.
\end{equation}
Notice that the $\eta$-coordinate is not bounded, hence $\eta_f\to\infty$. This has unpleasant consequences, for example the associated quiver
has a  gauge group that does not end, $\Pi_{k=1}^\infty SU(k N_6)$. In fact, there are no D6 brane sources, according to eq.(\ref{qd6good}). Similarly,  eqs.(\ref{qns5good}),(\ref{qd4good}) indicate a divergent number of five and four branes. The linking numbers do not satisfy eq.(\ref{consistency})
and the central charge in eq.(\ref{eqcc}), diverges as $\eta_f\to\infty$. The bad behaviour of the field theory observables is mirrored by a singularity in the background at $\sigma=1$. Still, some quantities may have an acceptable behaviour \footnote{We could regulate quantities using the Riemann $\zeta$-function $\zeta(s)=\sum_{k=1}^\infty \frac{1}{k^s}$. In fact, for a strictly infinite conformal quiver with gauge group $\Pi_{k=1}^\infty SU(k N_6)$ joined by bifundamental hypers, we have that $n_v=\sum_{k=1}^\infty (k^2 N_6^2 -1)$ and $n_h=\sum_{k=1}^\infty (k^2+k)N_6^2$. We obtain that 
\begin{equation}
\frac{a}{c} = \frac{5 n_v+n_h}{4 n_v+2 n_h}=\frac{\sum_{k=1}^\infty  6k^2 N_6^2 + k N_6^2-5}{\sum_{k=1}^\infty 6k^2 N_6^2+2k N_6^2 -4   }.\nonumber
\end{equation}
Using that $\zeta(-2)=0$, $\zeta(-1)=-\frac{1}{12}$ and $\zeta(0)\to\infty$, we find $\frac{a}{c}=\frac{5}{4}$. Satisfying the Hofman-Maldacena bound  \cite{Hofman:2008ar}.} .

These deficiencies might suggest that we should ignore the Sfetsos-Thompson solution as unphysical. But the background generated by $V_{ST}$ in eq.(\ref{sfetsosthompson}) has a very interesting property: the string theory sigma model is integrable on this background. This was shown in \cite{Borsato:2017qsx}. In particular, it was shown in \cite{Nunez:2018qcj}
that any other generic Gaiotto-Maldacena background as in eq.(\ref{10d}) leads to a non-integrable (and chaotic) sigma model for the string theory.

These ideas were exploited in \cite{Sfetsos:2014cea} to show that the Sfetsos-Thompson solution is a member of a family of integrable backgrounds. Interestingly, the geometry and fluxes produced by the potential $V_{ST}$ together with the definitions in eq.(\ref{10d}) were obtained in  \cite{Sfetsos:2010uq} by using  non-Abelian T-duality.  There are presently many new backgrounds that have been obtained using this powerful technique \cite{varios}.

It is in this sense that the Sfetsos-Thompson solution stands out as a paradigmatic example of non-Abelian T-duality as generating technique. While the conformal field theory obtained by following the prescription described in  Section \ref{GMgenerics} is not well defined \footnote{In \cite{Itsios:2017nou} the authors suggest that the system should be thought as a higher dimensional field theory with a conformal four dimensional defect.}, it was proposed in \cite{Lozano:2016kum} that the Sfetsos-Thompson solution should be embedded inside a 'complete' Gaiotto-Maldacena geometry, that regulates the background and solves the above mentioned problems of the CFT. The authors of \cite{Lozano:2016kum} suggested to consider the charge density in eq.(\ref{profile1}) as a regulator for $\lambda_{ST}$. Indeed, the solution in eq.(\ref{Vk}) with Fourier coefficients given in eq.(\ref{fourier1}) is proposed to be the potential from which to obtain the 'completed' background. This logic extended  successfully  \cite{extensions}-\cite{vanGorsel:2017goj}   to other backgrounds generated by non-Abelian T-duality. Below we comment on other ways to think about the Sfetsos-Thompson background and its associated CFT.

\subsubsection{A field theory view of the Sfetsos-Thompson background }
Let us add some comments about the field theoretical interpretation of the Sfetsos-Thompson background and non-Abelian T-duality (an operation on the string sigma model that generates a new background). We anticipate these comments to be adaptable to many other cases studied in \cite{extensions}.

Consider ${\cal N}=4$ Super-Yang-Mills. The bosonic part of the global 
symmetries is $SO(2,4)\times SO(6)$. We will use that $SO(6)\sim SU(2)_L\times SU(2)_R\times U(1)_r$. These symmetries are realised as isometries 
of the dual $AdS_5\times S^5$ background.  The non-Abelian T-dual transformation proposed by Sfetsos and Thompson in  \cite{Sfetsos:2010uq} picks the $SU(2)_L$ and operates on it. This operation preserves the $SO(2,4)$ as the $AdS_5$ part of the space is inert. The same happens to the $SU(2)_R\times U(1)_r$. Schematically the non-Abelian T-duality transforms
\begin{eqnarray}
& & AdS_5 +  d\alpha^2+\sin^2\alpha d\beta^2 +{\cos^2\alpha}d\Omega_3 \to  \nonumber\\
& &AdS_5 + d\alpha^2 +\sin^2\alpha d\beta^2 + \frac{d\rho^2}{\cos^2\alpha}+ \frac{\rho^2\cos^2\alpha }{\rho^2+\cos^4\alpha}d\Omega_2(\chi,\xi)\to \nonumber\\
& & AdS_5 +  \frac{1}{1-\sigma^2} (d\sigma^2+d\eta^2) +\eta^2 d\beta^2+ \frac{\eta^2(1-\sigma^2)}{4\eta^2+(1-\sigma^2)^2} d\Omega_2.\nonumber
\end{eqnarray}
In the last line we have changed variables $\sigma=\sin\alpha$ and $\rho\sim\eta$, to put the geometry in Gaiotto-Maldacena notation. The background is complemented by Ramond and Neveu-Scharz fields, for the details see for example  \cite{Lozano:2016kum}.

\begin{figure}[h!]
    \centering
    {{\includegraphics[width=12.5cm]{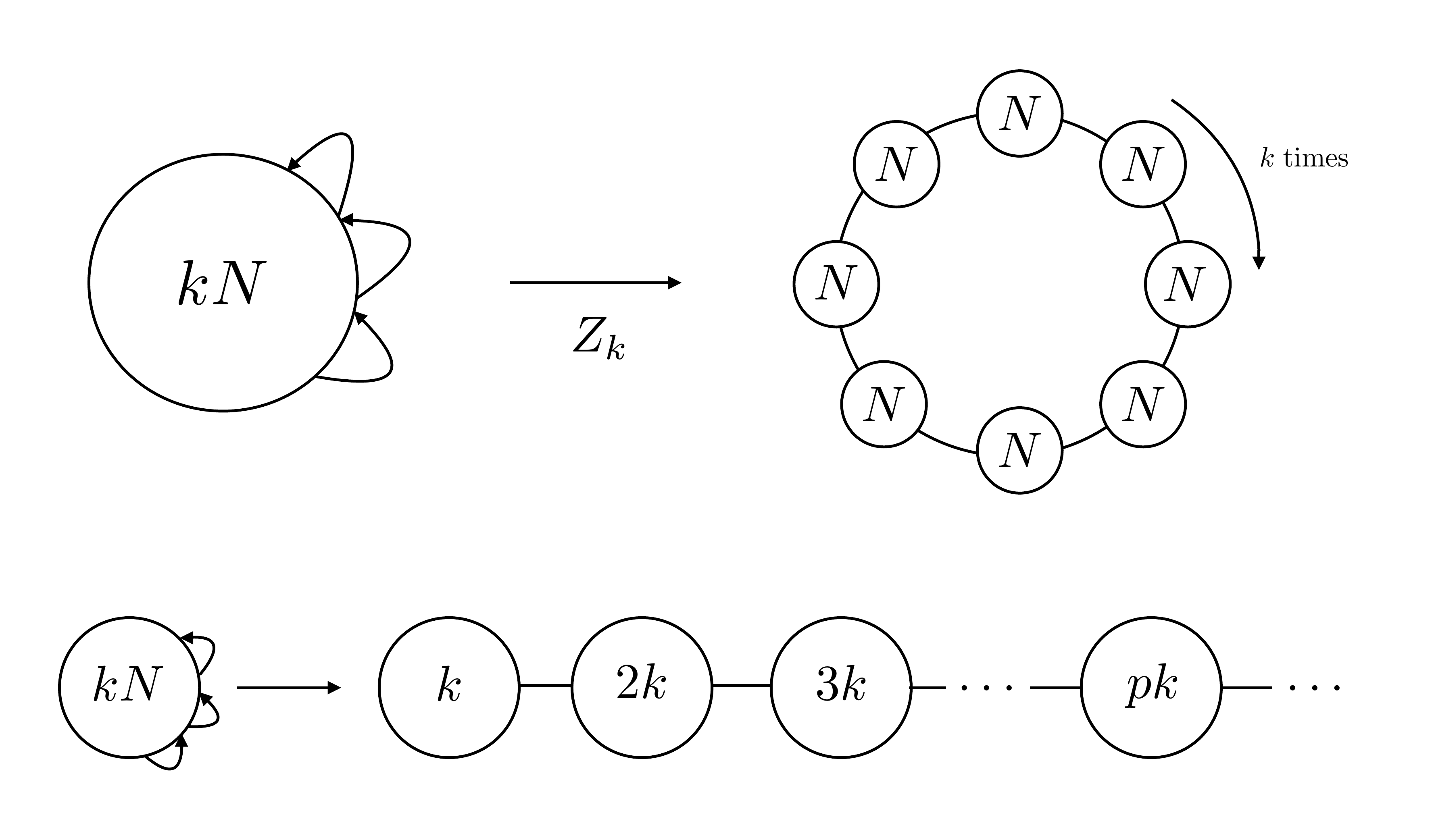} }}%
\caption{The two operations preserving conformality and $SU(2)_R\times U(1)_r$ as discussed in the text}
\label{CFTopx}
\end{figure}

The result is a background dual to an ${\cal N}=2$ SCFT,  with bosonic  isometries $SO(2,4)\times SU(2)_R\times U(1)_r$. One can imagine two operations on ${\cal N}=4$ SYM that  acting on $SU(2)_L$ produce an ${\cal N}=2$ SCFTs.  One is a modding by $Z_k$ and is represented in the top of Figure \ref{CFTopx}. The second is a higssing represented in the lower part of Figure \ref{CFTopx}.

The ranks of the gauge groups are determined by conformality. While the option on top of Figure \ref{CFTopx} is well defined, the one in the bottom runs into a problem as the quiver should extend infinitely. Another option is to end this linear quiver by the addition of a flavour group. This option is not available to the non-Abelian T-duality as it would imply the creation of an isometry $SU(kp+ k)$ and the presence of D6 sources to realise it. In the same vein, if we do not close the quiver, we eventually run-out of degrees of freedom to create a new gauge group, hence conformality  would be  compromised. The Sfetsos-Thompson solution reflects this by generating a singularity. Another alternative would be to start from the elliptic quiver on top of Figure \ref{CFTopx} and cut one bifundamental link. Then, distributing the degrees of freedom to enforce conformality in the linear quiver runs into the problems above discussed.
\\
Let us finally discuss a geometric aspect of the Sfetsos-Thompson background. We start by considering the derivative of the generic potential $\dot{V}(\sigma,\eta)$. Using eq.(\ref{Vk}) we compute
\begin{equation}
\dot{V}(\sigma,\eta)
= \sigma \partial_\sigma V(\sigma,\eta)= \sum_{k=1}^\infty c_k \sigma K_1(\frac{n\pi}{N_5}) \sin(\frac{n\pi}{N_5}\eta).\label{Vkdot}
\end{equation}
By Poisson summation, we rewrite this as \cite{ReidEdwards:2010qs},
\begin{equation}
\dot{V}(\sigma,\eta) = \frac{N_c}{2}\sum_{l=1}^{P} \sum_{m=-\infty}^{\infty} \int d\sigma \sigma \left[ \frac{1}{\sqrt{\sigma^2 +(\eta-\nu_l - m)^2 }     }-
\frac{1}{  \sqrt{ \sigma^2 +(\eta+ \nu_l - m)^2 }        } \right] . \label{Vksummn}
 \end{equation}
The values of the constants $\nu_l$ depend on the Fourier coefficients and can be found in  \cite{ReidEdwards:2010qs}.
\\
Of all the terms in the sum of eq.(\ref{Vksummn}),  we shall only keep the term $m=0$. We also approximate close to $\sigma=\eta=0$ to leading order both in $\sigma, \eta$. We find
\begin{equation}
\dot{V}(\sigma\to 0,\eta\to 0)\sim \dot{V}_{app}(\sigma,\eta)=\eta(c_1- c_2 \sigma^2)= \dot{V}_{ST}.\nonumber
 \end{equation}
 This is somewhat reminiscent of what occurs when lifting D2 branes to eleven dimensions \cite{Itzhaki:1998dd}. In that case, the correct solution is the one that contains  the  infinite number of 'images' just like eq.(\ref{Vkdot}) does.
The naive lifting of the D2 brane solution does not capture the full IR dynamics of D2 branes. By analogy this suggests that omitting 
the summation over the images in eq.(\ref{Vksummn}) misses the correct dynamics of the  CFT, that the completion in  \cite{Lozano:2016kum} provides.
\subsection{An interesting particular solution}
Around eq.(\ref{Vk}), we studied a general solution to the Laplace-like equation (\ref{toda})
with the boundary conditions of eq.(\ref{xxz}). This solution is the infinite superposition of functions of the type $V\sim K_0(\frac{n\pi \sigma}{N_5})\sin\frac{n\pi \eta}{N_5}$ with suitable coefficients. A natural question is what is the physical content of each term in this superposition. To answer this, we shall consider a solution to eq.(\ref{toda}) that is simply,
\begin{equation}
V(\sigma,\eta)=-K_0(\sigma)\sin\eta.
\label{papana}
\end{equation}
and study the background that this generates. In fact, by replacing in eq.(\ref{10d})-(\ref{definitions1}) we find, 
\begin{eqnarray}
& & 
\frac{ds_{10}^2}{L^2} = 4 \sigma \sqrt{\frac{K_2\left(  \sigma \right) }{K_0\left( \sigma \right) } } ds_{AdS_5}^2 + 2\frac{  \sqrt{K_0\left(  \sigma \right) K_2 \left( \sigma \right) } }{K1\left(\sigma\right)} (d \sigma^2 + d \eta^2) \nonumber\\
& & + 2  \frac{K_1\left(  \sigma \right) \sqrt{K_0\left(  \sigma \right) K_2\left(  \sigma \right) } \sin^2\eta}{ K_0\left( \sigma \right) K_2\left(\sigma\right)\sin^2\eta +  K_1^2\left(  \sigma \right) \cos^2 \eta }d\Omega^2 (\chi, \xi) 
+4 \sigma \sqrt{\frac{K_0\left(  \sigma \right)}{K_2\left( \sigma \right)} } d \beta^2, \nonumber\\
& & B_2=2 \alpha' \mu^2 \left(-\eta +  \frac{K_1^2\left(  \sigma \right) \sin \eta \cos \eta}{K_1^2\left(  \sigma \right) \cos^2 \eta + K_0\left(  \sigma \right) K_2\left(  \sigma \right) \sin^2 \eta}  \right) \sin \chi \, d\xi \wedge d\chi \, \nonumber\\
& &  C_1 = 2   \mu^4 \alpha'^{\frac{1}{2}} \frac{K_1^2\left(  \sigma \right) \cos \eta}{K_2\left( \sigma \right)} d \beta,\;\;\;e^{-2 \phi} = \frac{1}{2}  \mu^6 \sqrt{\frac{K_0\left(  \sigma \right)}{K_2^3\left(   \sigma \right)}} K_1\left(  \sigma \right) \left[K_1^2 \left( \sigma \right) \cos^2 \eta + K_0\left( \sigma \right) K_2\left(\sigma \right) \sin^2 \eta \right],\nonumber\\
& & A_3 =  -4  \alpha'^{\frac{3}{2}} \mu^6 \frac{K_0\left(  \sigma \right) K_1^2\left(  \sigma \right) \sin^3  \eta}{K_0\left(  \sigma \right) K_2\left( \sigma \right)\sin^2\eta + K_1^2\left( \sigma \right) \cos^2  \eta} \sin \chi \, d \xi \wedge d \chi \wedge d \beta.\label{configk0sin}
\end{eqnarray}
To get some intuition
about the physical meaning of this solution, we compare it with the background obtained in  eqs.(2.44)-(2.47) of the paper \cite{Lin:2005nh}. In fact, Lin and Maldacena describe there the configuration corresponding to type IIA Neveu-Scharz five branes on $R\times S^5$. The solution of eq.(\ref{configk0sin}) differs from the one in \cite{Lin:2005nh} by an 'analytic continuation' (that as explained in Section 3.1 of \cite{Lin:2004nb} changes $d\Omega_5\to AdS_5$ and  $-dt^2\to d\beta^2$).
This analytic continuation should also imply that the functions that  in eq.(\ref{configk0sin}) are $K_0(\sigma),K_1(\sigma),K_2(\sigma)$ (the modified Bessel functions of the second kind) turn into $I_0(\sigma),I_1(\sigma),I_2(\sigma)$  (the modified Bessel functions of the first kind) in eqs.(2.44)-(2.47) of \cite{Lin:2005nh}. 

This suggest that the solution of eq.(\ref{configk0sin}) represents NS five branes extended along $ AdS_5\times S^1_{\beta}$. The function $\lambda(\eta)=\sin\eta$ associated with the potential in eq.(\ref{papana}) does not have the characteristic of being a piece-wise continuous ensemble of straight lines as for example those in our examples of eqs.(\ref{profile1}),(\ref{profile2}) are. We may think about this background in eq.(\ref{configk0sin}) as one where the position of the D6 branes has been smeared and they are distributed along the whole $\eta$-direction. 

Analysing the asymptotics close to the position of the five branes, we find that the metric, dilaton and B-field read,
\begin{eqnarray}
& & ds^2(\sigma\to\infty)\sim 4\sigma(AdS_5+ d\beta^2) + d\sigma^2+ d\eta^2+\sin^2\eta d\Omega_2,\nonumber\\
& &e^{4\Phi}\sim e^{4\sigma} \sigma^2,\;\;\;\; B_2\sim (\eta-\cos\eta\sin\eta )d\Omega_2.\nonumber
\end{eqnarray}
We see that the integral  $\int H_3=N_5$ and that the dilaton diverges close to the five branes.

Interestingly, these solutions can offer a connection with the proposal of the paper \cite{Aharony:2015zea}, according to which (see page 33 in \cite{Aharony:2015zea}) any four dimensional CFT of the type we are studying contain, in a suitable limit of parameters, a decoupled sector that is dual to the 6d (0,2) theory on $AdS_5\times S^1$. 

Let us study some of the observables previously calculated. We  use  the solution corresponding to the first harmonic $V(\sigma,\eta)=-N_c K_0(\frac{\pi \sigma}{N_5}) \sin(\frac{\pi\eta}{N_5})$. Calculating with eqs.(\ref{qns5good}),(\ref{qd6good}),(\ref{qd4good}) and (\ref{eqcc}) we find,
\begin{equation}
Q_{NS}=N_5,\;\; Q_{D4}=\frac{2 N_5 N_6}{\pi}, \;\;\; Q_{D6}=\frac{2\pi N_6}{N_5},\;\;\;\; c=\frac{N_6^2 N_5}{8\pi}.\nonumber
\end{equation}
The particular solution studied should be thought as representing a situation where the D4 and D6 branes are smeared over the Hanany-Witten set up. We cannot identify a localised gauge or flavour group.

Just like the solution of eq.(\ref{papana}) could be thought as a 'smeared version' of the usual Gaiotto-Maldacena solutions with piece-wise continuous $\lambda(\eta)$, it would also be interesting to study the potential and associated charge density,
\begin{equation}
V(\sigma,\eta)=e^{-\eta}\left[c_1 J_0(\sigma) -\frac{\pi}{2} Y_0(\sigma) \right] +\log\sigma,\;\;\; \lambda(\eta)=1-e^{-\eta}.\nonumber
\end{equation}
as an approximation to the piece-wise continuous solution of \cite{Maldacena:2000mw}.
\\
To complement this study, in Appendix \ref{appendixBH} we present a new solution representing a black hole in a generic Gaiotto-Maldacena background and briefly discuss its thermodynamics.
\\
Let us now move to the second part of this work, where we study holographically the marginal deformation of these ${\cal N}=2$ SCFTs.

\section{ Part 2: marginal deformations of CFTs and holography}\label{sectionN=1}
The aim of this section is to start a discussion on marginal deformations of the ${\cal N}=2$ SCFTs studied above. The methods used to
find the holographic dual to these marginal deformations are those developed by Lunin and Maldacena \cite{Lunin:2005jy} and its extensions \cite{Gursoy:2005cn}, \cite{Gauntlett:2005jb}.
Let us start with a brief discussion of the field theory.  The aim now is to express a gamma deformed ${\cal N}=2$ SCFT in the language of ${\cal N}=1$ SCFT.

\subsection{Details about the deformation of the  CFT}\label{sectionCFT}
\begin{figure}[h!]
    \centering
    {{\includegraphics[width=12.5cm]{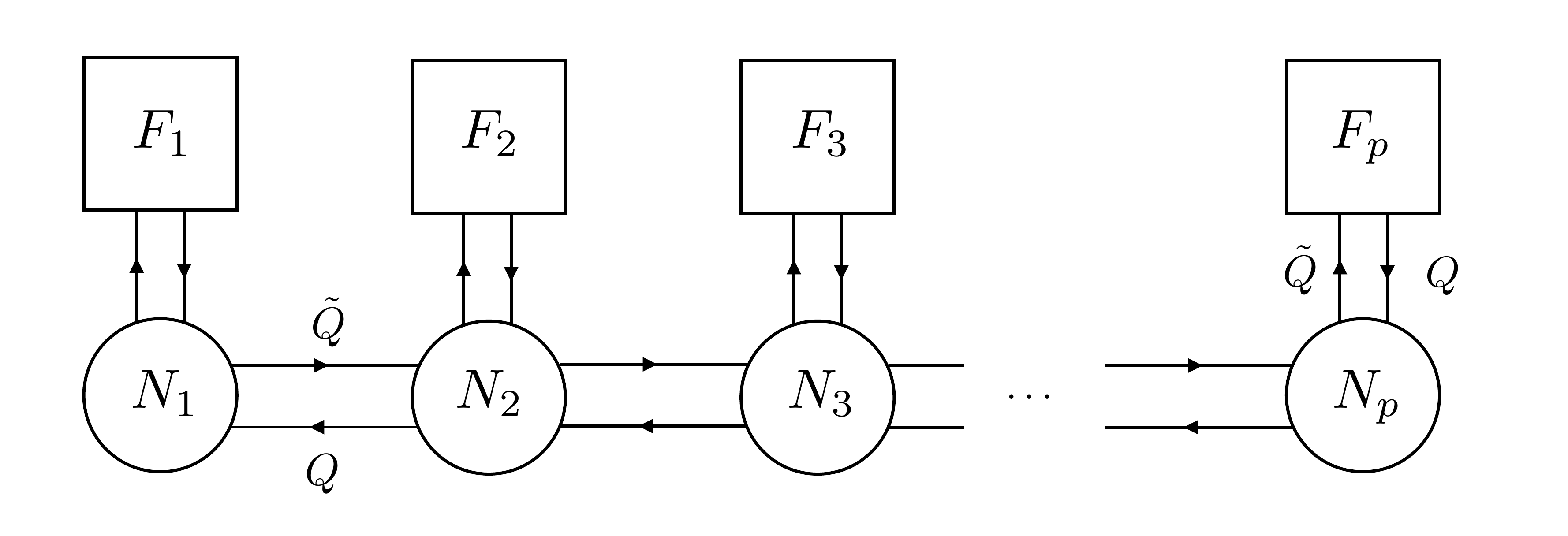} }}%
\caption{A generic ${\cal N}=1$ CFT.}
\label{figurexx}
\end{figure}

Consider a field theory like the one  represented in the quiver in the Figure \ref{figurexx}. There are gauge groups
$SU(N_1)\times SU(N_2)\times ....SU(N_P)$ with bifundamental fields in between the gauge groups and flavour groups $SU(F_1)\times ...\times SU(F_P)$. We are using the ${\cal N}=1$ notation, indicating an ${\cal N}=2$ hypermultiplet by two arrows. There  are also ${\cal N}=1$ adjoint fields associated with each gauge group. Expressing a generic ${\cal N}=2$ SCFT in terms of ${\cal N}=1$ multiplets  is useful when studying the marginal deformation.
%
\\
\\
Following the ideas of the papers \cite{Bah:2012dg}, \cite{Bah:2013aha}, 
we use that the R-symmetry mixes with 
the flavour symmetries. We propose the R-charges,
\begin{eqnarray}
& & R_{N=1}= R_0 +\frac{\epsilon}{2}F.\nonumber\\
& &R_{N=1}[Q]=R_{N=1}[\tilde{Q}]=\frac{1}{2}+ \frac{\epsilon}{2},\;\; R_{N=1}[\Phi]=1 -\epsilon,\;\;\; R[{\cal W}_\alpha]= 1.\label{chargesR}
\end{eqnarray}
 This is in line with the fact that the marginal
 deformation  does not change the number of degrees of freedom but just changes  the way in which the different fields interact.

To determine the value of $\epsilon$, we use a-maximisation \cite{Intriligator:2003jj}. The $a$ and $c$ central charges are 
\begin{eqnarray}
& & a(\epsilon)=\frac{3}{32\pi} \left[3 \textrm{Tr} R_{N=1}^3 -\textrm{Tr} R_{N=1}    \right],\;\;\; c(\epsilon)= \frac{1}{32\pi} \left[9
\textrm{Tr} R_{N=1}^3 -5\textrm{Tr} R_{N=1}    \right] .    \label{chargesepsilon}
\end{eqnarray}
For the quiver of the Figure \ref{figurexx}, we find that the contribution of 
the hypermultiplets $H=(Q,\tilde{Q})$  and the vector $V=({\cal W}_\alpha, \Phi)$  is,
\begin{eqnarray}
& & \textrm{Tr} R_{H}=2\times \frac{\epsilon-1}{2}\left(\sum_{j=1}^{P} N_j F_j  +\sum_{j=1}^{P-1}N_j N_{j+1}     \right)= n_{H}(\epsilon-1),\\
& & \textrm{Tr} R_{H}^3=2\times \frac{(\epsilon-1)^3}{8}\left(\sum_{j=1}^{P} N_j F_j  +\sum_{j=1}^{P-1}N_j N_{j+1}     \right)=n_H \frac{(\epsilon-1)^3}{4}.\nonumber\\
& & \textrm{Tr} R_{V}=\!\sum_{j=1}^P \!(N_j^2-1)(1-\epsilon)\!=\!n_V (1-\epsilon),\;\;\;\textrm{Tr} R_{V}^3=\!\sum_{j=1}^P\! (N_j^2-1)(1-\epsilon^3)\!=\!n_V (1-\epsilon^3).\nonumber
\end{eqnarray}
where $n_H$, $n_V$ is the total number of ${\cal N}=2$ hypermultiplets and vector multiplets in the quiver.
Using eq.(\ref{chargesepsilon}) we find
\begin{eqnarray}
& & a(\epsilon)=\frac{3}{32\pi} \left[ n_V \left( 3-3\epsilon^3 +\epsilon-1  \right)     + n_H\left(\frac{3(\epsilon-1)^3}{4}+1-\epsilon       \right)    \right]      \\   
& & c(\epsilon)= \frac{1}{32\pi} \left[9\left(n_V(1-\epsilon^3)+\frac{n_H}{4}(\epsilon-1)^3      \right) -5\Big( n_V (1-\epsilon) +n_H (\epsilon-1)    \Big)   \right] .     \nonumber
\label{chargesepsilon2}
\end{eqnarray}
We  maximise $a(\epsilon)$ and find  $\epsilon=\frac{1}{3}$. For the two charges $a,c$ above, we find the expression in eq.(\ref{centralchargeCFT})
$
a=\frac{5 n_V + n_H}{24\pi}$, $ c=\frac{2 n_V + n_H}{12\pi}$.

Hence, using eq.(\ref{chargesR}), for a marginal deformation that breaks from ${\cal N}=2$ to ${\cal N}=1$ SCFT, the R-charges  are given by
\begin{eqnarray}
& & R[Q]=R[\tilde{Q}]= R[\Phi]=\frac{2}{3},\;\;\;\; R[{\cal W}_\alpha]=1.\label{vava}
\end{eqnarray}
A superpotential term like 
\begin{equation}
W=h \sum_{j=1}^{P} \textrm{Tr} \left[ \Phi_j  Q_{j}\tilde{Q}_j  \right],\label{wsuperpot}
\end{equation}
has the correct $R$-charge R[W]=2 and the correct mass dimension (being $h$ dimensionless, hence marginal), satisfying $\textrm{dim}[W]=3=\frac{3}{2} R[W]$. Other possible gauge invariant operators, like ${\cal O}_1=\textrm{Tr} Q_j \tilde{Q}_j$ or ${\cal O}_2=\textrm{Tr} \Phi_j^2$ satisfy the unitarity bound
$1\leq \textrm{dim}~ {\cal O}=\frac{3}{2}R_{{\cal O}} $. 
\\
We find that the anomalous dimensions $\hat{\gamma}_{{fields}}$ are  vanishing $\hat{\gamma}_Q=\hat{\gamma}_{\tilde{Q}}=\hat{\gamma}_{\Phi}=0$. This in turn implies that the beta function of the couplings vanish $\beta_g\sim 3N- N_f(1-\hat{\gamma})=0$, $\beta_h=0-\hat{\gamma}_\Phi/2 -\hat{\gamma}_Q =0$. In this calculation, the adjoint chiral multiplets count as 'flavours' for a given gauge group. 
We can check that the CFTs we are dealing with above, do satisfy the bound $\frac{1}{2}\leq \frac{a}{c}\leq \frac{3}{2}$, in agreement with \cite{Hofman:2008ar}.
\\
\\
The marginal deformation is changing the products in the superpotential by powers of $e^{i R}$ (a combination of the R-charges of the fields participating in the interaction). There is not a RG-flow  taking place, but still we are breaking SUSY ${\cal N}=2\to{\cal N}=1$ via the interaction terms. No degrees of freedom are lost, as is supported by the calculation of the central charge, coincident with the ${\cal N}=2$ values. We just have different interactions between the fields and different  global symmetries.

Let us now discuss the holographic viewpoint of the above. We shall construct two different deformations of Gaiotto-Maldacena CFTs. They will be described by a parameter $\gamma$. We shall then calculate the central charge in each geometry finding the same result as in the parent ${\cal N}=2$ background. We will also compute the associated Page charges.

\subsection{Backgrounds  dual to  marginal deformations}
In this section we write the backgrounds constructed using various dualities. These backgrounds are proposed as duals to the ${\cal N}=1$ SCFTs  in the lines of what we discussed above. The details of the calculations are presented in the appendixes.
First we present a background in eleven dimensional supergravity and in Type IIA obtained using an $SL(3,R)$ transformation generalisation of the Lunin-Maldacena TsT \cite{Lunin:2005jy}.
Then we present a different solution obtained first by moving a generic Gaiotto-Maldacena background to Type IIB (via T-duality) and the performing a TsT transformation.
The outcome are two new families of solutions, one in M-theory/IIA, the other in Type IIB. They will be described in terms of a potential function $V(\sigma,\eta)$ satisfying a Laplace equation (\ref{toda}). For any solution to the Laplace equation with a given boundary condition, we generate a new solution in IIA/M-theory or in Type IIB.
\subsubsection{The $\gamma$-deformed backgrounds in eleven-dimensions and in Type IIA}\label{section3}
In this section, we shall present one possible $\gamma$-deformation of the Gaiotto-Maldacena backgrounds. We follow the formalism of  \cite{Gauntlett:2005jb}.
\\
Consider the eleven dimensional background in eq.(\ref{11d}) rewritten in the  form,
\be
\begin{split}\label{deformed}
&\quad ds^2=\mu^4\alpha' \left(\hat{\Delta}^{-1/6}g_{\mu\nu}dx^{\mu}dx^{\nu}+\hat{\Delta}^{1/3}M_{ab}\mathcal{D}\phi^{a}\mathcal{D}\phi^{b}\right), \quad a,b=1,2,3\\
C_3=&\kappa\left(C_{(0)}\mathcal{D}\phi^{1}\wedge\mathcal{D}\phi^{2}\wedge\mathcal{D}\phi^{3}+\frac{1}{2}C_{(1)ab}\wedge\mathcal{D}\phi^{a}\wedge\mathcal{D}\phi^{b}+C_{(2) a}\wedge\mathcal{D}\phi^{a}+C_{(3)}\right),\\
&\qquad \qquad \qquad \qquad  \qquad \mathcal{D}\phi^{a}=d\phi^{a}+A_{\mu}^{a}dx^{\mu}.
\end{split}
\e
All the coordinates are dimensionless quantities. We have $\phi^{1,2,3}=\xi,\beta, y$,  and 
\be\label{general11d}
\begin{split}
A_{\mu}^{a}=&0,\quad 
M_{ab}=\hat{\Delta}^{-1/3}\begin{pmatrix} 
F_3\sin^2\chi& 0 &0\\
0 & F_4+F_5\tilde{A}^2 &\tilde{A} F_5\\
0& \tilde{A} F_5&F_5
\end{pmatrix},
\quad \hat{\Delta}= F_3 F_4 F_5\sin^2\chi\\
&\quad \quad C_{(1)\xi\beta}= F_6\sin \chi d\chi, \quad C_{(1)\xi y}= F_7\sin \chi d\chi, \quad C_{(0)}=C_{(2)}=C_{(3)}=0\\
&\qquad \mu^4 \alpha'  \hat{\Delta}^{-1/6}g_{\mu\nu}dx^{\mu}dx^{\nu}=\kappa^{2/3}\left[4 F_1 ds_{AdS_5}^2 + F_2 (d\sigma^2 + d\eta^2)+F_3d\chi^2\right], 
\end{split}
\e
with $\kappa^{2/3}=\mu^4\alpha'$.
The functions $F_i$ and $\tilde{A}$ have been defined in eq.(\ref{Fixx}).

The background obtained  via an $SL(3,R)$ transformation with parameter $\gamma$ is constructed  following the rules of  \cite{Gauntlett:2005jb}.
We give details of the construction applied to this particular case in Appendix \ref{appendixD}. The resulting eleven dimensional solution is given by, 
\begin{eqnarray}\label{11ddef}
 \frac{ds^2}{\kappa^{2/3}}&=&\left(1+\gamma^2\hat{\Delta}\right)^{1/3}\left(4 F_1 ds_{AdS_5}^2+F_2(d\sigma^2+d\eta^2)+F_3 d\chi^2\right)+\nonumber\\
& &\left(1+\gamma^2\hat{\Delta}\right)^{-2/3}\left(F_3 \sin^2\chi d\xi^2 +F_4\tilde{\mathcal{D}}\beta^2+F_5\left(\tilde{\mathcal{D}}y+\tilde{A}\tilde{\mathcal{D}}\beta\right)^2\right),\nonumber\\
C_3&=&\kappa \left(\left(F_6\tilde{\mathcal{D}}\beta+F_7\tilde{\mathcal{D}}y\right)\wedge d\Omega_2(\chi,\xi)-\frac{\gamma \hat{\Delta}}{1+\gamma^2\hat{\Delta}}d\xi\wedge \tilde{\mathcal{D}}\beta\wedge \tilde{\mathcal{D}}y\right),
\end{eqnarray}
  where 
  \be
  \begin{split}
  \tilde{\mathcal{D}}\beta=d\beta-\gamma F_7\sin\chi d\chi, \quad   \tilde{\mathcal{D}}y=dy+\gamma F_6\sin \chi d\chi .
  \end{split} 
 \e
We have proved that this  is a solution of eleven-dimensional supergravity for any function $V(\sigma,\eta)$ solving eq.(\ref{toda}). Obviously, when $\gamma=0$ this background reduces to the one in eq. (\ref{11d}).

 We can write this family of solutions in Type IIA performing a reduction along the direction $y$---the details of this reduction are discussed in Appendix \ref{appendixreduction}---and write all functions in terms of those defined in eq.(\ref{definitions1}). The background in Type IIA is,
\begin{eqnarray}
& & ds_{10}^2 = \alpha' \mu^2 \Big[ 4 f_1 ds_{AdS_5}^2 +   f_2 (d\sigma^2 + d\eta^2) +f_3 d\chi^2 
 + \frac{f_3\sin^2\chi}{(1+\gamma^2 f_3f_4\sin^2\chi)} d\xi^2 +\nonumber\\
 & & \qquad \qquad\qquad    \frac{f_4}{(1+\gamma^2 f_3f_4\sin^2\chi)} (d\beta-\gamma f_5\sin\chi d\chi  )^2 \Big].\nonumber\\
& & e^{2\phi}= \frac{f_8}{(1+\gamma^2 f_3 f_4 \sin^2\chi)},\;\;\; C_1= \mu^4\sqrt{\alpha'}\left[f_6 d\beta +\gamma(f_7-f_5f_6)\sin\chi d\chi \right],\nonumber\\
& & B_2=\frac{\mu^2\alpha'}{(1+\gamma^2 f_3 f_4\sin^2\chi)}\left[f_5 d\Omega_2 -\gamma f_3f_4\sin^2\chi d\xi\wedge d\beta  \right],\nonumber\\
& &  A_3= \frac{\mu^6\alpha'^{3/2} }{(1+\gamma^2 f_3 f_4 \sin^2\chi)}f_7 d\beta\wedge d\Omega_2.
\label{gammaIIA}
\end{eqnarray}
As expected, when $\gamma=0$, we are back to the Gaiotto-Maldacena backgrounds in eqs.(\ref{10d})-(\ref{ns}).

In summary, we constructed a family of backgrounds with $SO(2,4)\times U(1)_\beta\times U(1)_\xi$ isometries. For any solution to the Laplace equation (\ref{toda}), we have a valid background. We have not checked the preservation of SUSY. The isometries suggest that the background preserves supersymmetry. One possible strategy to prove SUSY would be to put this background to the coordinates of \cite{Bah:2015fwa}, but finding such change of coordinates is not immediate. Nevertheless, given the arguments explained in \cite{Bashmakov:2017rko}, it seems likely that some amount of supersymmetry is preserved.
\\
We suggest that the integrability of the ${\cal N}=2$ Sfetsos-Thompson solution \cite{Sfetsos:2010uq} should translate into the integrability of the string sigma model in the background of eq.(\ref{gammaIIA}) for the case in which the functions $f_i$ are derived from the Sfetsos-Thompson potential in eq.(\ref{sfetsosthompson}). It would be interesting to find the Lax pair along the lines of \cite{Borsato:2017qsx}.
\subsubsection{The  gamma-deformed Type IIB backgrounds}\label{gammaIIBxx}
In this section we write the backgrounds obtained by moving the Gaiotto-Maldacena solutions to Type IIB via a T-duality and then performing a Lunin-Maldacena TsT transformation.

Let us apply a
  T- duality along the $\beta$ direction of the background in eq. (\ref{10d}). Using the Buscher rules
we find the T-dual NS sector, which reads
\begin{equation}\label{tIIB}
\begin{split}
ds^2=&\alpha' \mu^2 \left(4f_1 ds^{2}_{AdS_5}+f_2(d\sigma^2+d\eta^2)+f_3 (d\chi^2+\sin^2\chi d\xi^2)+f_4^{-1}\frac{d\beta^2}{\mu^4}\right),\\
&B_2=\alpha' \mu^2 f_5\sin \chi d\chi\wedge d\xi, \qquad e^{2\phi}=\frac{f_8}{\mu^2 f_4},
\end{split}\end{equation}
whilst the Ramond potentials and corresponding field strengths are
\begin{equation}\label{tIIBr}
\begin{split}
C_0=&\mu^4 f_6,\qquad C_2=\alpha'\mu^6 f_7\sin \chi d\chi\wedge d\xi\\
F_1=&dC_0, \quad F_3=dC_2-H_3C_0
\end{split}
\end{equation}
Let us apply now the TsT transformation to this solution. Following the rules 
 of the papers \cite{Lunin:2005jy,Gursoy:2005cn} (the details are given in Appendix \ref{appendixIIB}) we find the TsT transformed background
\begin{eqnarray}
\label{10ddef}
 & & ds^2=\alpha' \mu^2\left(4f_1 ds^{2}_{AdS_5}+f_2(d\sigma^2+d\eta^2)+f_3 d\chi^2\right.\nonumber\\
 & & + ~\left( \frac{1}{f_4+\gamma^2 f_3\sin^2\chi}\left(f_3f_4\sin^2\chi d\xi^2+(d\beta-\gamma f_5\sin\chi d\chi)^2\right)\right),\nonumber\\
& & e^{2\phi}=\frac{f_8}{\mu^2(f_4+\gamma^2 f_3\sin^2\chi)},\\
& & B_2=\alpha'\mu^2\left(\frac{\gamma f_3\sin^2\chi}{f_4+\gamma^2 f_3\sin^2\chi}(d\beta-\gamma f_5\sin\chi d\chi)\wedge d\xi+f_5\sin\chi d\chi\wedge d\xi,\right)\nonumber\\
& & C_0 =\mu^4f_6,\;\;C_2= \alpha'\mu^6 \left(\frac{\gamma f_6 f_3\sin^2\chi}{f_4+\gamma^2 f_3\sin^2\chi}(d\beta-\gamma f_5\sin\chi d\chi)\wedge d\xi+f_7\sin\chi d\chi\wedge d\xi\right), \nonumber
\end{eqnarray}
where $\gamma$ is the deformation parameter.  In addition, it is easily seen that after turning off the deformation parameter $\gamma$ the above background reduces to that in eqs. (\ref{tIIB}) and (\ref{tIIBr}). 
The same comments  as those written below eq.(\ref{gammaIIA}) apply here. We have shown that for any potential function satisfying eq.(\ref{toda}), the background of eq.(\ref{10ddef}) is solution of the Type IIB equations of motion. We have not explicitly checked the supersymmetry preservation, but the $SO(2,4)\times U(1)_\xi\times U(1)_\beta$  isometries suggest that some SUSY is preserved.
The construction of a Lax pair for the string sigma model on eq.(\ref{10ddef}), for the $f_i$ evaluated with the potential $V_{ST}$ in eq.(\ref{sfetsosthompson}) should be related to that in \cite{Borsato:2017qsx} via dualities.
\\
Let us calculate some observables of these backgrounds.
\subsubsection{Page charges and central charge}\label{PagegammaIIA}
We follow the treatment of Section \ref{pagechargesII} and compute the Page charges
of the backgrounds in eqs.(\ref{gammaIIA}), (\ref{10ddef}). For the Type IIA solutions in eq.(\ref{gammaIIA}), let us define the cycles,
\begin{eqnarray}
& &\Sigma_2=[\eta,\beta]_{\sigma=0},\;\;\;\hat{\Sigma}_2=[\eta,\chi]_{\sigma=0},\;\;\;
 \Sigma_3=[\eta,\chi,\xi]_{\sigma=\infty},\;\;\; \hat{\Sigma}_3=[\sigma,\beta,\xi].
\label{cyclesgamma}
\end{eqnarray}
We calculate the integrals
\begin{eqnarray}
& & Q_{NS5}=\frac{1}{2\kappa_{10}^2 T_{NS5}}\int_{\Sigma_3} H_3,\;\;\; \hat{Q}_{NS5}=\frac{1}{2\kappa_{10}^2 T_{NS5}}\int_{\hat{\Sigma}_3} H_3,\nonumber\\
& & Q_{D6}=\frac{1}{2\kappa_{10}^2 T_{D6}}\int_{\Sigma_2} F_2,\;\;\; \hat{Q}_{D6}=\frac{1}{2\kappa_{10}^2 T_{D6}}\int_{\hat{\Sigma}_2} F_2.\nonumber
\end{eqnarray}
The first and third integrals give the same results as in Section \ref{pagechargesII}, namely
\begin{equation}
Q_{NS5}=-\frac{2}{\pi}\mu^2 N_5,\;\;\;\; Q_{D6}=\mu^4 \left(\lambda'(\eta_f)-\lambda'(0)\right).
\end{equation}
As before, this implies the condition $\mu^2=\frac{\pi}{2}$. Hence $Q_{NS5}=N_5$ and, as before the definition $N_6=\frac{\pi^2}{4}N_c$ should be used.
The integral defining $\hat{Q}_{NS5}$ can be performed,
\begin{eqnarray}
& & \hat{Q}_{NS5}= \frac{1}{4\pi^2\alpha'} \mu^2\alpha' \gamma \int d\xi d\beta
 \int_{\sigma=0}^{\sigma=\infty} d\sigma \partial_{\sigma}\left[\frac{f_3f_4\sin^2\chi}{1+\gamma^2 f_3f_4 \sin^2\chi}  \right]=\nonumber\\
 & & \hat{Q}_{NS5}=-\frac{\mu^2}{\gamma}=\hat{N}_5.\label{nonzerogamma}
 \end{eqnarray}
 This implies a new quantisation condition $2\gamma \hat{N}_5=\pi$.  
 %
It may be confusing that in the limit of $\tilde{\gamma}\to 0$ the new charge of five branes diverges. But it should be observed that the component we are integrating to obtain $\hat{Q}_{NS5}$ is
vanishing in the limit $\tilde{\gamma}\to 0$.
Similarly, one can calculate $\hat{Q}_{D6}$,
 \begin{eqnarray}
 & & \hat{Q}_{D6}=\frac{1}{2\pi\sqrt{\alpha'}} \gamma \mu^4\sqrt{\alpha'}\int_0^\pi d\chi \sin\chi \int_0^{\eta_f} d\eta\partial_\eta[f_7(0,\eta)- f_5(0,\eta)f_6(0,\eta)]=\nonumber\\
& & \hat{Q}_{D6}=-\frac{\gamma \mu^4}{\pi} \left[f_7(0,\eta)- f_5 f_6(0,\eta)  \right
]_{\eta=0}^{\eta=\eta_f}=\gamma \frac{\pi}{2}N_5\lambda'(N_5).
\end{eqnarray}
For the solutions of Type IIB in eq.(\ref{10ddef}), we 
define the cycles,
\begin{eqnarray}
& & \Sigma_1=[\eta]_{\sigma=0},\;\;\; \Sigma_3=[\eta,\chi,\xi]_{\sigma\to\infty},\;\;\;\widehat{\Sigma}_3=[\sigma,\beta,\xi]_{\eta=\eta_0}.
\end{eqnarray}
 Using this, we calculate the following charges,
 \begin{eqnarray}
 & & Q_{D7}=\frac{1}{2\kappa_{10}^2 T_{D7}}\int_{\Sigma_1} F_1=\mu^4 (\lambda'(N_5)-\lambda'(0)),\label{cargasiib}\\
 & & Q_{NS5}=\frac{1}{2\kappa_{10}^2 T_{NS5}}\int_{\Sigma_3} H_3=\frac{\mu^2\alpha'}{4\pi^2\alpha'}\int d\Omega_2 \int_{0}^{N_5}\partial_\eta\left[ \frac{f_5f_4}{f_4 +\gamma^2 f_3\sin^2\chi}\right]=\frac{2\mu^2}{\pi^2}N_5.\nonumber\\
 & & \widehat{Q}_{NS5}=\frac{1}{2\kappa_{10}^2 T_{NS5}}\int_{\widehat{\Sigma}_3} H_3= \frac{\mu^2}{4\pi^2}\int d\xi d\beta \int_0^\infty \partial_\sigma\left[\frac{\gamma f_3\sin^2\chi}{f_4+\gamma^2 f_3\sin^2\chi} \right] d\sigma=\frac{\mu^2}{\gamma}=\widehat{N}_5.\nonumber
 \end{eqnarray}
 As in the Type IIA case, we see that a new set of NS-five branes appear and we need to impose that $\gamma=\frac{\pi}{2n}$.
\\ 
\\
\underline{{\it Central charge}}
\\
Let us now study the central charges. We follow the procedure
outlined in Section \ref{centralcharge}. 
For the Type IIA solutions, we  identify, from eq. (\ref{gammaIIA})
\begin{eqnarray}
& & \det[g_{int}]=(\alpha' \mu^2)^5 \frac{f_2^2 f_3^2 f_4 \sin^2\chi}{(1+\tilde{\gamma}^2 f_3 f_4 \sin^2\chi)^2},\\
& & a(R)= \alpha'\mu^2 4 f_1 R^2,\;\;\;\;\; e^{-4\phi}=\frac{(1+\tilde{\gamma}^2 f_3 f_4 \sin^2\chi)^2}{f_8^2}.\nonumber
\label{quantitiesgamma}
\end{eqnarray}
An straightforward computation  shows that the 
internal volume $V_{int}$ is,
\begin{equation}
V_{int}=\int d\eta d\sigma d\chi d\xi d\beta \sqrt{e^{-4\phi}\det[g_{int}] a(R)^3}= 64\pi^2\alpha'^4 \mu^8\int_{0}^{\eta_f} d\eta \int_0^{\infty} d\sigma \frac{f_1^{3/2} f_4^{1/2} f_2 f_3}{f_8}.\label{vintgamma}
\end{equation}
Using as above that $\dot{V}(\sigma\to\infty,\eta)=0$ and 
after some straightforward algebra we find that the internal volume in eq.(\ref{vintgamma})
is precisely equal to that in eq.(\ref{vintGM}). This implies, following the steps in eqs.(\ref{vintGM})-(\ref{eqcc}) that the
central charge for both backgrounds, the one in eqs.(\ref{10d}),(\ref{ns}) and that in eq.(\ref{gammaIIA}), is  the same and  given by eq. (\ref{eqcc}).  The same happens in Type IIB. This is in line with the fact that these solutions represent  CFTs that have the same number of degrees of freedom, but the interactions are slightly different.

These solutions are realising what we explained in Section \ref{sectionCFT}, namely they behave as ${\cal N}=1$ SCFTs with vanishing anomalous dimensions (they are 'finite SCFTs'). They have the same number of degrees of freedom that the parent ${\cal N}=2$ SCFTs have.
\\
In Appendix \ref{moreongammacft}, we discuss the role of the $\widehat{NS}$ five branes and propose a relation between the backgrounds in eqs.(\ref{gammaIIA}),(\ref{10ddef}) and brane box models.

\section{Conclusions and Future Directions}\label{conclusection}
In this work we have  presented several new entries in the dictionary between SCFTs in four dimensions and supergravity backgrounds with an $AdS_5$ factor.
New expressions were given,  calculating charges, number of branes and linking number of the branes composing the associated Hanany-Witten set-ups that encode the CFTs.
These expressions were written in terms $\lambda(\eta)$, the function fixing  a boundary condition of the Laplace equation, that encodes all the information of the supergravity background. We have tested these expressions in various examples of varying level of complexity and presented proofs for them, when available.
\\
We constructed holographic descriptions of marginal deformations of the ${\cal N}=2$ SCFTs above studied. New infinite families of solutions were constructed, again with all the information being encoded by a Laplace equation and its boundary conditions. New solutions were explored, observables calculated and CFT interpretation presented.
\\
It would be very interesting to repeat this type of calculation and derive analogous expressions for the observables for CFTs in diverse dimensions. 
\\
It would also be nice to study the integrability (or not) of the string sigma model on the backgrounds in Sections \ref{section3} and \ref{gammaIIBxx}, when evaluated on the potential in eq.(\ref{sfetsosthompson}).
\\
Another natural project would be to consider any of the CFT-supergravity background pairs presented here and deform them in such a way that a relevant operator acts on the CFT or the $AdS_5$ isometries are broken. The flow to the low-energy dynamics is surely very rich and depends on the details of the UV-CFT. Various new phenomena and entries in the supergravity-QFT dictionary will be encoded in these flows. We hope to report on these topics in the future.

\section*{Acknowledgments:} The input given by various colleagues, in many discussions and seminars shaped the findings and presentation of the topics of this paper.
We thank: Stefano Cremonesi, Timothy Hollowood, Daniel Thompson.
\\
CN is Wolfson Merit Research Fellow of the Royal Society.
\\
DR thanks The Royal Society UK and SERB India for financial support.

\appendix

\section{Physical Interpretation of $\lambda(\eta)$}\label{physicalintlambda}
The equation (\ref{toda}) and the conditions in eq.(\ref{boundaryconditions}) do not look
like the typical Laplace problem in two dimensions, but actually like a Laplace problem in three dimensions with a cyclic coordinate that does not belong to the space\footnote{In fact, the Laplace equation in an auxiliary space with metric $ds_3^2 = d\sigma^2+ d\eta^2+ \sigma^2 d\varphi^2$ for a function that is cyclic in the variable $\varphi$, $\nabla^2 V(\sigma,\eta)$ is eq.(\ref{toda}). }. Below, we show  that the interpretation of the quantity $\lambda(\eta)$ in eq.(\ref{lambdasum}) is precisely that of a charge density. To do this, we consider the solutions in the form of eq.(\ref{Vk}) and use an integral representation of the Bessel function $K_0(w_n \sigma)$,
\begin{equation}
K_0(w_n \sigma)= \int_0^\infty \frac{\cos (w_n\sigma t)}{\sqrt{t^2+1}} dt.
\end{equation}
Using the that $2 \cos x \sin y= \sin (x+y) - \sin(x-y)$,  the potential in eq.(\ref{Vk}) can be rewritten as,
\begin{eqnarray}
& & V(\sigma,\eta)=-\sum_{n=1}^{\infty} \frac{c_n}{2w_n}\left[\int_{0}^{\infty} \frac{\sin\left( w_n(\eta+\sigma t)\right)}{\sqrt{t^2+1}}dt   - \int_{0}^{\infty} \frac{\sin\left( w_n(-\eta+\sigma t)\right)}{\sqrt{t^2+1}}     dt \right]\\
& & V(\sigma,\eta)=-\sum_{n=1}^{\infty} \frac{c_n}{2w_n}\left[\int_{-\infty}^{\infty} \frac{\sin\left( w_n u \right)}{\sqrt{(u-\eta)^2+\sigma^2}} du\right].\nonumber
\end{eqnarray} 
 Now, exchanging the sum and the integral and using eq.(\ref{lambdasum}), we find
 \begin{equation}
 V(\sigma,\eta)=-\int_{-\infty}^{\infty} \frac{\lambda (  u)}{2 \sqrt{(u-\eta)^2+\sigma^2}} du=
-\int_{0}^{\infty} \frac{\lambda (  u)}{ \sqrt{(u-\eta)^2+\sigma^2}} du
\end{equation}
This precisely the electric potential produced by an odd-extended density of charge $\lambda$ along the $\eta$-axis, at some generic point $(\sigma,\eta)$. This makes clear the interpretation as an electrostatic problem.

\section{The 11d Supergravity-Type IIA connection}\label{appendixreduction}

In this appendix we start by connecting the ten dimensional background in eqs.(\ref{10d})-(\ref{ns})
with that in eq.(\ref{11d}), in other words, we 'oxidise' the ten dimensional Gaiotto-Maldacena background. We will pay special attention to the constants, $\mu, \alpha',\kappa$.

Start with eqs.(\ref{10d})-(\ref{ns}). We lift according to the usual prescription,
\begin{equation}
\begin{split}
ds_{11}^2 &= e^{-\frac{2}{3} \phi} ds_{10}^2 + e^{\frac{4}{3} \phi}(dx_{11} + C_{(1)})^2 \\
C_{3} &= A_{3} + B_{2} \wedge dx_{11} \, .
\end{split}
\label{baba}
\end{equation}
The dilaton given in eq.(\ref{ns}) can be re-written as,
\begin{eqnarray}
e^{-\frac{2}{3}\phi}=f_8^{-1/3}=\mu^2 \left(  \frac{4(2\dot{V}-\ddot{V})^3}{V''\dot{V}^2\Delta^2} \right)^{-1/6},\;\;\; e^{\frac{4}{3}\phi}=f_8^{2/3}=
\frac{1}{\mu^4} \left(  \frac{4(2\dot{V}-\ddot{V})^3}{V''\dot{V}^2\Delta^2} \right)^{1/3}.
\end{eqnarray}
Using eq.(\ref{baba}), we find the eleven dimensional metric to be,
\begin{eqnarray}
& & ds_{11}^2= \alpha' \mu^4  \left(  \frac{4(2\dot{V}-\ddot{V})^3}{V''\dot{V}^2\Delta^2} \right)^{-1/6} \Big[ 
  4 f_1 ds_{AdS_5}^2 +   f_2 (d\sigma^2 + d\eta^2) +f_3 ds_{S^2}^2(\chi,\xi) + f_4 d\beta^2   \Big] \nonumber\\
  & &  +\frac{1}{\mu^4} \left(  \frac{4(2\dot{V}-\ddot{V})^3}{V''\dot{V}^2\Delta^2} \right)^{1/3} (dx_{11} +\mu^4\sqrt{\alpha'} f_6d\beta)^2.
  \label{naba}
\end{eqnarray}
Now, the coordinates of the ten-dimensional part of the space are dimensionless.
On the other hand, the $x_{11}$-coordinate has length dimensions. We rescale (as $\partial_{x_{11}}$ is a Killing vector), $dx_{11}= dy \sqrt{\alpha'}\mu^4$ and we have,
\begin{eqnarray}
& & ds_{11}^2= \alpha' \mu^4  \left(  \frac{4(2\dot{V}-\ddot{V})^3}{V''\dot{V}^2\Delta^2} \right)^{-1/6} \Big[ 
  4 f_1 ds_{AdS_5}^2 +   f_2 (d\sigma^2 + d\eta^2) +f_3 ds_{S^2}^2(\chi,\xi) + f_4 d\beta^2   \Big] +\nonumber\\
  & &  {\alpha' \mu^4} \left(  \frac{4(2\dot{V}-\ddot{V})^3}{V''\dot{V}^2\Delta^2} \right)^{1/3} (dy + f_6d\beta)^2.
  \label{naba2}
\end{eqnarray}
Identifying $\mu^4\alpha'=\kappa^{2/3}$  and after simple algebra, we find the background in eq.(\ref{11d}). 

We can proceed similarly with the Kalb-Ramond fields,
\begin{eqnarray}
& & C_3=A_3+ B_2\wedge dx_{11}= \mu^6 \alpha'^{3/2} f_7 d\beta \wedge d\Omega_2 + \mu^2 \alpha' f_5 d\Omega_2 \wedge dx_{11}\nonumber\\
& & =  \mu^6 \alpha'^{3/2}\left[f_7 d\beta + f_5 dy      \right]\wedge d\Omega_2=\kappa \left[f_7 d\beta + f_5 dy      \right]\wedge d\Omega_2,
\end{eqnarray}
in coincidence with eq.(\ref{11d}), after using the definitions in eq.(\ref{Fixx}).

Following the same procedure, we connect the eleven dimensional background in eq.(\ref{11ddef}) with that in type IIA of eq.(\ref{gammaIIA}).

\section{Expansion of the various background functions}\label{appendixc}

Here, we write the expansions of the various functions appearing in the background for $\sigma\to 0$ using the potentials in eqs.(\ref{Vk})-(\ref{vtay}) and the expansion for $\sigma\to\infty$ using the expansion in eq.(\ref{Vk}).
\subsection{Expansion of the various background functions using the solution in eq.(\ref{Vk})}

We consider first the expressions in eq.(\ref{Vk}). We calculate,
 \begin{eqnarray}
& & \dot{V}(\sigma,\eta)=\sum_{n=1}^\infty
\frac{c_n}{w_{n}} (w_n \sigma)K_{1}(w_n \sigma)\sin (w_n \eta),\;\;\;
\dot{V}'(\sigma,\eta)=\sum_{n=1}^\infty
{c_n} (w_n \sigma)K_{1}(w_n \sigma)\cos (w_n \eta),\label{V1detail}\\
& & \ddot{V}(\sigma,\eta)=-\sum_{n=1}^\infty
\frac{c_n}{w_{n}} (w_n \sigma)^2 K_{0}(w_n \sigma)\sin (w_n \eta),\;\;\;
{V''}(\sigma,\eta)=\sum_{n=1}^\infty
{c_n}{w_{n}} (w_n )K_{0}(w_n \sigma)\sin (w_n \eta).\nonumber
\end{eqnarray}
Now, we use the previous expressions to compute,
\begin{eqnarray}
& & 2 \dot{V}-\ddot{V}=  \sum_{n=1}^\infty
\frac{c_n}{w_{n}} (w_n \sigma)^2K_{2}(w_n \sigma)\sin (w_n \eta)  ,\\
& & 2\dot{V}\dot{V}'= \sum_{n=1}^\infty \sum_{k=1}^\infty
\frac{c_n}{w_{n}} c_k (w_n \sigma) (w_k \sigma)K_{1}(w_n \sigma) K_1(w_k\sigma)\sin (w_n \eta)\cos(w_k\eta)      ,\nonumber\\
& & \Delta=  \left[\sum_{n=1}^\infty
\frac{c_n}{w_{n}} (w_n \sigma)^2 K_{2}(w_n \sigma)\sin (w_n \eta)   \right]  \times \left[ \sum_{k=1}^\infty
c_k  (w_k )K_{0}(w_k \sigma)\sin (w_k \eta)   \right]+ \nonumber\\
& &+   \left[ \sum_{n=1}^\infty
{c_n} (w_n \sigma)K_{1}(w_n \sigma)\cos (w_n \eta)\right]^2.\nonumber
\end{eqnarray}
To discuss expansions close to $\sigma=0$, we use
\begin{eqnarray}
& & z^2 K_2(z)\sim  2-\frac{z^2}{2}(3-4\gamma+\log 16-4\log z)   ,\\
& & z K_1(z)\sim 1+\frac{z^2}{4}(2\gamma-1-\log 4 +2 \log z)  +O(z^4)    ,\\
& & K_0(z)\sim  \log2-\gamma-\log z +\frac{z^2}{4}(1+\log 2-\gamma -\log z) . 
\end{eqnarray}
Using eq.(\ref{lambdasum}), we then find,
\begin{eqnarray}
& & 2 \dot{V}-\ddot{V}\sim 2\sum_{n=1}^{\infty} \frac{c_n}{w_n}\sin(w_n\eta)= 2 \lambda(\eta),\\
& & 2 \dot{V}\dot{V}'\sim 2 \sum_{n=1}^\infty \sum_{k=1}^\infty
\frac{c_n}{w_{n}} c_k \sin (w_n \eta)\cos(w_k\eta)= 2 \lambda(\eta) \lambda'(\eta)      ,\nonumber\\
& & \Delta\sim \log\sigma\to\infty.
\end{eqnarray}
The following combinations are useful. We study their $\sigma\to 0$ asymptotics,
\begin{eqnarray}
& & g_1=\frac{2\dot{V} \dot{V}'}{2\dot{V}-\ddot{V}}\sim \lambda'(\eta),\;\;\; g_2=2(\frac{\dot{V} \dot{V}'}{\Delta}-\eta)\sim-2\eta,\;\; g_3=-4 \frac{\dot{V}^2 V''}{\Delta}\sim -2 \lambda(\eta). \label{functionsuseful}
\end{eqnarray}

If we expand the potential function close to $\sigma\to\infty$
we use,
\begin{eqnarray}
& & z^2 K_2(z)\sim e^{-z}\sqrt{\frac{\pi z^3}{2}}, \;\;\; z K_1(z)\sim e^{-z}\sqrt{\frac{\pi z}{2}},\;\;\;
z^2 \partial_z K_1(z)\sim e^{-z}\sqrt{\frac{\pi z^3}{2}},\\
& & z^2 K_0(z)\sim e^{-z}\sqrt{\frac{\pi z^3}{2}},\;\;\; K_0(z)\sim e^{-z}\sqrt{\frac{\pi }{2z}},\\
& & 2\dot{V}-\ddot{V}\sim c_1 e^{-w_1 \sigma}\sqrt{\frac{\pi w_1 \sigma^3}{2}}\sin(w_1 \eta),\\
& & 2\dot{V}\dot{V}'\sim \frac{\pi c_1^2}{2w_1}\sin(w_1\eta)\cos(w_1\eta)e^{-2w_1 \sigma}(w_1\sigma),\\
& & \Delta\sim c_1^2\pi^2 w_1 \sigma e^{-2 w_1 \sigma}.\label{infiniteexpansions}
\end{eqnarray}



\section{How to count D4 branes?}\label{appendixd4}
In eq.(\ref{qd4good}) we presented a formula that counts the number of D4 branes in different Hanany-Witten set ups. This expression works nicely in the examples of eqs.(\ref{profile1})-(\ref{profile2}) and in those more elaborated examples studied in Appendix \ref{detailsCFT}.

\begin{figure}[h!]
    \centering
    {{\includegraphics[width=12.5cm]{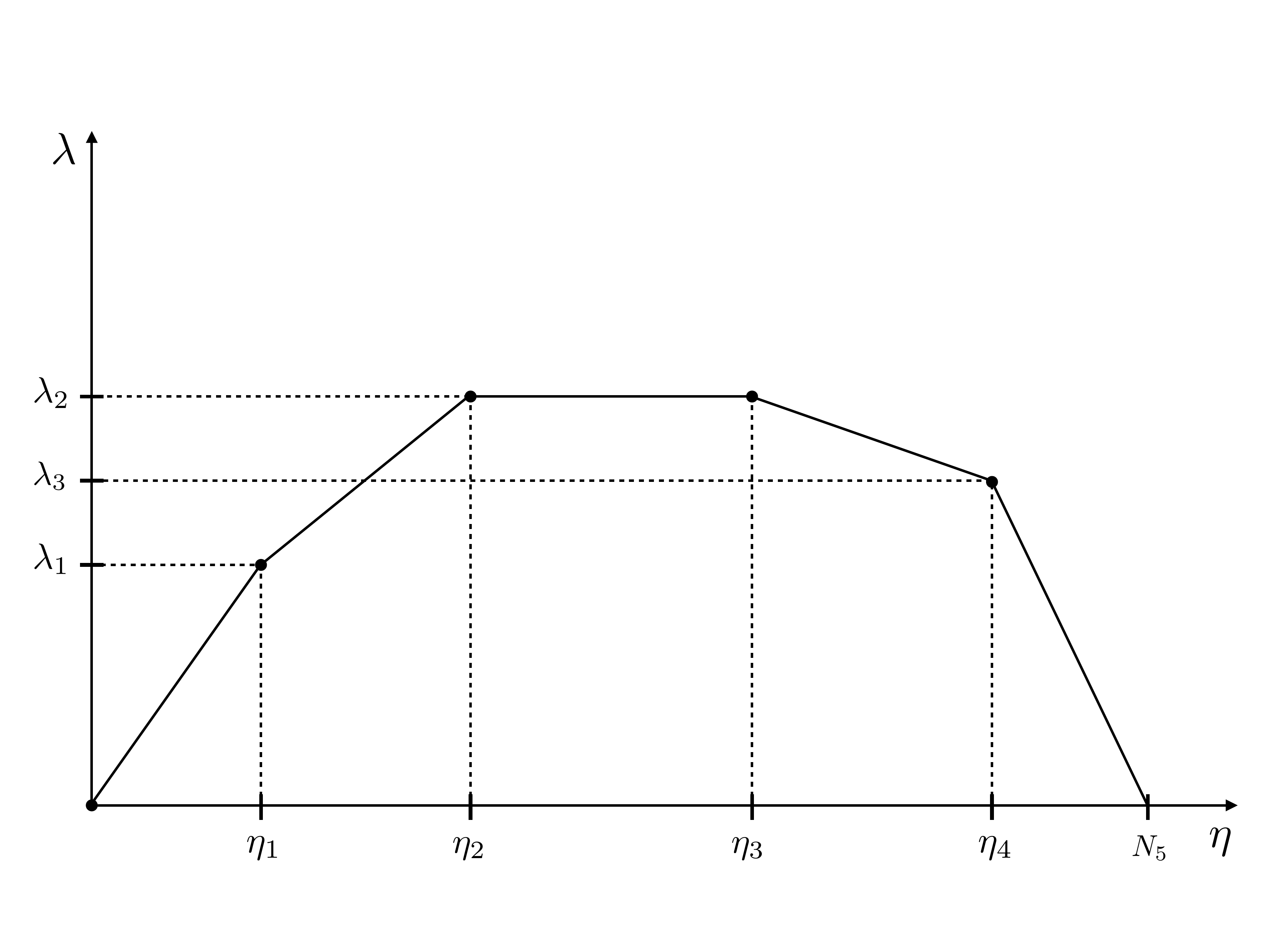} }}%

\caption{The charge density $\lambda(\eta)$ for the profile in eq.(\ref{profilegeneral}).}
\label{Figure-1-compact}
\end{figure}

Here, we  give a derivation of eq.(\ref{qd4good}) for a generic profile $\lambda(\eta)$.
In fact, consider an electrostatic charge profile
\be \label{profilegeneral}
\lambda(\eta)
                    =N_c\left\{ \begin{array}{ccrcl}
                      \frac{\lambda_1}{\eta_1} \eta & 0\leq\eta\leq \eta_1 \\
                      \lambda_1 +\left(\frac{\lambda_2-\lambda_1}{\eta_2-\eta_1}  \right)(\eta-\eta_1)& \eta_1\leq\eta\leq\eta_2\\
 \lambda_2 & \eta_2\leq\eta\leq\eta_3\\
 \lambda_2 +\left(\frac{\lambda_3-\lambda_2}{\eta_4-\eta_3}  \right)(\eta-\eta_3)& \eta_3\leq\eta\leq\eta_4\\
 \lambda_3 -\left(\frac{\lambda_3}{N_5-\eta_4}  \right)(\eta-\eta_4) & \eta_4\leq\eta\leq N_5.                   
                                             \end{array}
\right.
\e
As explained in the paper, we set $N_6=\frac{\pi^2}{4}N_c$. The charge profile is drawn in Figure \ref{Figure-1-compact}. 

\subsection{Number of D4 branes in the different intervals}
We shall count explicitly the number of D4 branes present in each interval. We will work out explicitly the counting in the five different intervals and  will check that this is coincident with the result of eq.(\ref{qd4good}).    

\begin{figure}[h!]
    \centering
    {{\includegraphics[width=12.5cm]{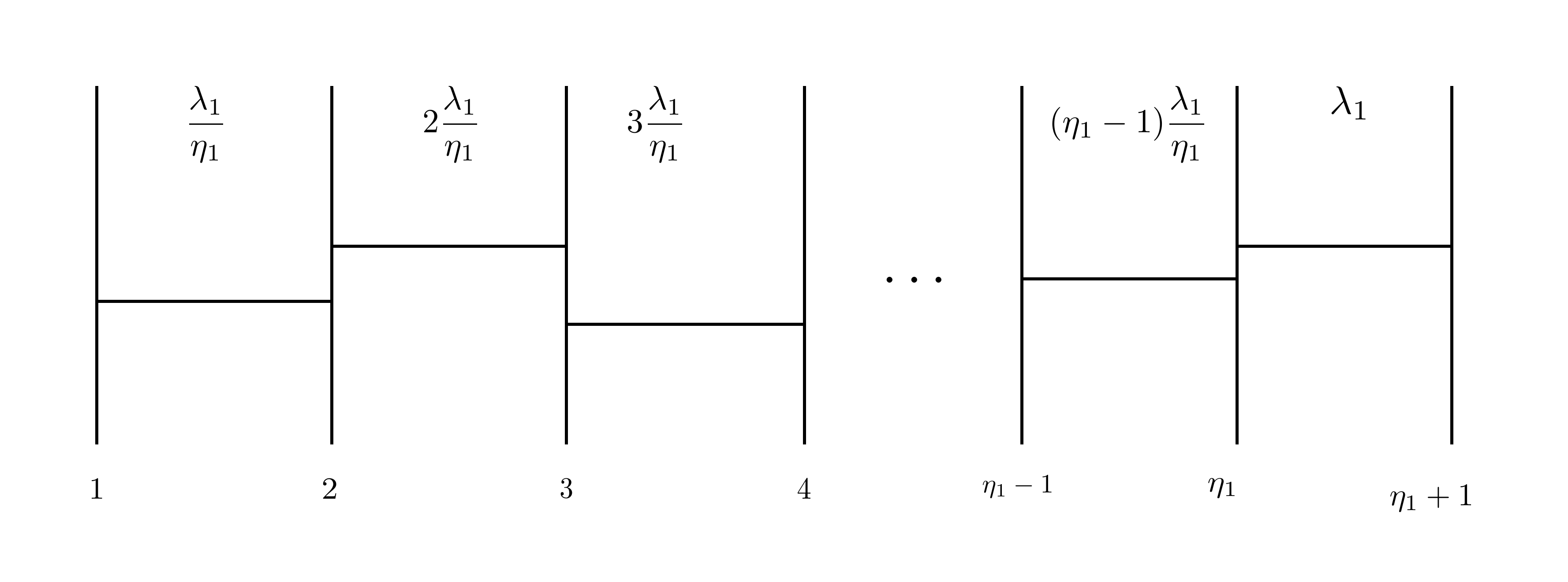} }}%

\caption{The Hanany-Witten set-up for the first interval $[0,\eta_1]$ of the profile in eq.(\ref{profilegeneral}). The number of branes should be multiplied by $N_6$}
\label{Figure-2-compact}
\end{figure}

\begin{figure}[h!]
    \centering
    {{\includegraphics[width=12.5cm]{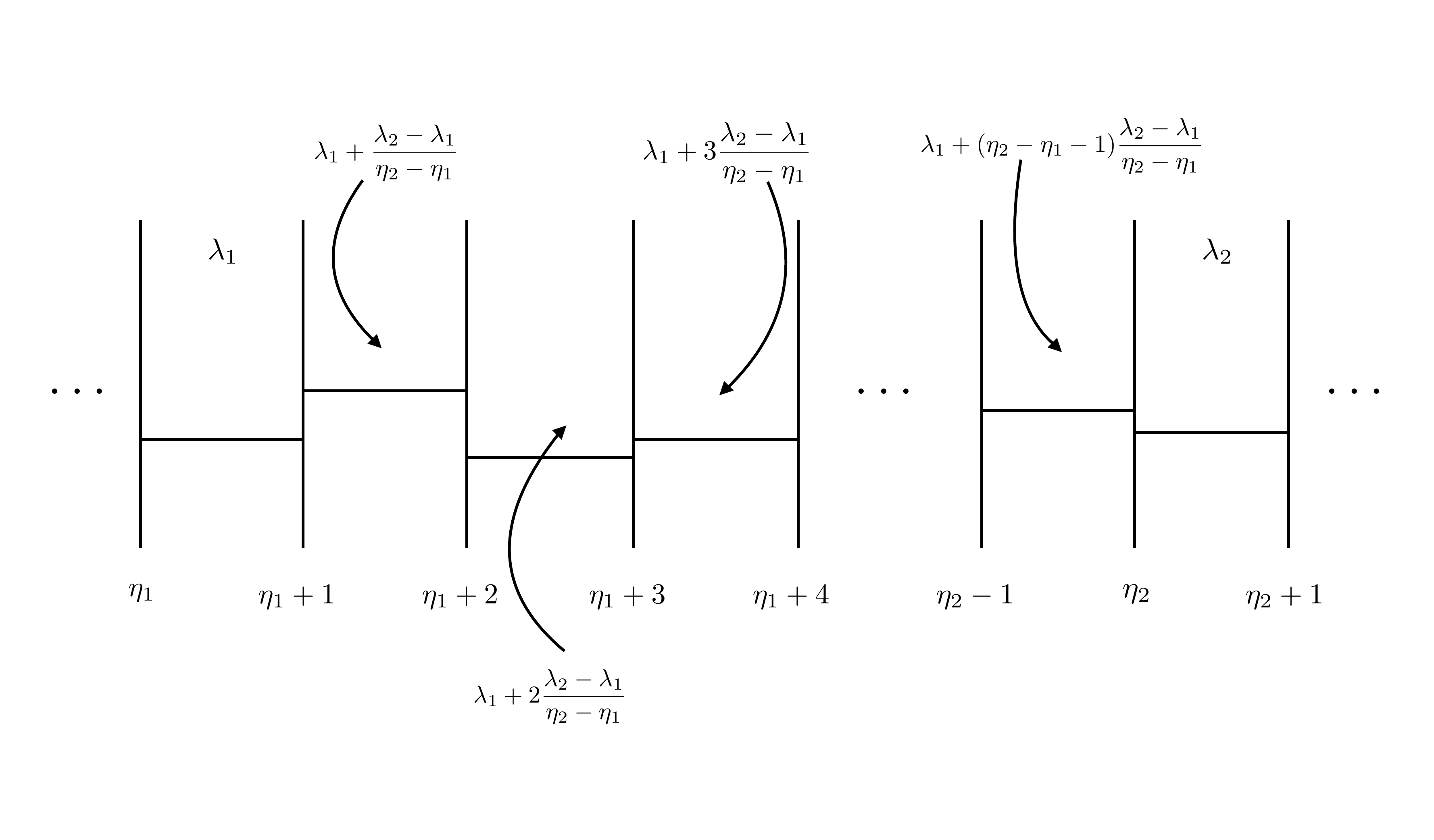} }}%

\caption{The Hanany-Witten set up corresponding to the second interval for the profile in eq.(\ref{profilegeneral}). The number of branes should be multiplied by $N_6$}
\label{Figure-3-compact}
\end{figure}

Consider the  portion of the Hanany-Witten set-up  shown\footnote{In what follows, we will not draw the D6-flavour branes, to avoid cluttering the figures} in Figure \ref{Figure-2-compact}. This corresponds to the first interval in the piecewise continuous function $\lambda(\eta)$ in eq.(\ref{profilegeneral}). We see that the number of D4 branes is
\begin{eqnarray}
& & N_{D4}=N_6 \frac{\lambda_1}{\eta_1}(1+2+3+4+....+\eta_1-1)+\lambda_1 N_6=\frac{\lambda_1 N_6}{2}(\eta_1+1).\label{intervalo1}
\end{eqnarray}
We now move to study the second interval $\eta_1\leq\eta\leq\eta_2$. In this case relevant part of the quiver and Hanany-Witten set-up are drawn in Figure \ref{Figure-3-compact}. We count explicitly and find 

\begin{figure}[h!]
    \centering
    {{\includegraphics[width=12.5cm]{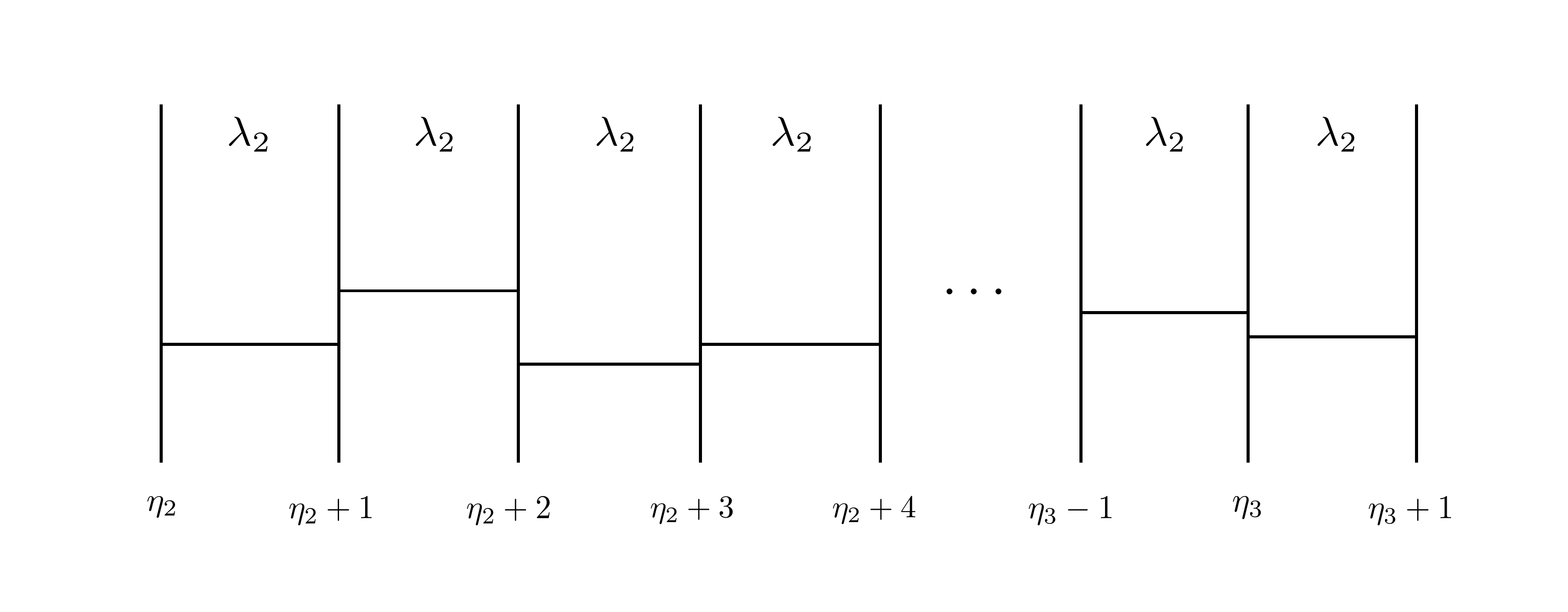} }}%

\caption{The Hanany-Witten set up corresponding to the third interval for the profile in eq.(\ref{profilegeneral}). The number of branes should be multiplied by $N_6$}
\label{Figure-4-compact}
\end{figure}

\begin{eqnarray}
& & N_{D4}= N_6 \left(\left[\lambda_1+ (\frac{\lambda_2-\lambda_1}{\eta_2-\eta_1}) \right] + \left[\lambda_1+ 2(\frac{\lambda_2-\lambda_1}{\eta_2-\eta_1}) \right] +....+\left[\lambda_1+ (\eta_2-\eta_1-1)(\frac{\lambda_2-\lambda_1}{\eta_2-\eta_1}) \right] +\lambda_2\right) \nonumber\\
& & =N_6\left( \sum_{r=1}^{\eta_2-\eta_1-1}\left[\lambda_1+ \frac{\lambda_2-\lambda_1}{\eta_2-\eta_1} r\right] \right)+ N_6\lambda_2=N_6 \frac{(\eta_2-\eta_1)(\lambda_1+\lambda_2)}{2}+\frac{\lambda_2-\lambda_1}{2} N_6.\label{intervalo2}
\end{eqnarray}
In the $[\eta_2,\eta_3]$ interval, whose Hanany-Witten set up is drawn in Figure \ref{Figure-4-compact} we find
\begin{equation}
N_{D4}= N_6\lambda_2 \sum_{r=1}^{\eta_3-\eta_2} 1=N_6 \lambda_2 (\eta_3-\eta_2).
\label{intervalo3}
\end{equation}

\begin{figure}[h!]
    \centering
    {{\includegraphics[width=12.5cm]{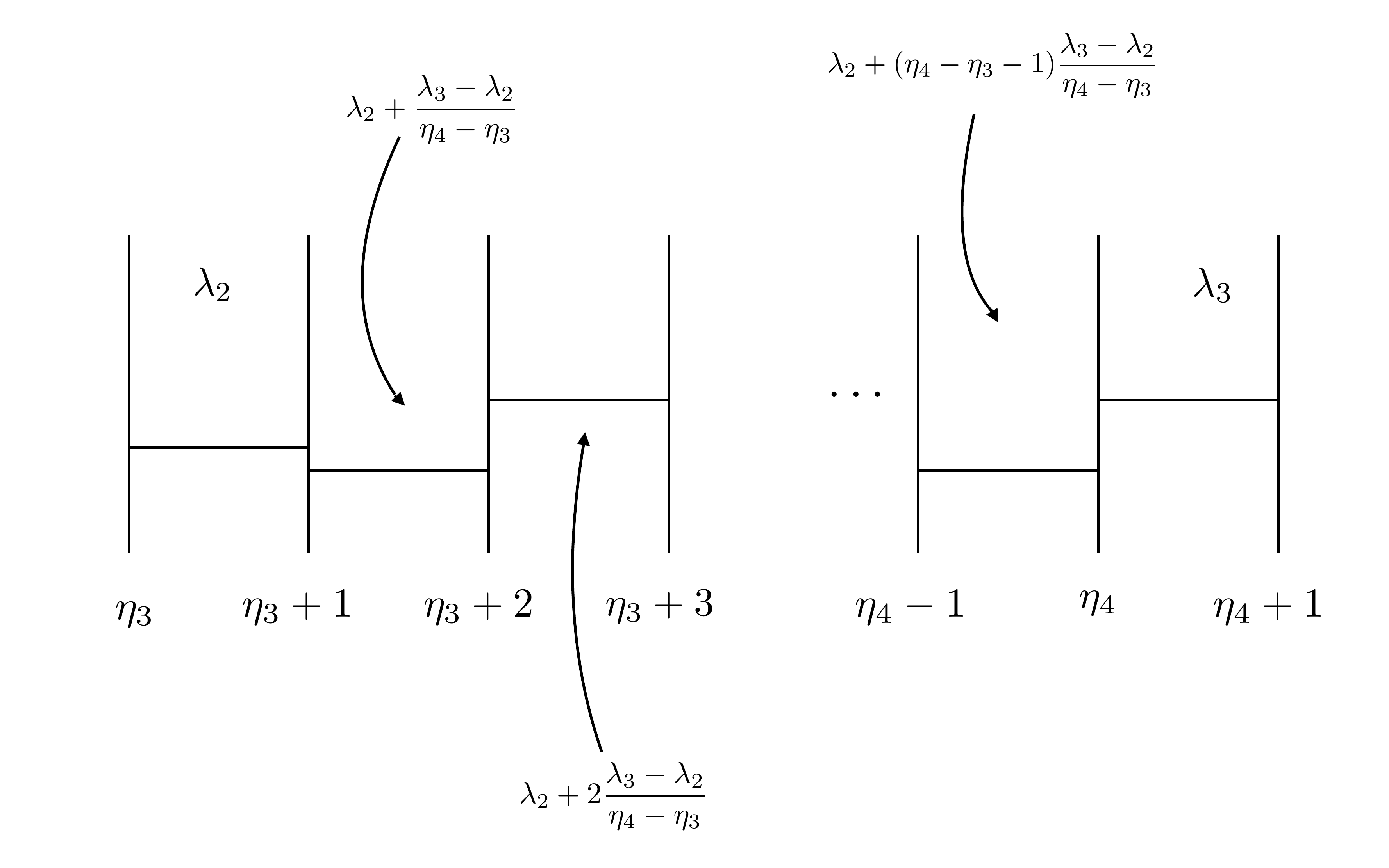} }}%

\caption{The Hanany-Witten set up corresponding to the fourth interval for the profile in eq.(\ref{profilegeneral}). The number of branes should be multiplied by $N_6$}
\label{Figure-5-compact}
\end{figure}

The rest of the intervals will work similarly to what we show above.
In fact, in the interval $[\eta_3,\eta_4]$---whose brane set-up is depicted in Figure \ref{Figure-5-compact} we find,
\begin{equation}
N_{D4}=N_6\left( \sum_{r=1}^{\eta_4-\eta_3-1} \left[\lambda_2+ r\frac{(\lambda_3-\lambda_2)}{(\eta_4-\eta_3)}\right]\right)+N_6\lambda_3=N_6\frac{(\lambda_2+\lambda_3)(\eta_4-\eta_3)}{2}
+\frac{(\lambda_3-\lambda_2)}{2}N_6.\label{intervalo4}
\end{equation}

\begin{figure}[h!]
    \centering
    {{\includegraphics[width=12.5cm]{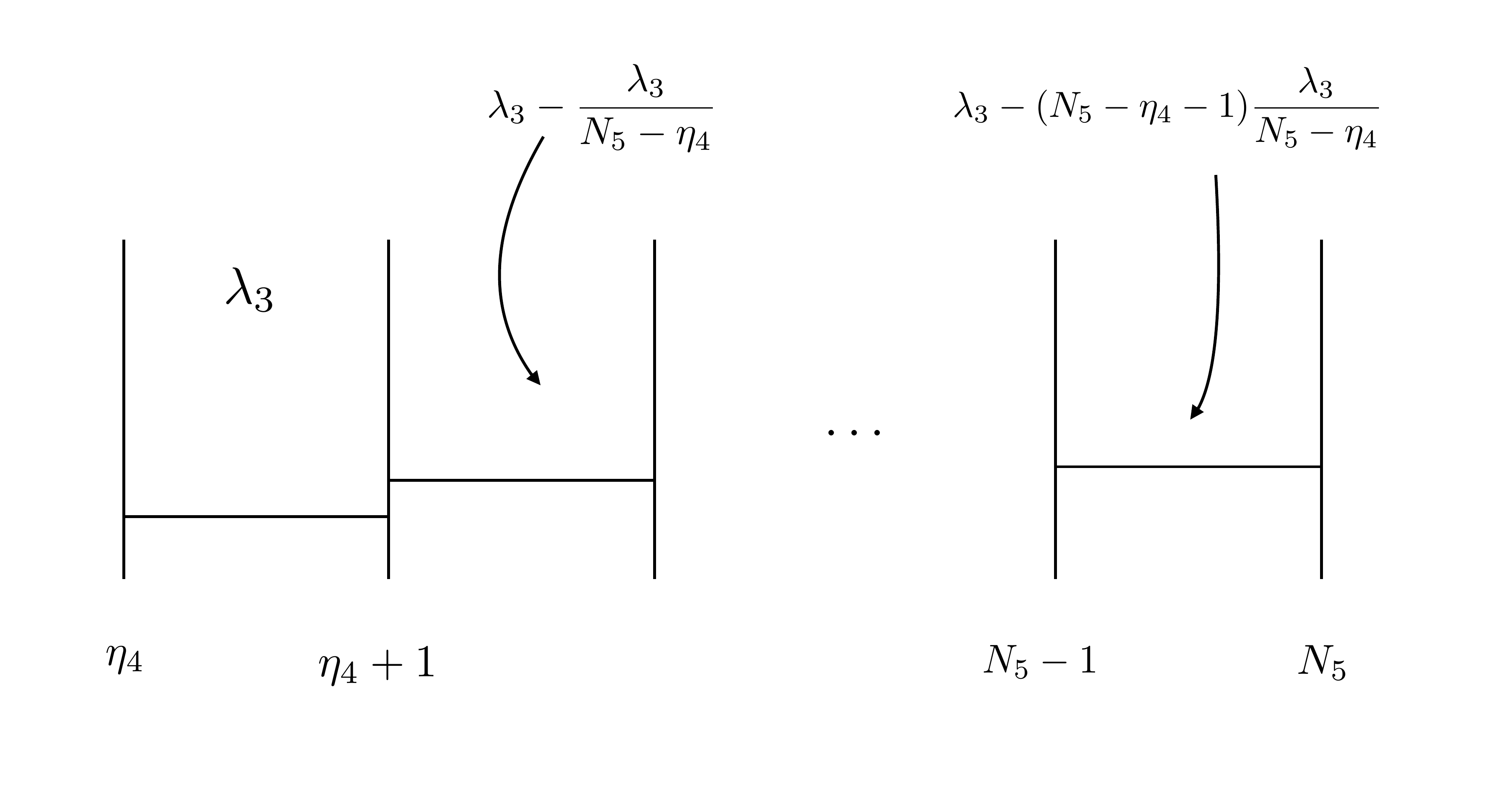} }}%

\caption{The Hanany-Witten set up corresponding to the last interval for the profile in eq.(\ref{profilegeneral}). The number of branes should be multiplied by $N_6$}
\label{Figure-55-compact}
\end{figure}

For the $[\eta_4,N_5]$ interval, corresponding to the brane set-up of Figure \ref{Figure-55-compact}, we have,
\begin{equation}
N_{D4}=N_6\sum_{r=1}^{N_5-\eta_4-1}\left[ \lambda_3 -\frac{\lambda_3}{N_5-\eta_4}r\right]=\frac{N_6\lambda_3}{2}(N_5-\eta_4)-\frac{N_6\lambda_3}{2}.\label{intervalo5}
\end{equation}

Summing the results for the five intervals in eqs(\ref{intervalo1})-(\ref{intervalo5}), we find
\begin{eqnarray}
& & N_{D4}=\frac{N_6}{2}\left[\lambda_1\eta_1 +(\lambda_2+\lambda_1)(\eta_2-\eta_1)+2\lambda_2(\eta_3-\eta_2)+(\lambda_2+\lambda_3)(\eta_4-\eta_3)+\lambda_3(N_5-\eta_4)       \right]\nonumber\\
& &=\frac{2\mu^6}{\pi}\int_0^{N_5}\lambda(\eta) d\eta.
\end{eqnarray}
This result is obtained for a generic Gaiotto-Maldacena charge profile, like that in eq.(\ref{profilegeneral}), hence justifying the validity of eq.(\ref{qd4good}).
\subsection{A derivation for the formula in eq.(\ref{qd4good}) }
In this section we will provide a derivation for the formula counting the number of D4 branes, see eq.(\ref{qd4good}). To this end consider a non-trivial profile for the function $\lambda(\eta)$ respecting the boundary conditions stated in eq.(\ref{boundaryconditions}). Let us write the function $\lambda$,
\be\label{piecewiselambda}
\lambda(\eta)
                    =N_6\left\{ \begin{array}{ccrcl}
                      \frac{\lambda_1}{\eta_1} \eta & 0\leq\eta\leq \eta_1 \\
                      \lambda_1 +\left(\frac{\lambda_2-\lambda_1}{\eta_2-\eta_1}  \right)(\eta-\eta_1)& \eta_1<\eta\leq\eta_2\\
 \vdots\\
 \lambda_{n-1} -\left(\frac{\lambda_n-\lambda_{n-1}}{\eta_{n}-\eta_{n-1}}  \right)(\eta-\eta_{n-1}) & \eta_{n-1} <\eta\leq \eta_n.                   
                                             \end{array}
\right.
\e
Notice that in order to satisfy the boundary conditions in eq.(\ref{boundaryconditions}) we must choose $\lambda_n =\lambda_0= 0$. Following the previous section, it is not difficult to see that the counting of D4 branes of the Hanany-Witten set up  can be done in the following way\footnote{Notice that this last formula acquire a precise meaning only after the sum over $r$ is carried out.}
\be
Q_{D4}=N_6 \sum_{s=1}^n \sum_{r=1}^{\eta_s-\eta_{s-1}} \left( \lambda_{s-1} + \frac{\lambda_s-\lambda_{s-1}}{\eta_s-\eta_{s-1}} r \right)  \, .
\e
The first sum explicitly leads to the following result
\be
Q_{D4}=N_6 \sum_{s=1}^n \left( \frac{\lambda_{s-1}-\lambda_{s}}{2} \right) + N_6 \sum_{s=1}^n \frac{\lambda_{s} + \lambda_{s-1}}{2}(\eta_s-\eta_{s-1})  \, .
\e
The first sum amounts to computing the difference $\lambda_0-\lambda_n=0$ (because of the boundary conditions). We end up with the following result
\be
Q_{D4}= N_6 \sum_{s=1}^n \frac{\lambda_{s} + \lambda_{s-1}}{2}(\eta_s-\eta_{s-1}) \, .
\e
Taking the continuous limit (i.e. sending $n$ to infinity and taking infinitesimal the distance $\eta_s-\eta_{s-1}$) the approximation becomes exact and we get the formula in eq.(\ref{qd4good}),
\be
Q_{D4}= N_6 \int_0^{N_5} \lambda(\eta) d \eta \, ,
\e
where we have made the identification $\eta_n \equiv N_5$.
\subsection{Counting of D6 branes}
The D6 branes appear every time we change intervals in eq.(\ref{profilegeneral}). In fact, whenever the derivative $\lambda'(\eta)$ shows a discontinuity, this indicates the presence of D6 branes.
The number is precisely the one needed to satisfy that every gauge groups $SU(\lambda_i)$ has $2\lambda_i$ flavours. We can count the changes in slope for each interval in the profile of eq.(\ref{profilegeneral}). We find,
\begin{eqnarray}
& & Q_{D6}^{(1)}=N_6\left(\frac{\lambda_2-\lambda_1}{\eta_2-\eta_1}-\frac{\lambda_1}{\eta_1}   \right),\;\;\; Q_{D6}^{(2)}=N_6\left(0-\frac{\lambda_2-\lambda_1}{\eta_2-\eta_1}   \right)          ,\nonumber\\
& & Q_{D6}^{(3)}=N_6\left(\frac{\lambda_3-\lambda_2}{\eta_4-\eta_3}-0   \right),\;\;\; Q_{D6}^{(4)}=N_6\left(-\frac{\lambda_3}{N_5-\eta_4}-\frac{\lambda_3-\lambda_2}{\eta_4-\eta_3}   \right)          ,\nonumber\\
& &Q_{D6}^{total}= \sum_i Q_{D6}^{(i)}= N_6 \left[ \frac{\lambda_3}{n_5-\eta_4}+\frac{\lambda_1}{\eta_1}\right]=-\mu^4N_c (\lambda'(N_5)-\lambda'(0)).
\end{eqnarray}
This shows the validity of eq.(\ref{qd6good}).

\section{Entanglement Entropy}\label{appendixEE}
The calculation of the Entanglement Entropy for a square region was studied in various papers. General formulas are presented in \cite{Klebanov:2007ws}, \cite{Kol:2014nqa}. In fact, following those papers, one finds expressions for the (density of) Entanglement Entropy $S_{EE}$ in terms of the length of a region $L$, by solving a minimisation problem for an eigth-surface exploring the bulk, as a function of the turn-around point in the bulk $R_*$. We have,
\begin{eqnarray}
& & \frac{2G_{10}}{V_2}S_{EE}= \int_{R_*}^{\infty} dR {\cal H}(R) \sqrt{\frac{b(R)}{{\cal H}(R) -{\cal H}(R_*)}} -\int_0^{\infty} dR \sqrt{b(R) {\cal H}(R)},\nonumber\\
& & L(R_*)= 2\sqrt{{\cal H}(R_*)}\int_{R_*}^\infty dR  \sqrt{\frac{b(R)}{{\cal H}(R) -{\cal H}(R_*)}}.
\end{eqnarray}
Here, the functions $b(R)$ and ${\cal H}(R)=V_{int}^2$ are the same ones appearing when studying the central charge, see eqs.(\ref{manaba}),(\ref{vintxx}). Changing variables to $R= R_* v$ and using the explicit expressions $b(R)=\frac{1}{R^4}$, ${\cal H}={\cal N}^2 R^6$, we find
\begin{eqnarray}
& & \frac{2G_{10}}{V_2}S_{EE}={\cal N} R_*^2 \Big(\int_{1}^{\infty} dv \frac{v^4}{\sqrt{v^6-1}}-\int_0^{\infty} dv v\Big)=\hat{q} ~{\cal N} R_*^2,\nonumber\\
& & L(R_*)= \frac{2}{R_*}\int_{1}^\infty dv \frac{1}{\sqrt{v^4(v^6-1)} }= 2\sqrt{\pi}\frac{\Gamma(\frac{2}{3})}{\Gamma(\frac{1}{6})}\frac{1}{R_*}.
\end{eqnarray}
Finally, using the values for $G_{10}$ and $\mu$ found above, we obtain,
\begin{equation}
{\frac{S_{EE}}{V_2}= \frac{\hat{q}\pi^4 (\Gamma(2/3))^2 }{4 (\Gamma(1/6))^2} \frac{1}{L^2}\int_0^{N_5} \lambda^2(\eta) d\eta= \frac{\hat{q}\pi^2 \left(\Gamma(2/3)\right)^2 N_5^3}{2 \left(\Gamma(1/6)\right)^2} \frac{1}{L^2} \sum_{m=1}^{\infty}\frac{c_m^2}{m^2}.}
\label{EEfinal}
\end{equation} 
This is the result expected for a CFT (the $L^{-2}$ dependence). The dynamics is in the integral of $\lambda^2$ or in the sum of harmonics. This will distinguish different CFTs.

\

\section{General ${\cal N}=2$ quivers and matching of observables}\label{detailsCFT}
In this appendix we work out the field theory and dual gravity Page charges, linking numbers and central charge for various quivers,  genricaly more elaborated than those in the main part of this work.
\subsection{First example}
\begin{figure}[h!]
    \centering
    {{\includegraphics[width=12.5cm]{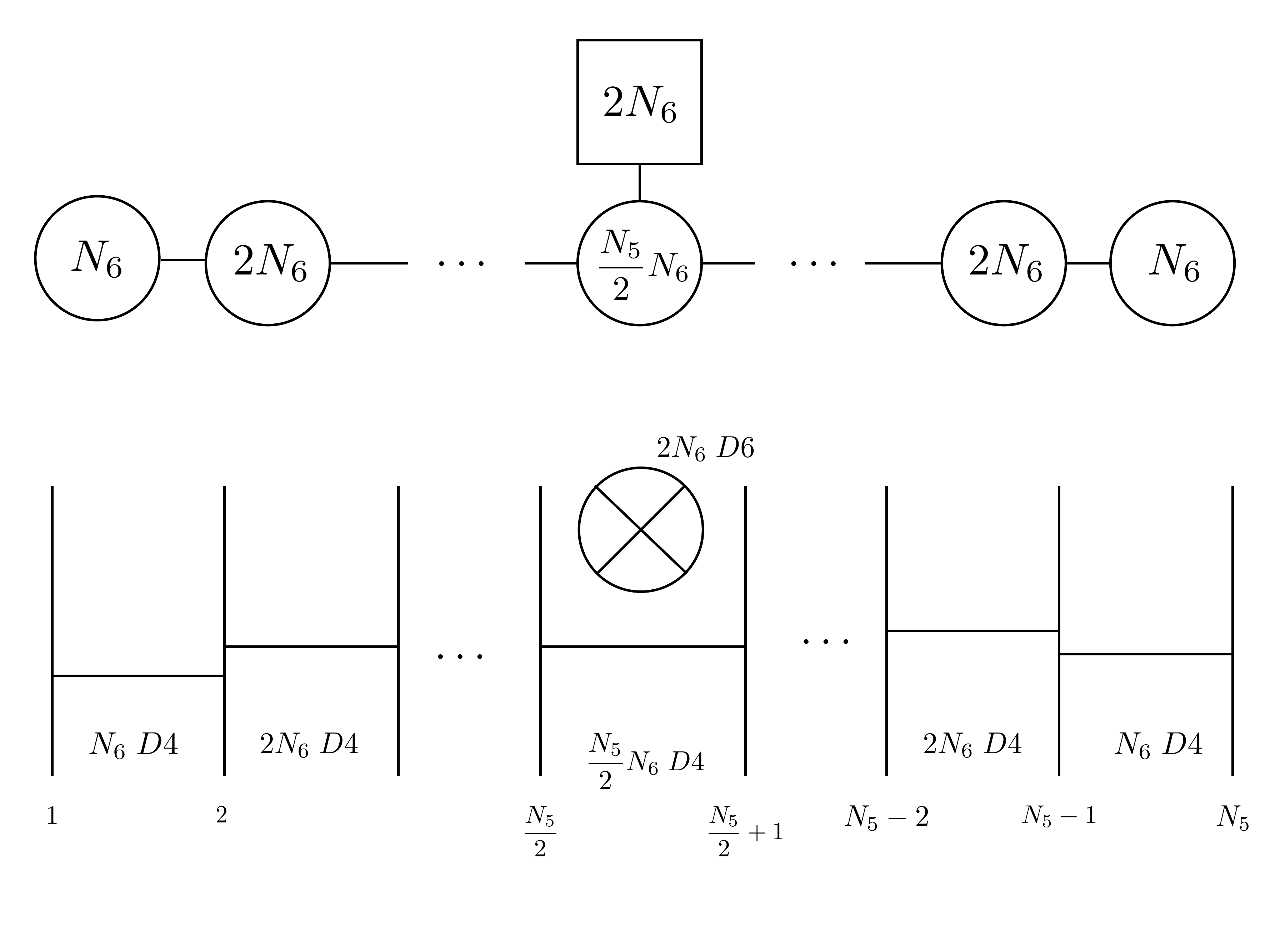} }}%

\caption{The quiver and Hanany-Witten set-up for the profile in eq.(\ref{profile3}).}
\label{Figure3}

\end{figure}

Let us start with a $\lambda$-profile given by,
\be \label{profile3}
\lambda(\eta)
                    =N_c\left\{ \begin{array}{ccrcl}
                       \eta & 0\leq\eta\leq \frac{N_5}{2} \\
                        (N_5-\eta) & \frac{N_5}{2}\leq\eta\leq N_5
                                             \end{array}
\right.
\e
The associated quiver 
and the Hanany-Witten set up are in Figure \ref{Figure3},
The number of D4 and D6 branes is ,
\begin{eqnarray}
& & N_{D4}=   \sum_{r=1}^{\frac{N_5}{2}} r N_6  +\sum_{r=1}^{\frac{N_5}{2}-1} N_6(\frac{N_5}{2}-r)= \frac{N_6 N_5^2}{4},\;\;\;\; N_{D6}=   2 N_6      .
\label{d4d6quiver3}
\end{eqnarray}
We can count the number of vectors and hypers and calculate the central charge,
\begin{eqnarray}
& & n_v=  \sum_{r=1}^{\frac{N_5}{2}} r^2 N_6^2-1  +\sum_{r=1}^{\frac{N_5}{2}-1} N_6^2(\frac{N_5}{2}-r)^2-1= \frac{N_6^2 N_5^3}{12} +\frac{N_5}{6}(N_6^2-6) +1  ,\nonumber\\
& & n_h=    \sum_{r=1}^{\frac{N_5}{2}-1} r(r+1) N_6^2  +N_5 N_6^2+ \sum_{r=0}^{\frac{N_5}{2}-1} N_6^2(\frac{N_5}{2}-r)(\frac{N_5}{2}-r-1)\! =\!\frac{N_6^2 N_5}{12}(N_5^2+8),\nonumber \\
& & c=    \frac{1}{48\pi}( N_6^2 N_5^3 +4 N_5(N_6^2-2) +8)\sim \frac{N_6^2 N_5^3}{48\pi}      .\label{quiver3}
\end{eqnarray}
We can check these values by performing the holographic calculations in eqs.(\ref{qd4good}),(\ref{qd6good}),(\ref{eqcc}). We find,
\begin{eqnarray}
& & N_{D4}=   \frac{2}{\pi}\mu^6 \int_{0}^{\eta_f} \lambda(\eta)d\eta= \frac{N_6 N_5^2}{4}, \;\;\;\; N_{D6}=-\mu^4 (\lambda'(\eta_f) -\lambda'(0)) = 2 N_6, \nonumber\\
& &  c=\frac{2}{\pi^4}\mu^{14}\int_0^{\eta_f}\lambda^2(\eta) d\eta=   \frac{N_6^2 N_5^3}{48\pi}         .
\end{eqnarray}
In agreement with the CFT values. 

Let us now compute the linking numbers for the Hanany-Witten set up in Figure \ref{Figure3}.
 Using the definition in eq. \eqref{linkingshw} we find
\begin{equation}\label{c8}
\begin{split}
K_{i}=&-N_6, \quad i=1,2...,N_5,\\
L_j=&N_{5}/2, \quad j=1,2,...,2N_6.
\end{split}
\end{equation}
We can easily see that eq. \eqref{consistency} is satisfied. Moreover, in the supergravity side we compute the linking numbers of the NS5 and D6 branes 
using eqs. \eqref{linkingNSgrav} and \eqref{linkingD6grav} and the $\lambda$ profile in eq. \eqref{profile3}. We find 
\begin{equation}
\sum_{i=1}^{N_5}K_i=\frac{2}{\pi}\mu^6\lambda'(\eta_f)\eta_f=-\frac{2}{\pi}\mu^6N_cN_5\equiv -N_6 N_5=-\sum_{i=1}^{ 2N_6}L_i.
\end{equation}

\subsection{Second example}
\begin{figure}[h!]
    \centering
    {{\includegraphics[width=12.5cm]{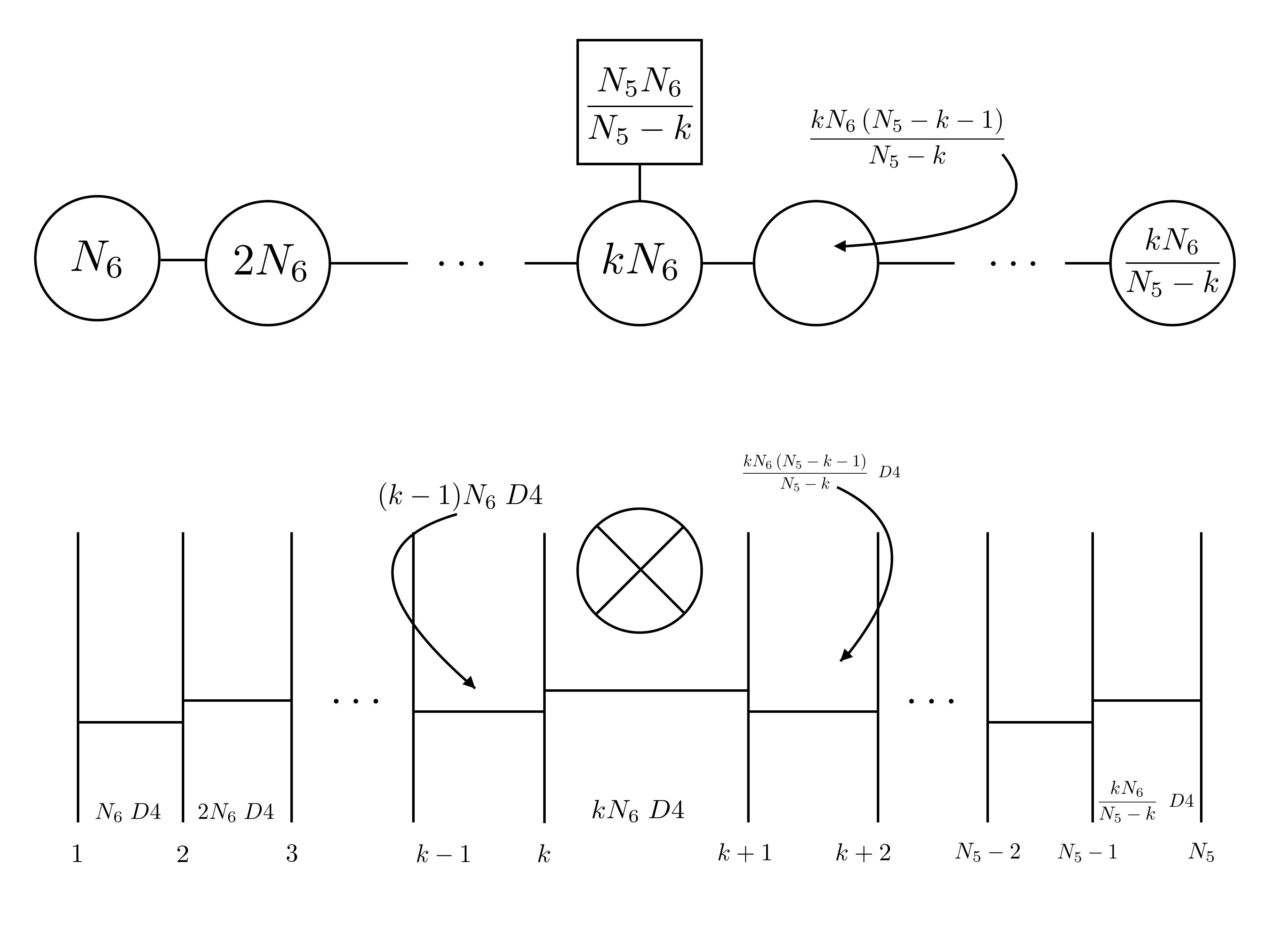} }}%
\caption{The quiver and Hanany-Witten set-up for the profile in eq.(\ref{profile4}).}
\label{Figure4}

\end{figure}

The  $\lambda$-profile is given by,
\be \label{profile4}
\lambda(\eta)
                    =N_c\left\{ \begin{array}{ccrcl}
                       \eta & 0\leq\eta\leq K \\
                       \frac{K(N_5-\eta)}{(N_5-K)} &K \leq\eta\leq N_5 .                                             \end{array}
\right.
\e
The associated quiver and the Hanany-Witten set up are drawn in Figure \ref{Figure4},
The number of D4 and D6 branes is,
\begin{eqnarray}
& & N_{D4}=   \sum_{r=1}^{K} N_6 r +\sum_{r=1}^{N_5-K-1} \frac{K N_6 (N_5-K-r)}{N_5-K}=\frac{N_6 N_5 K}{2} ,\;\;\;\;N_{D6}=  \frac{N_5 N_6}{(N_5-K)} \nonumber     .
\label{d4d6quiver4}
\end{eqnarray}
We can count the number of vectors and hypers and calculate the central charge,
\begin{eqnarray}
& & n_v=  \sum_{r=1}^{K} r^2 N_6^2 -1  +\sum_{r=1}^{N_5-K-1} \frac{K^2 N_6^2}{(N_5-K)^2}(N_5-K-r)^2-1  ,\nonumber\\
& &
n_v=\!\frac{1}{6(N_5-K)}\!\left[2 K^2 N_5^2 N_6^2 + K N_5 (N_6^2+6)-2 K^3 N_5 N_6^2 -6 N_5(N_5-1) - 6 K \right],\nonumber\\
& & n_h=   \left(\sum_{r=1}^{K} r(r+1) N_6^2\right)  +\left(\frac{K N_6^2 N_5}{N_5-K} +\frac{K^2 N_6^2}{N_5-K}(N_5-K-1)\right) \nonumber\\
& & + \left( \sum_{r=1}^{N_5-K-2} \frac{K^2 N_6^2}{(N_5-K)^2} (N_5-K-r)(N_5-K-r-1)\right)\nonumber\\
& &=\frac{N_6^2 K}{3(N_5-K)}\left[5 N_5- K^2(N_5+3) + K(N_5^2+3N_5-3)  \right].\nonumber\\
& & c=\frac{1}{12(N_5-K)}\left[K^2 N_6^2(N_5^2+N_5-1)+ 2 K(N_6^2 N_5+ N_5-1) - K^3 N_6^2(N_5+1) +2 N_5(N_5-1)  \right]\nonumber\\
& & c  \sim \frac{K^2 N_6^2 N_5}{12\pi}        .\label{quiver4}
\end{eqnarray}
We can check these values by performing the holographic calculations in eqs.(\ref{qd4good}),(\ref{qd6good}),(\ref{eqcc}). We find,
\begin{eqnarray}
& & N_{D4}= \frac{2\mu^2}{\pi}\mu^4 \int_{0}^{\eta_f} \lambda(\eta)d\eta= \frac{N_6 N_5 K}{2} ,\;\;\;N_{D6}=   \frac{N_6 N_5}{N_5-K}   ,\nonumber\\
& & c=  \frac{2\mu^{14}}{\pi^4}\int_0^{\eta_f}\lambda^2(\eta) d\eta= \frac{K^2  N_6^2 N_5}{12\pi}               .
\end{eqnarray}
The associated linking numbers for the Hanany-Witten set up in Figure \ref{Figure4} are 
\begin{equation}\label{c18}
\begin{split}
K_{i}=&-\frac{K N_6}{N_5-K}, \quad i=1,2...,N_5\\
L_j=&K, \quad j=1,2,...,\frac{N_5 N_6}{N_5-K}.
\end{split}
\end{equation}
We can easily see that eq. \eqref{consistency} is satisfied. Using the $\lambda$ profile in eq. \eqref{profile4}
and the expressions in eqs. \eqref{linkingNSgrav} and \eqref{linkingD6grav} the linking numbers of the NS5 and D6 branes are
\begin{equation}
\sum_{i=1}^{N_5}K_i=\frac{2}{\pi}\mu^6\lambda'(\eta_f)\eta_f=\frac{2}{\pi}\mu^6\frac{K N_c N_5}{K-N_5}\equiv \frac{k N_6 N_5}{K-N_5}=-\sum_{i=1}^{ N_5 N_6/N_5-K}L_i
\end{equation}

\subsection{Third example}
\begin{figure}[h!]
    \centering
    {{\includegraphics[width=12.5cm]{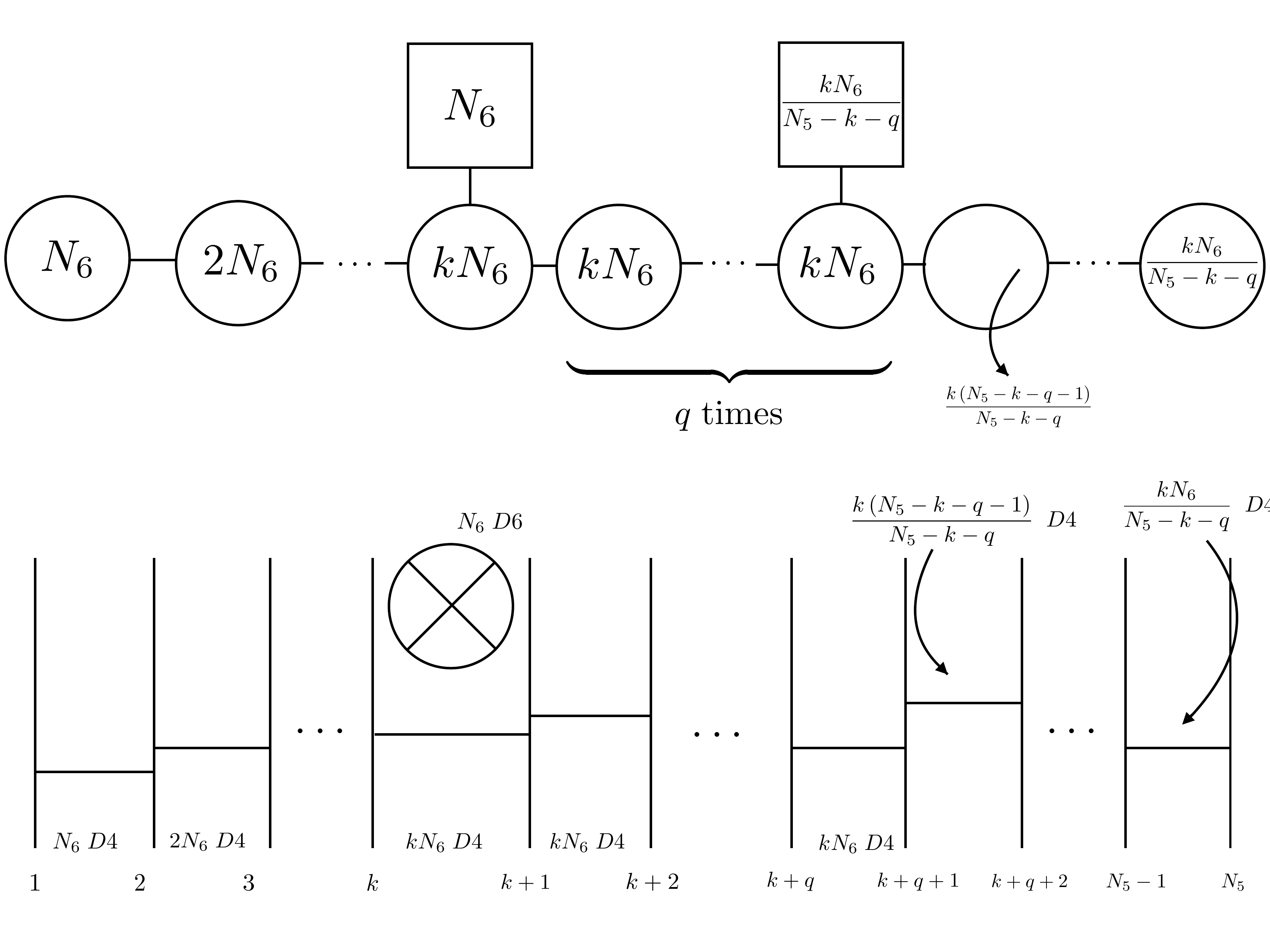} }}%
\caption{The quiver and Hanany-Witten set-up for the profile in eq.(\ref{profile5}).}
\label{Figure5}

\end{figure}

The $\lambda$-profile is given by,
\be \label{profile5}
\lambda(\eta)
                    =N_c\left\{ \begin{array}{ccrcl}
                       \eta & 0\leq\eta\leq K \\
                       K &K \leq\eta\leq K+q \\
                       K\frac{( N_5-\eta)}{N_5-K-q} & (K+q)\leq\eta\leq N_5
                                             \end{array}
\right.
\e
The associated quiver and the Hanany-Witten set up can be seen in Figure \ref{Figure5},
The number of D4 and D6 branes is ,
\begin{eqnarray}
& & N_{D4}= \left( \sum_{r=1}^K N_6 r \right) +K q N_6+ \sum_{r=1}^{N_5-K-q-1}\frac{N_6 K}{N_5-K-q} (N_5-K-q-r)\nonumber\\
& &= \frac{K N_6}{2}(N_5+q)     ;\nonumber\\
& & N_{D6}=   \frac{(N_5-q)N_6}{N_5-K-q}      .
\label{d4d6quiver5}
\end{eqnarray}
We can count the number of vectors and hypers and calculate the central charge,
\begin{eqnarray}
& & n_v=   \left( \sum_{r=1}^{K} r^2 N_6^2 -1 \right)+q (K^2N_6^2 -1) +\sum_{r=1}^{N_5-K-q-1} \frac{K^2 N_6^2(N_5-K-q-r)^2}{(N_5-K-q)^2}    \nonumber\\
& & =\frac{1}{6}\left(6+ 2 N_5 (K^2 N_6^2-3)+ K N_6^2 (1+4q+\frac{K}{N_5-K-q})   \right),\nonumber\\
& & n_h=   \left(\sum_{r=1}^{K} r(r+1) N_6^2\right)  +\left(K^2 N_6^2 q+ K N_6^2 +\frac{K^2 N_6^2 }{N_5-K-q} \right)+\nonumber\\
& & +\sum_{r=1}^{N_5-K-q-1}\frac{K^2 N_6^2}{(N_5-K-q)^2}(N_5-K-q-r)(N_5-K-q-r-1)   =\nonumber\\
& & n_h=\frac{K^2N_6}{3}\left(5+ K(N_5+2q +\frac{5}{N_5-K-q})  \right),\\
& & c=  \frac{1}{12\pi}\left[  2+ 2 K N_6^2+ N_5(K^2 N_6^2 -2) + 2 K^2 N_6^2(q+\frac{1}{N_5-K-q})\right]   \sim \frac{K^2 N_6^2 N_5}{12\pi} \nonumber    .\label{quiver3}
\end{eqnarray}
We can check these values by performing the holographic calculations in eqs.(\ref{qd4good}),(\ref{qd6good}),(\ref{eqcc}). We find,
\begin{eqnarray}
& & N_{D4}= \frac{K N_6}{2}(N_5+q)   ,\;\;\; N_{D6}=\frac{N_6(N_5-q)}{N_5-K-q},\;\;\;\; c=\frac{K^2 N_6^2 N_5}{12\pi}                 .
\end{eqnarray}

The linking numbers for the Hanany-Witten set up in Figure \ref{Figure5} are 
\begin{equation}\label{c27}
\begin{split}
K_{i}=&-\frac{KN_6}{N_5-K-q}, \quad i=1,2...N_5\\
L_j=&K, \quad j=1,2,...,N_6\\
L_n=&K+q, \quad n=1,2,...,KN_6/(N_5-K-q)
\end{split}
\end{equation}
We can easily see that eq. \eqref{consistency} is satisfied. The linking numbers of the NS5 and D6 branes 
using eqs.  \eqref{linkingNSgrav} and \eqref{linkingD6grav} and the $\lambda$ profile in eq. \eqref{profile5} are 
 
\begin{equation}
\sum_{i=1}^{N_5}K_i=\frac{2}{\pi}\mu^6\lambda'(\eta_f)\eta_f=\frac{2}{\pi}\mu^6\frac{K N_c N_5}{K+q-N_5}\equiv \frac{K N_6 N_5}{K+q-N_5}=-\left(\sum_{j=1}^{N_6}L_{j}+\sum_{n=1}^{ N_5 N_6/N_5-K-q}L_n\right)
\end{equation}

\section{Black Holes in Gaiotto Maldacena Backgrounds}\label{appendixBH}

In this section we will consider the generic Gaiotto-Maldacena class of geometries given in eq.(\ref{10d}) with a Schwarzschild black hole profile solution in the AdS sector. In particular, the background metric reads
\begin{equation}\label{10dblackening}
\frac{ds_{10}^2}{\alpha' \mu^2} = 4 f_1 \left( -r^2 g(r)dt^2 + \frac{dr^2}{r^2 g(r)} + r^2 d \vec{x}^2 \right) +  \frac{ds_{\text{int}}^2}{\alpha' \mu^2} \, ,
\end{equation}
where, as in eq.(\ref{10d}), $ds_{\text{int}}^2$ is given by
\begin{equation}
\frac{ds_{\text{int}}^2}{\alpha' \mu^2} = f_2 (d\sigma^2 + d\eta^2) +f_3 ds_{S^2}^2(\chi,\xi) + f_4 d\beta^2 \, ,
\end{equation}
while $g(r)$ is the blackening factor whose precise form is determined by the equations of motion. The functions $f_i (i=1 \dots 4)$ are still given in eq.(\ref{definitions1}), while $\vec{x}$ is a vector in $\mathbb{R}^3$.

The dilaton equation of motion gives a simple equation for the function $g(r)$,
\begin{equation}\label{blackening}
r^2 g''(r)+10 r g'(r)+20 g(r)-20 = 0 \, .
\end{equation}
The general solution for the equation (\ref{blackening}) is 
\begin{equation}
g(r) = 1 - \frac{c_1}{r^4} + \frac{c_2}{r^5} \, .
\end{equation}
The Einstein equations for the background metric (\ref{10dblackening}) force $c_2$ to be zero,  leaving $c_1$ undetermined. As usual, the potential $V(\sigma, \eta)$ appearing in the various functions $f_i$ still satisfies the same Laplace-like equation (\ref{toda}). In order to have a sensible black hole profile for the generic class of geometries we are considering, we will set $c_1$ to be $r_h^4$, with $r_h$ being the size of the horizon. The blackening factor $g(r)$ then takes the standard form
\begin{equation}
g(r) = 1 - \frac{r_h^4}{r^4} \, .
\end{equation}

It is now straightforward to compute the temperature of such a black hole. This is given by the general formula
\begin{equation}\label{temperature}
T = \frac{1}{2 \pi} \sqrt{- \frac{1}{4} g^{tt} g^{rr} (\partial_r g_{tt})^2} \, .
\end{equation}
Evaluating (\ref{temperature}) on the background (\ref{10dblackening}) we get
\begin{equation}
T = \frac{r_h}{\pi} \, .
\end{equation}

Let us now compute the entropy $S$ for this back hole solution. This is given by the standard BH relation 
\begin{equation}
S = \frac{A}{4} \, ,
\end{equation}
where $A$ is the area of the  black hole horizon. This reads
\begin{equation}
A = \int d^8 x \sqrt{\tilde{g}_8} \, ,
\end{equation}
where $d^8 x = d^3 \vec{x} d \sigma d \eta d \chi d \xi d \beta$ and $\tilde{g}_8$ is the determinant of the eight-dimensional subspace in Einstein frame.  It is easy to see that $S$ is given by
\begin{equation}\label{entropy}
S = 16 \pi^2 Vol(\mathbb{R}^3) r_h^3 \int d \sigma d \eta \sqrt{e^{-4 \phi} f_1^3 \det{g_{\text{int}}}} \, , 
\end{equation}
where $\det{g_{\text{int}}} = f_2^2 f_3^2 f_4$. Notice that the integrand in  eq.(\ref{entropy}) is the same as that in  eq.(\ref{eqcc}), and the one  studied in Appendix \ref{appendixEE}.  

in conclusion, being both the entropy and the central charge extensive quantities, and so counting degrees of freedom of the theory, they have the same dependence.

\section{Detailed construction of the deformed backgrounds}\label{appendixD}
In this appendix, we give details about the construction of our new backgrounds in Section \ref{sectionN=1}.
\subsection{The construction in eleven dimensions}
Here, we will derive the gamma-deformed background of Section \ref{section3} following the rules discussed in \cite{Gauntlett:2005jb}. Let us define the doublet
\begin{equation}
B^a=
\begin{pmatrix}
    A^a  \\
    -\frac{1}{2} \epsilon^{abc} C_{(1)bc} \\
\end{pmatrix} \, ,
\end{equation}
where $A^a$ and $C_{(1)bc}$ are defined in eq.(\ref{general11d}). For this particular background $C_{(2)}$ and $g_{\mu \nu} dx^{\mu}dx^{\nu}$ are invariant under gamma-deformation, while  $C_{(3)}$ is identically vanishing and therefore not subjected to any transformation. A non trivial transformation can possibly affect $A_{1}$, $C_{(0)}$ and $C_{(1)ab}$ as we discuss below.

According to the rules of \cite{Gauntlett:2005jb}, the doublet $B^a$ defined above transforms under gamma deformation in the following way
\begin{equation}
B^a \rightarrow \Lambda^{-T} B^a\, ,
\end{equation}
where $\Lambda \in SL(2, \mathbb{R})$ given by
\begin{equation}
\Lambda=
\begin{pmatrix} 
1 & 0 \\
\gamma & 1 
\end{pmatrix} \, .
\end{equation}
Here $\gamma$ is the parameter of the deformation. It is not difficult to see that the only (eight-dimensional) vector transforming is $A^a$. It transforms in the following way
\begin{equation}
A^a \rightarrow A^a = \frac{1}{2} \gamma \epsilon^{abc} C_{(1)bc}
\end{equation}
and in particular we have
\begin{equation}
A^1=0 \, , \quad A^2= -\gamma C_{(1)\xi y} \equiv -\gamma \kappa F_7 \sin \chi d \chi \, , \quad A^3= \gamma C_{(1)\xi \beta} \equiv \gamma \kappa F_6 \sin \chi d \chi \, .
\end{equation}
Moreover the $\tau$ parameter, defined as $\tau \equiv - C_{(0)}+i \hat{\Delta}^{1/2}$, undergoes a non trivial transformation given by $\tau \rightarrow \tau/(1 + \gamma \tau)$. This in turn implies
\begin{equation}
\hat{\Delta} \rightarrow \frac{\hat{\Delta}}{(1+\gamma^2 \hat{\Delta})^2}\, , \quad C_{(0)} \rightarrow - \frac{\gamma \hat{\Delta}}{1+\gamma^2 \hat{\Delta}} \, .
\end{equation}
Inserting these new definitions for the fields into the general eq. (\ref{deformed}) the background metric and the three-form $C_{3}$ take the form
\begin{equation}
\begin{aligned}
 \frac{ds^2}{\kappa^{2/3}}=&\left(1+\gamma^2\hat{\Delta}\right)^{1/3}\left(4 F_1 ds_{AdS_5}^2+F_2(d\sigma^2+d\eta^2)+F_3 d\chi^2\right)+\\
&\left(1+\gamma^2\hat{\Delta}\right)^{-2/3}\left(F_3 \sin^2\chi d\xi^2 +F_4\tilde{\mathcal{D}}\beta^2+F_5\left(\tilde{\mathcal{D}}y+\tilde{A}\tilde{\mathcal{D}}\beta\right)^2\right),\\
C_3=&\kappa \left(F_6\tilde{\mathcal{D}}\beta+F_7\tilde{\mathcal{D}}y\right)\wedge d\Omega_2(\chi,\xi)-\frac{\gamma \hat{\Delta}}{1+\gamma^2\hat{\Delta}}d\xi\wedge \tilde{\mathcal{D}}\beta\wedge \tilde{\mathcal{D}}y \, ,
\end{aligned}
\end{equation}
consistent with eq. (\ref{11ddef}).


\subsection{The TsT transformation of the Gaiotto-Maldacena solution in type IIB}\label{appendixIIB}

The purpose of this Appendix is to provide the details of the  construction of the TsT transformed GM solution studied in Section \ref{gammaIIBxx} following \cite{Lunin:2005jy}. The starting point is the type IIB 
solution in eq.(\ref{tIIB}) obtained by performing a T-duality on the GM solution of eq. (\ref{10d}) along the isometric $ \beta $ direction,
\begin{equation}\label{IIB}
\begin{split}
ds^2=&\alpha'\mu^2\left(4f_1 ds^{2}_{AdS_5}+f_2(d\sigma^2+d\eta^2)+f_3 (d\chi^2+\sin^2\chi d\xi^2)+f_4^{-1}d\beta^2\right),\\
&\qquad \qquad B_2=\mu^2 \alpha' f_5\sin \chi d\chi\wedge d\xi, \qquad e^{2\phi}=\frac{f_8}{\mu^2 f_4},\\
&\quad \qquad \qquad C_0=\mu^4 f_6,\qquad C_2=\mu^6 \alpha'f_7\sin \chi d\chi\wedge d\xi
\end{split}
\end{equation}

Moreover, the most generic configuration in IIB supergravity takes the form 
\begin{equation}
\begin{split}
ds^2_{IIB}=&\alpha'\mu^2\left(\frac{F}{\sqrt{\Delta}}(D\varphi^{1}-C D\varphi^{2})^{2}+F\sqrt{\Delta}(D\varphi^2)^2 +\left(\frac{e^{2\phi /3}}{F^{1/3}} \right)\mathcal{G}_{\mu \nu}dX^{\mu}dX^{\nu}\right),\\\ 
B=&\alpha'\mu^2\left(B_{12}D\varphi^1\wedge D\varphi^2+ \left[ B_{1\mu}(D \varphi^1)+B_{2\mu}(D\varphi^2)\right] \wedge dX^{\mu}-\frac{1}{2}A_{\mu}^{m}B_{m\nu}dx^{\mu}\wedge dx^{\nu}\right.\\
& \left.+\frac{1}{2}\tilde{b}_{\mu\nu} dx^{\mu}\wedge dx^{\nu}\right),\qquad e^{2 \phi_B}=e^{2 \phi},\\
C_{2}=&\alpha'\mu^6\left(C_{12}D\varphi^1\wedge D\varphi^2+ \left[ C_{1\mu}(D \varphi^1)+C_{2\mu}(D\varphi^2)\right] \wedge dX^{\mu}-\frac{1}{2}A_{\mu}^{m}C_{m\nu}dx^{\mu}\wedge dx^{\nu}\right.\\
& \left.+\frac{1}{2}\tilde{c}_{\mu\nu} dx^{\mu}\wedge dx^{\nu}\right),\qquad C_0 =\mu^4  A_0, \\
C_{4}=&\alpha'^2\mu^8\left(-\frac{1}{2}(\tilde{d}_{\mu\nu}+B_{12}\tilde{c}_{\mu\nu}-\epsilon^{mn}B_{m\mu}C_{\mu\nu}-B_{12}A_{\mu}^{m}C_{m\nu})dx^{\mu}\wedge  dx^{\nu}\wedge D\varphi^1 \wedge D\varphi^{2}\right.\\
&\left.+\frac{1}{6}(C_{\mu\nu\lambda}+3(\tilde{b}_{\mu\nu}+A_{\mu}^1B_{1\nu}-A_{\mu}^2B_{2\nu})C_{1\lambda})dx^{\mu}\wedge dx^{\nu}\wedge dx^{\lambda}\wedge  D\varphi^1+\right.\\
&\left.+d_{\mu_1\mu_2\mu_3\mu_4} dx^{\mu_1}\wedge dx^{\mu_2}\wedge dx^{\mu_3}\wedge dx^{\mu_4}+\hat{d}_{\mu_1\mu_2\mu_3} dx^{\mu_1}\wedge dx^{\mu_2}\wedge dx^{\mu_3}\wedge D\varphi^{2}\right),
\label{IIBS}
\end{split}
\end{equation}
where the indices $m,n=1,2$ and all the quantities above defining the fields in the solution are dimensionless quantities. The coordinates  $\varphi^{1,2}$ are the two isometric coordinates associated with the two-torus and 
\begin{equation}
D\varphi^1=d\varphi^1+\mathcal{A}^{(1)}_{\mu}dx^{\mu},\qquad D\varphi^2=d\varphi^2+\mathcal{A}^{(2)}_{\mu}dx^{\mu}.
\end{equation}

For the solution in eq. (\ref{IIB}) we identify $\varphi^1 = \beta ,~~\varphi^2 = \xi$. A direct comparison between (\ref{IIB}) and (\ref{IIBS}) leads to the following identifications
\begin{equation}
\begin{split}
 &\mathcal{G}_{\mu \nu}dX^{\mu}dX^{\nu} = (e^{-2 \phi/3}F^{1/3})\left(4f_1 ds^{2}_{AdS_5}+f_2(d\sigma^2+d\eta^2)+f_3 d\chi^2\right),\\
F =& \sqrt{\frac{f_3}{f_4}}\sin\chi ,~~\sqrt{\Delta}=\sqrt{f_3 f_4}\sin\chi ,~~e^{2 \phi_B}=e^{2 \phi}=\frac{f_8}{\mu^2 f_4},\\
&B_{2 \chi}=-f_5 \sin\chi ,~~C_{2\chi}=- f_7 \sin\chi ,~~C_0 =A_0 =  f_6,
\end{split}
\end{equation}
with the remaining quantities in the solution set to zero. 
We are now in a position to apply the standard TsT transformation rules \cite{Lunin:2005jy} to the type IIB background expressed above in eq.(\ref{IIB}). The $ SL(3,\mathbb{R}) $ transformation is applied with,
\[
   \Lambda=
  \left[ {\begin{array}{ccc}
   1 & \gamma & 0 \\
   0 & 1 & 0 \\
   0 & 0 & 1 \\
  \end{array} } \right]
\]
We then group the different components of the fields in the solution of eq. (\ref{IIBS}) according to their transformation under $ SL(3,\mathbb{R}) $. For the scalar sector,
he transformed fields are given in terms of the following matrix elements  \cite{Gursoy:2005cn},
\begin{equation}
\begin{split}
g^T_{11}=&\frac{e^{-\phi/3}}{F^{1/3}}\sqrt{1+ \gamma^2 F^2},~~g^T_{12}=\frac{\gamma e^{-\phi / 3}F^{5/3}}{\sqrt{1+ \gamma^2 F^2}},~~g^T_{22}=\frac{e^{-\phi/3}F^{2/3}}{\sqrt{1+\gamma^2 F^2}}\\
&\qquad \qquad  g^T_{31}=\frac{e^{2\phi/3}A_0}{F^{1/3}},~~g^T_{32}=0,~~g^T_{33}=\frac{e^{2\phi/3}}{F^{1/3}},\label{metric}
\end{split}
\end{equation}
 
In particular, the metric components and the dilaton transform according to
\begin{eqnarray}
F'&=& \frac{g^T_{22}}{g^T_{11}}=\frac{F}{1+\gamma^2 F^2}=\frac{ \sqrt{f_3 f_4}\sin\chi}{f_4+\gamma^2 f_3 \sin^2\chi},~~\Delta^{(TsT)}=\Delta =f_3 f_4 \sin^2 \chi\nonumber\\
e^{2\phi'}&=&\left( \frac{g^T_{33}}{g^T_{11}}\right) ^2=\frac{e^{2\phi}}{1+ \gamma^2 F^2}=\frac{f_8}{\mu^2 (f_4+\gamma^2 f_3 \sin^2\chi)}.
\end{eqnarray} 
Moreover, the non-zero components of the NS two-form,
\begin{equation}
B'=B'_{12}(D \varphi^1)'\wedge (D \varphi^2)'+B'_{2 \chi}(D \varphi^2)'\wedge d \chi, 
\end{equation}
have the following transformation rules
\begin{equation}
B'_{12}=\frac{g^T_{12}}{g^T_{11}}=\frac{\gamma F^2}{1+\gamma^2 F^2}=\frac{\gamma f_3 \sin^2\chi}{f_4+\gamma^2  f_3 \sin^2\chi},~~B'_{2\chi}=B_{2\chi}=- f_5 \sin\chi,
\end{equation}
whilst 
\begin{equation}
(D \varphi^1)' =d \beta + (\mathcal{A}^{1}_{\chi})'d\chi= d \beta -\gamma f_5 \sin\chi d\chi ,~~(D \varphi^2)'=d \xi.
\end{equation}

The RR potentials, on the other hand, could be formally expressed as,
\begin{eqnarray}
A'_0 &=& \left( \frac{g^T_{22}g^T_{11}}{g^T_{33}}\right)^{1/2}g^T_{31}=A_0 =  f_6 \nonumber\\
C'_{2}&=&C'_{12}(D \varphi^1)'\wedge (D \varphi^2)'+C'_{2 \chi}(D \varphi^2)'\wedge d \chi
\end{eqnarray} 
where the components of the 2-form RR potential transform as
\begin{eqnarray}
C'_{12}&=&A'_0 B'_{12} - g^T_{32}g^T_{22}g^T_{11}=\frac{\gamma  f_3 f_6 \sin^2\chi}{f_4+\gamma^2 f_3 \sin^2\chi},\nonumber\\C'_{2 \chi}&=&C_{2 \chi}=- f_7 \sin\chi.
\end{eqnarray}

The TsT transformed solution is given by eq. (\ref{IIBS}) by replacing the original fields by the transformed ones. The final result is the one given in eq. (\ref{10ddef}) in the main text.



\subsection{More comments about the CFTs}\label{moreongammacft}
In contrast with the ${\cal N}=2$ SUSY system in eqs.(\ref{10d})-(\ref{toda}), one characteristic  of the backgrounds in eqs.(\ref{gammaIIA}) and (\ref{10ddef}) is the presence of two types of Neveu-Schwarz five brane charges. In fact, as we calculated in Sections \ref{PagegammaIIA}, we find that aside from the $N_5$ NS-five branes, a new charge 
$\widehat{Q}_{NS5}$ is present only after the gamma-deformation takes place.

In the ${\cal N}=2$ system of Gaiotto-Maldacena, D6 sources act as flavour branes while the  D4s are colour branes. After the gamma-deformation we encounter both D4 branes in Type IIA and D5 branes in Type IIB realising the colour group. In Type IIA D6 or D7 branes in Type IIB give place to the flavour group. This is reminiscent of the so called {\it Brane-Box models}. Let us first study the Type IIB version.
\begin{table}[h!]
\centering
 \begin{tabular}{c |c c c c cc cc c c}
  & $R^{1,3}$  & $x_4$ & $x_5$ & $x_6$ & $x_7$& $x_8$& $x_9$ \\ [0.5ex] 
 \hline 
 NS & $\line(1,0){10}$ & $\line(1,0){10}$ & $\line(1,0){10}$ & $A$ & $O$ & $O$ & $O$  \\ 
  $\widehat{\textrm{NS}}$ & $\line(1,0){10}$ & $A{}$ & $O$ & $\line(1,0){10}$ & $\line(1,0){10}$ & $O$ & $O$  \\
  D5 & $\line(1,0){10}$ & $\line(1,0){10}$ & $O$ & $\line(1,0){10}$ & $O$ & $O$ & $O$  \\ 
\end{tabular}
\caption{Brane set up for the original brane box model. The $A$'s  denote that the branes can be placed at an arbitrary 
position in the corresponding direction  whilst the circle means all branes sit at the same point.}
\label{figurita1}
\end{table}

Introduced in \cite{Hanany:1998ru} and further studied in \cite{Hanany:1998it}, the brane boxes consist of a Type IIB  array of $k$ NS-five branes, $\hat{k}$ $\hat{NS}$ five branes and D5  branes. The positions of the branes is given in Table \ref{figurita1}. In these set-ups the D5 branes fill the $(x_4,x_6)$ plane. We have $(k+1)\times(\hat{k}+1)$ boxes, of which $(k-1)\times(\hat{k}-1)$ have finite area corresponding with the gauge groups with non-zero gauge couplings, according to
\begin{equation}
\frac{1}{g_{4,\alpha,\hat{\alpha}}^2}=\frac{Vol[Box]}{g_6^2}=\frac{(x_{4,\alpha+1}- x_{4,\alpha})(x_{6,\hat{\alpha}+1} -x_{6,\hat{\alpha}})}{g_s \alpha'}.\nonumber
\end{equation}
The gauge group is $G=\Pi_{\alpha=1}^{k-1} \Pi_{\hat{\alpha}=1}^{\hat{k}-1}SU(k_{\alpha,\hat{\alpha}})$, 
being $k_{\alpha,\hat{\alpha}}$ the number of D5 branes in the $[\alpha,\hat{\alpha}]$ box. 
The flavour group is represented  by  semi-infinite D5 branes in the boundaries of the system. By a Hanany-Witten move they  transform
into D7 branes. There are three types of fields for each box, called $H,V,D$ that connect boxes along the horizontal, vertical and diagonal directions respectively. 
In the case of finite theories with vanishing beta functions and anomalous dimensions
the superpotential is cubic and schematically of the form ${\cal W}=h Tr [HVD]$---see \cite{Hanany:1998ru} for details.
Comparing with our set-up in Section \ref{gammaIIBxx}, we see that the systems share common characteristics. 

On the other hand, the system in Section \ref{section3} can be put in correspondence with the work \cite{Hanany:1998it}. The system contains two types of NS-five branes and D4 branes, by a Hanany-Witten move also flavour D6 branes appear, as in Section \ref{section3}. The Hanany-Witten set-up is shown in Table \ref{figurita2}.

\begin{table}[h!]
\centering
 \begin{tabular}{c ||c c c c cc cc c c}
  & $R^{1,3}$  & $x_4$ & $x_5$ & $x_6$ & $x_7$& $x_8$& $x_9$ \\ [0.5ex] 
 \hline\hline 
 NS & $\line(1,0){10}$ & $\line(1,0){10}$ & $\line(1,0){10}$ & $O$ & $O$ & $O$ & $O$  \\ 
 \hline\hline 
  $\overline{\textrm{NS}}$ & $\line(1,0){10}$ & $O$ & $O$ & $O$ & $O$ & $\line(1,0){10}$ & $\line(1,0){10}$  \\
 \hline\hline 
  D4 & $\line(1,0){10}$ & $O$ & $O$ & $\line(1,0){10}$ & $A$ & $O$ & $O$  \\ 
\end{tabular}
\caption{Brane content of the Type IIA brane box model. $A$'s denote that the branes can be placed at an arbitrary position in the corresponding direction whilst the circle means all branes sit at the same point.}
\label{figurita2}
\end{table}


\end{document}